\documentclass[12pt]{article}
\usepackage[T1]{fontenc}
\usepackage[utf8]{inputenc}
\usepackage{authblk}
\usepackage{booktabs}
\usepackage{setspace}

    \usepackage{amsfonts}
    \usepackage{amssymb}
    \usepackage{amsbsy}
    \usepackage{amsmath}
    \usepackage{amsthm}
    \usepackage{bbm}
    \usepackage{bm}
    \usepackage{mathtools}
    \usepackage{caption}
    \usepackage{subcaption}    \newcommand\independent{\protect\mathpalette{\protect\independenT}{\perp}}
\def\independenT#1#2{\mathrel{\rlap{$#1#2$}\mkern2mu{#1#2}}} 
\newcommand{\lspace}[1]{\renewcommand{\baselinestretch}{#1} \small\normalsize}

    \newcounter{subreq}[req]
	
	\makeatletter
	\renewcommand{\p@subreq}{\thereq}
	\makeatother

    \newtheorem{assum}{Assumption}
\newcounter{subassumption}[assum]
	\renewcommand{\thesubassumption}{(\textit{\roman{subassumption}})}
	\makeatletter
	\renewcommand{\p@subassumption}{\theassum}
	\makeatother
	\newcommand{\subasu}{
  \refstepcounter{subassumption}%
  \thesubassumption~\ignorespaces}

	\newtheorem*{obs*}{Observation}

    \usepackage[flushleft,online,para]{threeparttable}
	\usepackage{graphics}
\usepackage[vlined,linesnumbered,ruled,resetcount]{algorithm2e}
\SetKwInOut{Initialization}{Initialization}

    \graphicspath{{Graphs/}}
\usepackage{adjustbox}
\usepackage{multirow}

    \usepackage{tikz}
    \usetikzlibrary{arrows}
    \usetikzlibrary{decorations.pathreplacing}
    \usepackage{pgfplots}
\usepackage{indentfirst}             
\usepackage{floatrow}
    \usepackage{tabularx}
        \newcolumntype{Z}{>{\centering\arraybackslash}X}
        \newcolumntype{L}{>{\raggedright\arraybackslash}X}
    \usepackage{dcolumn}
        \newcolumntype{d}[1]{D{.}{.}{#1}}
    \usepackage{rotating}                     
    \usepackage{lscape}
    \usepackage{pdflscape}   
\usepackage{authblk}

	\usepackage{csquotes}
	\usepackage[round]{natbib}
	\usepackage{xurl}

    \usepackage{xcolor}
        \definecolor{darkblue}{rgb}{0,0,0.4}
    \usepackage{hyperref}
        \hypersetup{
            colorlinks = true,
            linkcolor = darkblue,
            citecolor = darkblue,
            pdfborder = 0 0 0,
            pdfdisplaydoctitle = true,
            pdfhighlight = /N,
            pdfpagelayout = OneColumn,
            pdfpagemode = UseNone,
            pdfstartview = {FitH},
            pdfauthor = {{AB, CF \& JR}},
            pdftitle = {{null}},
            pdfsubject = {{}}
        }

    \usepackage[textsize=footnotesize, colorinlistoftodos, textwidth=4cm, obeyDraft]{todonotes}
    \usepackage{geometry}
    \usepackage{setspace}
    \usepackage[bottom, multiple]{footmisc}  
    \usepackage{verbatim}
    \usepackage[normalem]{ulem}     
    \usepackage{mathpazo}

    \usepackage[toc,page]{appendix}
	\usepackage{lipsum}

\usepackage{minitoc}

\begin{document}

\vspace{-.5in}

\title{Multi-cell experiments for marginal treatment effect estimation of digital ads\thanks{\enspace This paper was previously circulated under the title ``A multi-cell experimental design to recover policy relevant treatment effects, with an application to online advertising.'' We thank the editor, Jean-Pierre Dub\'{e}, the associate editor, and three anonymous reviewers for detailed comments and feedback; Ivan Canay, Gast\'{o}n Illanes, Garrett Johnson, TI Kim, and Ilya Morozov; and seminar participants at Michigan Ross, MIT Sloan, the 2023 Bass FORMS Conference, the 2023 Marketing Science Conference, the 2023 Workshop on Institutions, Economic Behaviour and Economic Outcomes, the 2023 12th Triennial Choice Symposium, and the 2023 Annual MIT Conference on Digital Experimentation for helpful comments. Gordon holds concurrent appointments at Northwestern and as an Amazon Scholar. This paper describes work performed at Northwestern and is not associated with Amazon. E-mail addresses for correspondence: caio.waisman@kellogg.northwestern.edu, b-gordon@kellogg.northwestern.edu.}}

\author{Caio Waisman \qquad Brett R. Gordon \\
Kellogg School of Management\\
Northwestern University \\
}

\date{May 28, 2024}

\maketitle
\thispagestyle{empty}

\lspace{1}

\begin{abstract}

\noindent Randomized experiments with treatment and control groups are an important tool to measure the impacts of interventions. However, in experimental settings with one-sided noncompliance, extant empirical approaches may not produce the estimands a decision-maker needs to solve their problem of interest. For example, these experimental designs are common in digital advertising settings, but typical methods do not yield effects that inform the intensive margin---how many consumers should be reached or how much should be spent on a campaign. We propose a solution that combines a novel multi-cell experimental design with modern estimation techniques that enables decision-makers to solve problems with an intensive margin. Our design is straightforward to implement and does not require additional budget. We illustrate our method through simulations calibrated using an advertising experiment at Facebook, demonstrating its superior performance in various scenarios and its advantage over direct optimization approaches.

\vspace{.2in}
\noindent


\end{abstract}

\newpage
\setcounter{page}{1}
\lspace{1.1}

\doparttoc 
\faketableofcontents 
\part{} 

\vspace{-0.7in}

\section{Introduction}

Randomized experiments with treatment and control groups are frequently used to measure the causal effects of interventions. However, decision-makers often need more than just these measurements to inform specific decisions. 

Consider a firm that wants to measure the effectiveness of its digital ad campaigns but faces limited control over treatment assignment. A common experimental approach is to randomize eligibility for treatment, assigning some consumers as eligible or ineligible to see the ads. This method results in one-sided noncompliance: not all eligible users will see the ads, while ineligible users cannot see them at all. This approach is popular for measuring online advertising effects \citep{johnson2022} and is also used in economics, political science, and medicine.\footnote{Experiments with one-sided noncompliance have been used in the context of: online A/B tests \citep{dys2021}; clinical trials \citep{sz1991}; breast self-examination treatments \citep{mifb2004}; interventions to incentivize voter turnout \citep{ggn2003}; effects of job training \citep{sbm2008} and job assistance \citep{cdgrt2013} programs; the impacts of access to microcredit \citep{cddp2015}; the effects of deworming drugs on children's health and education \citep{mk2004}; and housing voucher policies \citep{chk2016}.} 

Such an experimental design provides valuable information, namely the average treatment effect on the treated (ATT). The ATT quantifies the effect of the treatment on the observable subset of units that did receive treatment; alternatively, it quantifies the loss that would have been experienced had the experiment not been conducted. Compared against costs, it helps a decision maker assess whether the current policy is beneficial. 

However, this treatment effect parameter is of limited assistance when it comes to intensive margin decisions, such as helping an advertiser decide how many consumers to reach or how large a budget to set.

In this paper, we propose an approach that allows the researcher (and the decision-maker) to obtain the information necessary to make these decisions. Our approach combines a novel multi-cell experimental design with modern estimation techniques to recover the marginal treatment effect (MTE) function. This function allows us to inform these decisions and to recover the most common treatment effect parameters of interest, including the ATT. We illustrate our approach through a series of simulations that are calibrated using an advertising experiment at Facebook so that they represent real world environments as close as possible. We show that our proposed experimental design yields accurate estimates of the MTE function. 

To introduce our setting, we consider an advertiser deciding what fraction of users to reach with advertising from among a target audience.\footnote{In Section \ref{sec:dp2} we explain why this problem is equivalent to one where the advertiser sets a budget level.} Using this decision problem, we describe the typical experimental design with one-sided noncompliance and show that the MTE function is the object necessary to solve this decision problem. 

Our empirical approach to recover the MTE function is inspired by the estimation method in \cite{bmw2017}, who show how to obtain a polynomial approximation using a discrete instrumental variable that generates variation in the probability of treatment. However, unlike \cite{bmw2017}, our empirical approach is specifically structured to take advantage of an experimental design under one-sided noncompliance that, by construction, yields a suitable instrumental variable.\footnote{In fact, direct application of \cite{bmw2017} to a single-cell experiment with one-sided noncompliance is infeasible because the data contain too few moments to recover the MTE approximation. In Appendix D, we explain how our experimental design solves this underidentification problem.}

Specifically, our design first randomly allocates units across $C$ cells, and then units are once again randomly split into test and control groups within each cell. Consistent with typical limitations---especially those in online advertising---around treatment assignment in practice, each cell features an experiment with one-sided noncompliance. The advertiser, or the ad platform acting on their behalf, sets exposure rates across cells to generate the necessary variation. We show how this multi-cell design yields a sufficient number of moments to approximate the MTE function using a polynomial of degree $C$. 

We apply our method to data generating processes (DGPs) calibrated to an advertising experiment at Facebook.\footnote{We could apply our method to data from a multi-cell experiment if we had access to such data.} We consider cubic polynomials for the MTE function and show that our method can perform well in approximating it.

To determine the optimal fraction of target consumers to expose, we propose a Bayesian decision theoretic framework that accounts for uncertainty due to estimation. We find that our approach succeeds in virtually eliminating any losses in expected profits across different DGPs. 

An alternative strategy for this decision problem is direct budget optimization, which could be achieved by randomizing budget levels across cells and tracking profits, circumventing the need to perform causal inference. While simpler, direct optimization fails to exploit the underlying structure of the expected revenue function, namely its connection to the MTR functions. The MTR functions are structural and thus more stable objects because they reflect the relationship between ad exposure and user behavior. This renders our proposed approach more robust to changes in the cost function.\footnote{The assumption of stable MTR functions most reasonably applies to a future campaign with the same audience targeting criteria on the same ad platform. However, it is less likely to hold across ad platforms or under significantly different targeting parameters.}

To facilitate our method's adoption, Section \ref{sec:implem} discusses various considerations and obstacles that might arise in the practical implementation of our design. We start with a discussion of the model's inputs and the requirements that the experimentation platform must satisfy. Next, we highlight that many campaigns generate multiple ad exposures for a given user, whereas our model assumes binary treatment. This leads to a failure of the exclusion restriction necessary for our model to approximate the MTE. We address this issue by extending our model to accommodate multiple exposures in Appendix C. We also discuss the conditions under which ignoring multiple exposures, as we do in the body of the paper, may not significantly impact the quality of the approximation. Finally, we provide some intuition on generating the necessary variation in the propensity score---the probability of treatment given eligibility---across experimental cells to obtain the approximation. Even though researchers cannot directly control treatment assignment, they can influence the probability of treatment by adjusting the budget per user. Although generating variation in the propensity score is straightforward, we are unable to provide concrete guidance on how \textit{best} to choose the number of cells $C$ or the budget per cell, except under strong functional form assumptions. Critically, our method does not require an increase in the overall budget compared to a single-cell test.

Our paper makes three contributions. First, we contribute to the broad literature on estimating treatment effect parameters in experiments with one-sided noncompliance. In particular, we develop an experimental design that is built to leverage modern estimation techniques when only eligibility to receive treatment can be randomized. These techniques have their origins in the work of \cite{bm1987} and \cite{hv2005}, who showed identification of the MTE function using a continuous instrumental variable with observational data. More recently, recognizing that instruments are often discrete, \cite{bmw2017} showed how to recover polynomial MTE functions, or, equivalently, how to recover a polynomial approximation to the MTE function, whereas \cite{mst2018} showed how to obtain partial identification of the MTE function.\footnote{A common alternative is to impose specific structure on the DGP, often via a normality assumption, which aids in identification and estimation of the MTE function. We discuss this approach in more detail in Appendix C.} Neither study considers how these methods can be used in combination with experimental data specifically, and in particular when the design of the experiment can be altered to enhance estimation. 

This is our primary contribution: to tailor the experimental design to exploit these estimation methods. Importantly, our proposed design can potentially be used in any situation where a design with one-sided non-compliance can be implemented, not just in the context of online advertising, which may be of independent interest to researchers and decision-makers.

Second, we add to the expanding literature on estimating online advertising effects. Much of this work focuses on recovering the intent-to-treat (ITT) or ATT parameters using experiments with one-sided noncompliance.\footnote{Examples include \cite{lr2014, bgkrs2015, jlr2016, jln2017a, jlr2017, gordon_zettelmeyer_2019, snk2019, bb2021, gnn2021}; and \cite{gordon_moakler_zettelmeyer_2022}.} Obtaining such estimates is useful to document advertising effects and to inform an advertiser's extensive margin decision of whether to advertise (a ``go/no go'' decision). However, an advertiser is unable to apply these estimates to choose the intensive margin of how many consumers to reach with advertising. A recent exception is \cite{hm2022}, who propose an asymmetric budget splitting design to measure the returns to advertising. This work is distinct from ours in that it randomizes the budget levels across treatments without an explicit control group, and then estimates the returns to ad spend using a linear regression. Our paper makes a contribution by helping to fill this gap in the literature using an approach that embeds a causal inference setup in the advertiser's optimization problem. 

To the best of our knowledge, the only other paper that uses the MTE framework in marketing is \cite{dmrsy2022}, who apply \cite{mst2018} to data from a promotion targeting experiment with two-sided noncompliance conducted with a hotel chain. Like us, the authors outline a precise decision problem and extrapolate over the MTE to solve it, although with a different estimation approach. \cite{dmrsy2022} condition on the experimental design and explore how different identifying assumptions combined with alternative estimators affect optimal decisions. We view our approach, which does not condition on the experimental design but rather tailors it based on the estimator of interest and under minimal assumptions, as complementary to theirs.

Third, our paper is related to work that examines an advertiser's decision problem. Early work in this area sought to determine the optimal budget allocation given an aggregate advertising response model \citep{sethi1977, ha1982, simon1982, bb1988}. More recent work studies this problem in online advertising settings \citep{prs2017, bfpp2019, zhyzxy2019, gswznl2021}. However, none of these papers have causal inference in mind. \cite{wsnl2022} provides a framework to recover treatment effect parameters that account for parallel experimentation by competitors to inform an advertiser's extensive margin decision. A different strand of this literature connects causal inference with advertising decisions, specifically a firm's optimal bidding strategy in real-time bidding (RTB) environments \citep{lewis_wong_2018,wnc2022}. Neither of these papers obtain the MTE, which is unnecessary for the decision problems they study. With the MTE, we can solve a broader set of advertising decision problems, though our method does not account for other experimentation costs that would be relevant in RTB settings. Furthermore, we show how the practitioner can adopt a Bayesian decision theoretic framework in a straightforward manner to solve these problems while accounting for estimation uncertainty.

The rest of this paper proceeds as follows. Section \ref{sec:setting} introduces the typical experimental design through the advertiser's decision problem and shows this design does not provide the information needed to solve this problem. Section \ref{sec:empirical} presents our empirical strategy that consists of a novel multi-cell experimental design and an estimation technique to recover an approximation to the MTE function. Section \ref{sec:emp_app} uses data from Facebook advertising experiments to illustrate the benefits of our methodology. Section \ref{sec:implem} presents a discussion of several practical considerations and challenges when implementing our design in practice. Section \ref{sec:conc} concludes.

\section{Setting} \label{sec:setting}

We consider a firm's decision to select the optimal fraction of a target audience they should advertise to in order to maximize expected profits. We focus on this problem because it enables us to introduce our model, describe the typical experimental design with one-sided noncompliance, and explain why the treatment effect parameter obtained from this design (the ATT) does not suffice for the decision-maker to solve this problem.

\subsection{Firm's advertising problem} \label{sec:dp1}

A firm wishes to choose the \textit{fraction} of consumers from a target audience in a specific advertising market (e.g., Facebook) to reach with advertising to maximize expected profit. As we show below, this decision is equivalent to choosing an advertising budget to reach a given proportion of consumers. However, presenting the advertiser's problem treating this fraction as the decision variable, instead of the budget, yields a simpler analysis.

For ease of exposition, we treat all consumers within the target audience segment as \textit{observationally equivalent}. That is, all their characteristics that are observed by the platform and the advertiser, which can be encapsulated in a vector, $X$, are the same across these consumers---the target segments themselves can be defined by such variables. Alternatively, all the statements throughout can be interpreted as conditional on $X$. We show how to explicitly incorporate $X$ into the analysis and presentation in Appendix C.

Let $D$ be an indicator for whether a unit (consumer) is treated, $Y_1$ be the outcome when $D=1$, and $Y_0$ be the outcome when $D=0$. The observed outcome can be written as:
\begin{align}\label{eq:obs_out}
	Y &= DY_1 + (1-D)Y_0 \ . 
\end{align}
In our setting, $D$ represents exposure to the advertising campaign, and we assume consumers are exposed to at most one ad. Although exposure frequency is typically higher in practice, the extant literature has found, at best, weak effects of repeated exposures on purchases, which is our outcome of interest in our simulations. Furthermore, as we discuss in Section \ref{sec:implem:multi} and show in Appendix C, ignoring exposure frequency may have negligible effects on MTE estimation and budget recommendations.

Let $\phi$ be the fraction of units exposed to the treatment. Let the cost of treating a fraction $\phi$ of units be given by a known cost function, $\kappa(\phi)$. These costs represent the (expected) cost for impressions that the platform delivers to the advertiser. In practice, we expect $\kappa(\phi)$ to be convex and increasing to capture the notion that reaching the marginal consumer becomes more expensive as overall campaign reach increases. The advertiser's expected profit maximization problem is:
\begin{align}\label{eq:dec_problem1}
	\max_{\phi\in[0,1]} \left ( \delta \times \left \{ \phi \mathbb{E} \left [Y \middle \vert  D=1 \right ] + (1-\phi) \mathbb{E} \left [Y \middle \vert  D=0 \right ] \right \} - \kappa(\phi) \right ) 
\end{align}
where $\delta$ is a known constant that converts outcomes into monetary amounts.

In the context of online advertising, treating $\delta$ and $\kappa(\cdot)$ as known quantities is reasonable. Advertisers know $\delta$ because it represents the value of an online conversion for their business. Platforms, including Facebook, often present $\kappa(\cdot)$, or its inverse, to advertisers when they set up their campaigns.\footnote{The Meta Campaign Planner provides a forecast of audience size as a function of the campaign's total budget (\url{https://www.facebook.com/business/help/907925792646986?id=842420845959022}). Other ad platforms provide similar campaign planning tools, e.g., TikTok (\url{https://ads.tiktok.com/help/article/reach-frequency-campaign-forecaster?redirected=2}) and The Trade Desk (\url{https://www.thetradedesk.com/us/our-platform/dsp-demand-side-platform/plan-campaigns}). All links accessed on 12/23/2023.} These objects are not specific to the experiment; they exist in the normal advertising campaign environment. They are, however, specific to a platform, such that the cost and value of reaching users on Facebook may differ from those on, say, TikTok.

To solve the optimization problem in (\ref{eq:dec_problem1}), the advertiser needs to compute unknown conditional expectations. We now discuss how these objects can be estimated.

\subsection{Experimental design}

There are several methods to estimate the conditional expectations in expression (\ref{eq:dec_problem1}) from data; arguably, one preferred way to collect these data is by running an experiment, ideally one in which treatment itself is randomly assigned to the experimental units. However, often the experimenter, in this instance the advertiser, does not fully control treatment assignment and therefore cannot randomize it. The most common solution is to instead randomize eligibility to receive treatment, which is the experimental design we address.\footnote{If an experiment is infeasible, the advertiser could apply a model to observational data. However, lacking an exogenous source of variation on treatment, it may be difficult to reliably estimate treatment effect parameters due to unobservable confounds that are correlated with both treatment and outcomes \citep{gordon_zettelmeyer_2019,gordon_moakler_zettelmeyer_2022}.}

Let $Z$ be an indicator for whether the unit is \textit{eligible} to receive treatment, which we assume is randomly assigned. Following \cite{hv2005}, treatment is given by:
\begin{align}\label{eq:sel_eq}
D&=\mathbbm{1} \left \{p(Z)\geq U \right \},
\end{align}
where $p(\cdot)$ governs the process of selection into treatment. While $p(\cdot)$ is common to all users within an eligibility condition, variation across users in the unobservable $U$ creates user-specific exposure to ads. Consequently, $U$ can be interpreted as the ease with which the advertiser can expose the user to their ads due to unaccounted factors, such as an individual-specific propensity to be active on the platform. Since the objects in equations (\ref{eq:obs_out}) and (\ref{eq:sel_eq}) vary across individuals, this is a model with essential heterogeneity in the sense of \cite{huv2006}.

Following \cite{mst2018}, we maintain the following standard assumption.
\begin{assum}\label{assum:basic} \hfill\break
	\subasu $U \independent Z$, where $\independent$ denotes statistical independence. \label{assum:uncorr} \\
	\subasu $\mathbb{E} \left [Y_d \middle \vert Z,U \right ] = \mathbb{E} \left [Y_d \middle \vert U \right ]$ and $\mathbb{E} \left [Y_d^2 \right ]< \infty$ for $d\in\{0,1\}$. \label{assum:excl} \\
	\subasu $U$ is continuously distributed. \label{assum:cont} 
\end{assum}

Assumptions \ref{assum:uncorr} and \ref{assum:excl} require $Z$ to be exogenous with respect to the selection and outcome processes, thereby characterizing it as a valid instrumental variable for the treatment indicator, $D$.\footnote{As noted in the Introduction, this exclusion restriction is likely to fail in the presence of multi-valued treatments, such as multiple ad impressions. Section \ref{sec:implem:multi} briefly discusses this issue and Appendix C presents an extended version of the model that resolves this issue by explicitly conditioning on the number of prior treatments.} Assumption \ref{assum:uncorr} holds by construction in our setting due to the randomized experimental design. Given Assumption \ref{assum:uncorr}, \cite{vytlacil2002} showed that the assumption that the index of the selection is additively separable, as in equation (\ref{eq:sel_eq}), is equivalent to the monotonicity condition from \cite{ia1994}. Finally, Assumption \ref{assum:cont} is a weak regularity condition that allows us to normalize $U \sim \textrm{Uniform}(0,1)$.  Under these conditions, this model is equivalent to that of \cite{ia1994}. Assumption \ref{assum:basic} allows us to define the \textit{propensity score} as the probability of treatment given eligibility:
\begin{align} \label{eq:pscore}
p(Z) = \Pr \left (D=1 \middle \vert Z \right ) .
\end{align}
Since $D$ and $Z$ are observed in the data, it is straightforward to estimate $p(\cdot)$.\footnote{In reality, we recognize that platforms allow advertisers to target users based on observables $X$, but these platforms often do not report their results at the same granular level. In this case, the platform could implement our empirical methodology on the advertiser's behalf, or the platform could report the appropriate objects after integrating over $X$.}

Figure \ref{fig:exp_design} illustrates this experimental design, which can be viewed as a single cell. Within this cell, units are randomly assigned to $Z=1$ or $Z=0$. Since $p(1) \in (0,1)$, some, but not all units that are eligible to receive treatment are actually treated ($D=1$)---left column of Figure \ref{fig:exp_design}---, and because $p(0)=0$, none of the units that are ineligible to receive treatment are treated ($D=0$)---right column of Figure \ref{fig:exp_design}. 

\begin{figure}
   \centering
\begin{tikzpicture}[scale=.6]
\draw [fill=gray!100] (-4,0) to (0,0) to (0,-4) to (-4,-4);
\draw [fill=gray!25] (-4,-4) to (0,-4) to (0,-8) to (-4,-8);

\draw [ultra thick][-] (-4,0) -- (4,0);
\draw [dashed, ultra thick][-] (-4,-4) -- (0,-4);
\draw [ultra thick][-] (-4,-8) -- (4,-8);
\draw [ultra thick][-] (-4,0) -- (-4,-8);
\draw [ultra thick][-] (0,0) -- (0,-8);
\draw [ultra thick][-] (4,0) -- (4,-8);

\node[align=center, above] at (-2,0)
{$Z=1$};
\node[align=center] at (-2,-2)
{$D=1$};
\node[align=center] at (-2,-6)
{$D=0$};
\node[align=center] at (2,-4)
{$D=0$};
\node[align=center, above] at (2,0)
{$Z=0$};
\end{tikzpicture}
   \caption{Single-cell experiment with one-sided noncompliance}
        \label{fig:exp_design}
\end{figure}

This setup corresponds to an experiment with \textit{one-sided noncompliance}, a typical experimental design in many online advertising settings. Under standard conditions, any experimental design that features a binary treatment and a valid binary instrument can identify a local average treatment effect (LATE) parameter. However, one-sided noncompliance gives us the ability to estimate another important treatment effect parameter, the average treatment effect on the treated (ATT), defined as $\text{ATT}\equiv \mathbb{E} \left [Y_1 - Y_0 \middle \vert D=1 \right ]$, because it implies that $\text{ATT}=\text{LATE}$. 

Much of the recent literature on advertising measurement stops once a focal treatment effect parameter has been recovered. However, work in this area has been less focused on connecting those estimates to advertising decisions. This motivates our interest in the advertiser's decision problem, which we return to next.

\subsection{Revisiting the firm's advertising problem}\label{sec:dp2}

Using equations (\ref{eq:obs_out}) and (\ref{eq:sel_eq}) and the normalization that $U\sim U(0,1)$, we can rewrite the firm's optimization problem in (\ref{eq:dec_problem1}) as:
\begin{align}\label{eq:dec_problem2}
	\max_{\phi\in[0,1]} \left ( \delta  \times \left \{ \phi \mathbb{E} \left [Y_1 \middle \vert  U \leq \phi \right ] + (1-\phi) \mathbb{E} \left [Y_0 \middle \vert  U \geq \phi \right ] \right \} - \kappa(\phi) \right ). 
\end{align}
As we show in Appendix A, it follows that:
\begin{align}\label{eq:exp1}
      \mathbb{E} \left [Y_1 \middle \vert  U \leq \phi \right ]=\int_0^{\phi} m_1(u) \frac{1}{\phi} du \quad \text{ and } \quad \mathbb{E} \left [Y_0 \middle \vert  U > \phi \right ]= \int_{\phi}^1 m_0(u) \frac{1}{1-\phi} du,
\end{align}
where we used that $f(u)=1$ since $U$ follows a standard uniform distribution. The functions $m_d(u)$, where $d\in\{0,1\}$, are defined as $\mathbb{E} \left [Y_d \middle \vert U= u \right ]$. These functions are known as the marginal treatment response (MTR) functions.

As we also show in Appendix A, plugging the expressions in equation (\ref{eq:exp1}) back into equation (\ref{eq:dec_problem2}) allows us to rewrite the advertiser's decision problem as:
\begin{align}\label{eq:dec_problem3}
	&\max_{\phi\in[0,1]} \left ( \delta  \times  \int_0^{\phi} \text{MTE}(u)   du  - \kappa(\phi) \right ),	
\end{align}
where we defined the marginal treatment effect (MTE) function as:
\begin{align}\label{eq:mte}
\text{MTE}(u) \equiv \mathbb{E} \left [Y_1 - Y_0 \middle \vert U=u \right ] = m_1(u) - m_0(u) .
\end{align}
The MTE can be interpreted as the expected treatment effect at a particular (marginal) realization of the unobservable $U = u$. One of the benefits of this function is that, as shown, for example, in \cite{hv2005}, it can be used to obtain most treatment effect parameters of interest, such as the average treatment effect (ATE).

Assuming that the cost function $\kappa(\cdot)$ is continuous, then by the extreme value theorem the function being optimized in (\ref{eq:dec_problem3}) achieves its maximum in the interval $[0,1]$. However, the MTE function must be known for the advertiser to find the maximum.\footnote{The formulation of the optimization problem in terms of the MTE function, as shown in equation (\ref{eq:dec_problem3}), is not novel. It is analogous to how \cite{chv2010} defined their policy relevant treatment effect (PRTE) function and to Theorem 1 from \cite{su2020}, which represents the social welfare function in terms of the MTE function and generalizes a result from \cite{kt2018} by endogenizing treatment. Unlike these studies, however, our objective function corresponds to profits, not to measures of welfare.}

With a solution to this optimization problem in hand, $\phi^*$, we can determine the firm's optimal budget for advertising as $\kappa(\phi^*)$. Our specification for the decision problem can also accommodate an exogenous budget by adding a constraint that $\kappa(\phi)$ must not exceed. 

Importantly, the object the firm requires to solve their decision problem is the MTE function itself---this function is the object we aim to estimate. In turn, the ATT, which we can recover from data collected from the experimental design outlined earlier, is insufficient for the firm to make this decision. 

\section{Empirical approach} \label{sec:empirical}

Our goal is to recover credible estimates of the MTE function because it can be used to obtain multiple treatment effect parameters, including the ATT, and because it is an input to solve multiple decision problems, such as the one we presented above.

In this section, we first present our proposed multi-cell experimental design. Second, we discuss how to connect the data generated from this design to the MTR functions, which allow us to recover an approximation to the MTE function. Third, we explain our approximation strategy, which is motivated by and leverages the techniques in \cite{bmw2017}---henceforth ``BMW''. Fourth, we show how to use the approximations to solve a Bayesian version of the decision-maker's advertising problem from Section \ref{sec:dp2}.

\subsection{A multi-cell experimental design}

In our multi-cell design, first units are randomly divided across $C$ cells and then, given assignment to cell $c$, are randomly split into test and control groups within each cell. We define $\mathcal{C} = 1,\ldots,c,\dots,C$ to indicate assignment to cell $c$ and $Z_c$ as the indicator for treatment eligibility of an experimental unit from cell $c$. All these within-cell experiments feature one-sided noncompliance, so $\Pr \left (D=1 \middle \vert Z_c=0 \right )=0$ for all $c$. Notice that this is equivalent to randomly allocating some users to a single control cell and others to one of $C$ cells, with the latter users all being eligible to be exposed to ads.

We maintain the following assumption:
\begin{assum}\label{assum:exp} \hfill\break
	\subasu $\Pr (Z_c = z | \mathcal{C}=c) \in (0,1) $ for all $c$ and all $z$. \label{assum:exp1} \\
	\subasu $p(Z_c=1)\equiv\Pr \left (D=1 \middle \vert Z_c=1 \right ) \in (0,1)$ for all $c$. \label{assum:exp2} \\
 	\subasu $p(Z_c=1) \neq p(Z_{c^{\prime}}=1)$ for all $c \neq c^\prime$. \label{assum:exp3} 
\end{assum}

Assumptions \ref{assum:exp1} and \ref{assum:exp2} are innocuous because the experimenter can design the experiment so that they necessarily hold. First, recall that the experimenter has full control over the test/control split within each cell, and so can guarantee that $\Pr (Z_c = z | \mathcal{C}=c)$ is always strictly between 0 and 1. Second, consider cases in which the probability of treatment conditional on eligibility is either 0 or 1. If $p(Z_c=1)=1$, the endogeneity problem is resolved because eligibility to receive treatment becomes equivalent to exposure to treatment itself. In turn, if $p(Z_c=1)=0$, this exercise becomes meaningless because it implies that it is impossible for units to receive the treatment under consideration.

Assumption \ref{assum:exp3} requires that the probability of treatment conditional on eligibility varies across cells. The extent to which the experimenter is able to induce this variation is context-specific.\footnote{In settings where treatment is solely an active choice by the experimental unit, this might be more difficult to achieve. For example, when treatment is enrollment in a job training program, the decision of whether to enroll in the program is entirely the individual's choice. The experimenter can vary incentives for the individual to enroll in the program, but their effectiveness is a priori unknown.} For instance, in online advertising, treatment is exposure to ads, which is determined through auctions. The advertiser, as the experimenter, can influence treatment compliance---the exposure rate---by changing the average budget per user. The higher it is, the more likely the user is to be exposed to the ad. With a multi-cell experiment, this variation can be obtained by simply allocating the budget across cells appropriately (see Section \ref{sec:implem:testing} for more discussion). Hence, our analysis in terms of exposure rates can be seen as choosing a cell-specific budget per user; as our expected profit maximization problem shows, there is a direct correspondence between the two approaches.

As we show in Section \ref{sec:appr_met}, Assumption \ref{assum:exp} is crucial for BMW's method to be implementable in the context of our multi-cell design. On the other hand, with a single-cell experiment with one-sided noncompliance, the application of BMW's method requires the imposition of an additional constraint to alleviate an underidentification problem (see Appendix D). 

\subsection{Data generated from the multi-cell design}\label{sec:data_multicell}

We follow the literature on estimation of the MTE function and focus on the following moments:
\begin{align}\label{eq:info}
    \psi_{dzc} \equiv \mathbb{E} \left [ Y \middle \vert D=d, Z_c=z_c, \mathcal{C}=c \right ],
\end{align}
where $d \in \{0,1 \}$, $z_c \in \{0,1 \}$, and $c=1,\dots,C$. These moments are nonparametrically identified. To see how they provide information about the MTE function, we rely on the definition of treatment in equation (\ref{eq:sel_eq}) and the expressions in equation (\ref{eq:exp1}) to obtain:
\begin{align}\label{eq:info1}
    \psi_{1zc}&=\mathbb{E} \left [ Y \middle \vert D=1, Z_c = z_c, \mathcal{C}=c \right] \nonumber \\
    &= \mathbb{E} \left [ Y_1 \middle \vert U \leq p(z_c),Z_c = z_c \right] \nonumber \\
    &= \frac{1}{p(z_c)} \int_0^{p(z_c)}m_1(u)du
\end{align}
and
\begin{align}\label{eq:info0}
    \psi_{0zc}&=\mathbb{E} \left [ Y \middle \vert D=0, Z_c = z_c, \mathcal{C}=c \right] \nonumber \\
    &= \mathbb{E} \left [ Y_0 \middle \vert U > p(z_c),Z_c = z_c \right] \nonumber \\
    &= \frac{1}{1-p(z_c)} \int_{p(z_c)}^1m_0(u)du.
\end{align}
Hence, we have a known relationship between identified moments and the underlying MTR functions, $m_0(u)$ and $m_1(u)$, which we can then leverage to obtain information about the MTE function.

At first, it might seem like the multi-cell design generates $3 C$ different moments because $d \in \{0, 1\}$, $z_c \in \{0, 1\}$ and $d = 1$ only if $z_c=1$ would imply three moments per cell. However, notice that $p(Z_c = 0) = 0$ for all $c=1,\dots,C$. From equation (\ref{eq:info0}), this implies that 
\begin{align}\label{eq:info00}
 \psi_{00c} &= \mathbb{E} \left [ Y \middle \vert D=0,Z_c=0,\mathcal{C}=c \right ] \nonumber \\
    &=\mathbb{E} \left [ Y_0 \middle \vert U > 0, Z_c = 0 \right ] \nonumber \\
    &= \int_0^1 m_0(u)du \\
    &\equiv \psi_{00} \quad \textrm{for all } c. \nonumber
\end{align}
Hence, the multi-cell design generates $2C + 1$ different moments. Next, we show that these moments are sufficient to construct a polynomial approximation to the MTE function.

\subsection{Approximation method}\label{sec:appr_met}

BMW show that if an instrument, $Z$, takes $C$ different values, each associated with a propensity score that is strictly between 0 and 1, then we can approximate the MTR functions, $m_{d}(u)$, with a polynomial of degree $C-1$ provided that the propensity scores are also different from one another.

We adapt this approach to our multi-cell experimental design using the data to fit polynomial approximations to the MTR functions. When $d=1$,  we observe $C$ different values for $\psi_{1zc}$ from equation (\ref{eq:info1}). When $d=0$, we observe $C+1$ different values for $\psi_{0zc}$, with $C$ values from equation (\ref{eq:info0}) and one value from equation (\ref{eq:info00}).

Given the variation in the observed moments and in the propensity score, we consider the following polynomial approximations of the MTR functions:
\begin{align}\label{eq:app_mtr_til}
    \tilde{m}_1(u;\lambda_1)=\sum_{c=0}^{C-1}\lambda_{1c} u^c \quad \text{ and } \quad \tilde{m}_0(u;\lambda_0)=\sum_{c=0}^{C}\lambda_{0c} u^c ,
\end{align}
where it should be noted that the approximation when $d=0$ is of one higher degree compared to $d=1$. 
Plugging (\ref{eq:app_mtr_til}) back into the right-hand side of equations (\ref{eq:info1}) and (\ref{eq:info0}), we obtain the following approximations to the moments:
\begin{align}\label{eq:par_psi_til1}
    \tilde{\psi}_{1zc} & \equiv \frac{1}{p(z_c)} \int_0^{p(z_{c})} \sum_{c^{\prime}=0}^{C-1}\lambda_{1c^{\prime}} u^{c^{\prime}} du  \nonumber \\
    &= \sum_{c^{\prime}=0}^{C-1}\lambda_{1c^{\prime}} \frac{1}{p(z_{c})} \int_{0}^{p(z_{c})} u^{c^{\prime}} du \nonumber \\
    &= \sum_{c^{\prime}=0}^{C-1}\lambda_{1c^{\prime}} \left( \frac{p(z_c)^{c^{\prime}}}{c^{\prime}+1} \right)
\end{align}
and 
\begin{align}\label{eq:par_psi_til0}
    \tilde{\psi}_{0zc} &\equiv \frac{1}{1-p(z_c)} \int_{p(z_c)}^1 \sum_{c^{\prime}=0}^{C}\lambda_{0c^{\prime}} u^{c^{\prime}} du  \nonumber \\
    &= \sum_{c^{\prime}=0}^{C}\lambda_{0c^{\prime}} \frac{1}{1-p(z_c)} \int_{p(z_c)}^1 u^{c^{\prime}} du \nonumber \\
    &= \sum_{c^{\prime}=0}^{C}\lambda_{0c^{\prime}} \left( \frac{\sum_{s=0}^{c^{\prime}} p(z_c)^s}{c^{\prime}+1} \right)
\end{align}
for all $c \in C$. We can stack these terms and represent (\ref{eq:par_psi_til1}) and (\ref{eq:par_psi_til0}) in matrix form:
\begin{align}\label{eq:par_lambda_mat1}
     \underbrace{\begin{bmatrix}
    \tilde{\psi}_{111} \\ \tilde{\psi}_{112}\\ \vdots \\  \tilde{\psi}_{11C} 
    \end{bmatrix}}_{\tilde{\psi}_1} = 
    \underbrace{\begin{bmatrix}
    1 & \frac{p(z_1)}{2} & \dots & \frac{p(z_1)^{C-1}}{C} \\ 1 & \frac{p(z_2)}{2} & \dots & \frac{p(z_2)^{C-1}}{C} \\ \vdots & \vdots & \ddots & \vdots \\ 1 & \frac{p(z_C)}{2} & \dots & \frac{p(z_C)^{C-1}}{C}
    \end{bmatrix}}_{P_1}
    \underbrace{\begin{bmatrix}
    \lambda_{10} \\ \lambda_{11} \\ \vdots \\ \lambda_{1,C-1} 
    \end{bmatrix}}_{\lambda_1}
\end{align}
and 
\begin{align}\label{eq:par_lambda_mat0}
     \underbrace{\begin{bmatrix}
   \tilde{\psi}_{00} \\  \tilde{\psi}_{011} \\ \tilde{\psi}_{012}\\ \vdots \\  \tilde{\psi}_{01C} 
    \end{bmatrix}}_{\tilde{\psi}_0} = 
    \underbrace{\begin{bmatrix}
    1 & \frac{1}{2} & \dots & \frac{1}{C+1} \\
    1 & \frac{1+p(z_1)}{2} & \dots & \frac{1+p(z_1)+\dots+p(z_1)^{C}}{C+1} \\ 1 & \frac{1+p(z_2)}{2} & \dots & \frac{1+p(z_2)+\dots+p(z_2)^{C}}{C+1} \\ \vdots & \vdots & \ddots & \vdots \\ 1 & \frac{1+p(z_C)}{2} & \dots & \frac{1+p(z_C)+\dots+p(z_C)^{C}}{C+1}
    \end{bmatrix}}_{P_0}
    \underbrace{\begin{bmatrix}
    \lambda_{00} \\ \lambda_{01} \\ \vdots \\ \lambda_{0,C} 
    \end{bmatrix}}_{\lambda_0}.
\end{align}

Provided that the matrices $P_1$ and $P_0$ from equations (\ref{eq:par_lambda_mat1}) and (\ref{eq:par_lambda_mat0}) are invertible, we can compute $\lambda_1$ and $\lambda_0$ by replacing $\tilde{\psi}_1$, $\tilde{\psi}_0$, $P_1$ and $P_0$ with their observed counterparts from equation (\ref{eq:info}): $\lambda_1 =P_1^{-1} \tilde{\psi}_1$ and $\lambda_0 =P_0^{-1} \tilde{\psi}_0$. The invertibility of $P_1$ and $P_0$ is ensured by Assumption \ref{assum:exp}. Having recovered the $\lambda$s that parameterize the approximation to the MTR functions, we can obtain an approximation to the MTE function by equation (\ref{eq:mte}) and compute approximations to other treatment effect parameters of interest. 

\subsection{Utilization for decision-making: A Bayesian approach}\label{sec:bayes}

The approximation method described above allows us to estimate the parameters $\lambda_1$ and $\lambda_0$ from data. These estimates can then be used for decision-making, for instance, through the optimization problem given in Section \ref{sec:dp2}.

To see this more clearly, we plug (\ref{eq:app_mtr_til}) back into (\ref{eq:dec_problem3}), which yields the following approximated version of the firm's optimization problem:
\begin{align}\label{eq:dec_prob_app}
\max_{\phi\in[0,1]} \left ( \delta \times \left [ \sum_{c=0}^{C-1}\lambda_{1c} \frac{\phi^{c+1}}{c+1} - \sum_{c=0}^{C}\lambda_{0c} \frac{\phi^{c+1}}{c+1} \right ]-\kappa(\phi) \right ).
\end{align}

A naive approach would be to plug estimates of $\lambda_1$ and $\lambda_0$, say, $\hat{\lambda}_1$ and $\hat{\lambda}_0$ into (\ref{eq:dec_prob_app}) and solve for the optimal $\phi$. However, this plug-in approach ignores the uncertainty around the estimates $\hat{\lambda}_1$ and $\hat{\lambda}_0$, which should be accounted for when solving a statistical decision theory problem. Even though there are many different criteria to solve such problems, we adopt a Bayesian approach due to its convenience. This approach first integrates the objective function with respect to the unknown parameters ($\lambda_1$ and $\lambda_0$) using their posterior distribution given the data, and then solves the resulting optimization problem.\footnote{The decision maker could potentially incorporate uncertainty in $\kappa(\cdot)$ as well.} 

To be precise, denote this posterior distribution by $f\left (\lambda_1,\lambda_0 \middle \vert \text{data} \right )$. By adopting a Bayesian approach we solve the following problem:
\begin{align}\label{eq:dec_prob_bayes}
&\max_{\phi\in[0,1]} \left ( \delta \times \int_{\lambda_1,\lambda_0} \left [ \sum_{c=0}^{C-1}\lambda_{1c} \frac{\phi^{c+1}}{c+1} - \sum_{c=0}^{C}\lambda_{0c} \frac{\phi^{c+1}}{c+1} \right ]f\left (\lambda_1, \lambda_0 \middle \vert \text{data} \right)d\lambda_1 d\lambda_0 -\kappa(\phi) \right ) = \nonumber \\
&\max_{\phi\in[0,1]} \left ( \delta \times \left [ \sum_{c=0}^{C-1} \mathbb{E} \left [ \lambda_{1c} \middle \vert \text{data} \right ] \frac{\phi^{c+1}}{c+1} - \sum_{c=0}^{C} \mathbb{E} \left [ \lambda_{0c} \middle \vert \text{data} \right ] \frac{\phi^{c+1}}{c+1} \right ]-\kappa(\phi) \right ).
\end{align}
Hence, this new objective function depends solely on the posterior expected $\lambda$s given the data, which is a consequence of our approximation being linear in these parameters.

Deriving $f\left (\lambda_1,\lambda_0 \middle \vert \text{data} \right )$ directly, and thus $\mathbb{E} \left [ \lambda_{1} \middle \vert \text{data} \right ]$ and $\mathbb{E} \left [ \lambda_{0} \middle \vert \text{data} \right ]$, can be challenging. Nevertheless, it is straightforward to: derive the posterior distribution of $\psi$ and $p$ given the data; take draws from this distribution; apply (\ref{eq:par_lambda_mat1}) and (\ref{eq:par_lambda_mat0}) using these draws to obtain draws from $f\left (\lambda_1,\lambda_0 \middle \vert \text{data} \right )$; use these new draws to compute $\mathbb{E} \left [ \lambda_{1} \middle \vert \text{data} \right ]$ and $\mathbb{E} \left [ \lambda_{0} \middle \vert \text{data} \right ]$; and then solve the decision problem in (\ref{eq:dec_prob_bayes}). 

We can obtain the posterior of $p$ through a simple Beta-Bernoulli specification. In turn, the posterior of $\psi$ will depend on the nature of the potential outcomes. For example, if outcomes are continuously distributed, then a Normal-Gamma specification can be a convenient way to model their distribution. In our simulations, the outcome variable is binary, so we also use a Beta-Bernoulli specification. Notice that this approach places priors on $\psi$ and treats \textit{average} treatment effects as common across all users; however, it does not assume or impose that the treatment effects themselves are constant. 

We provide details of this procedure in Appendix E. The approach is sequential: First, we obtain draws of $p$, and then conditional on them, we draw $\psi$. This avoids the feedback issue raised by \cite{zwywcd2013}.

\section{Calibrated simulations} \label{sec:emp_app}

We illustrate the value of our proposed multi-cell experimental design through a series of simulations calibrated to an online advertising experiment at Facebook. We follow this simulation approach because we do not have data from a multi-cell experiment but want them to be as realistic as possible. Specifically, we use the results from a single-cell experiment with one-sided noncompliance to calibrate a set of data generating processes (DGPs). We use these DGPs to simulate what our proposed multi-cell design would have produced had it been used instead of the typical single-cell design. The results confirm that our design enables the practitioner to approximate the underlying MTE function well.

We use the approximations of the MTE function to derive the implied solutions to the optimization problem from equation (\ref{eq:dec_prob_bayes}). We compare both the quality of our MTE approximation and the implied optimal exposure rates to those with direct expected budget optimization. This approach experimentally varies the budget across cells, obtains the expected revenue function, and then approximates expected profits to select the optimal exposure rate. This strategy is appealing for its simplicity because there is no need to estimate the MTR functions, but as we show in a series of  examples, our MTE approach is likely more robust. Overall, our approach yields the solution that best approximates the true optimal solution, and, consequently, yields the lowest loss in expected profits.

In presenting these simulations, we do not claim to provide an exhaustive demonstration of our method's performance. Given the single-cell nature of the Facebook experiment, there are infinitely many parameters that we \textit{could} have chosen. Since we lack a real-world experiment, we are limited in our ability to determine the most ``reasonable'' true DGP and therefore cannot truly assess the quality of our approximations, which are conditional on our assumed DGPs. Another shortcoming is that the Facebook experiment on which these simulations are based allowed users to receive multiple ad exposures, whereas our model assumes treatment is binary. If exposures beyond the first have significant effects, the exclusion restriction in our model fails to hold. We discuss implications and overview a solution in Section \ref{sec:implem:multi} with details in Appendix C.

\subsection{Data and simulation approach} \label{sec:data_sim}

Our simulation exercise is based on one of the 15 large-scale online advertising experiments (or ``studies'') at Facebook used in \cite{gordon_zettelmeyer_2019}, to which we direct the reader for more details on the experiments and underlying data.

In what follows, we focus solely on Study 4, which featured a retailer hoping to drive purchase outcomes on its website. The experiment involved about 25 million users, with $\Pr(Z=1) = 0.7$ being the share allocated to the test group. Like the other experiments, Study 4 was a single-cell experiment with one-sided noncompliance. As such, we observe the ATT and the expectations $\psi_{11} = 0.00079$, $\psi_{01} = 0.00025$, and $\psi_{00} = 0.00033$, which correspond to the moments associated with the three regions in Figure \ref{fig:exp_design}. The exposure propensity in the test group is $p(Z=1) = 0.37$. Together, these objects contain all the information we use to calibrate MTE functions. 

In short, we proceed as follows. First, we specify the following MTR functions:
\begin{align*}
m_1(u) & = m_{10} + m_{11}u + m_{12}u^2 \\
m_0(u) & = m_{00} + m_{01}u + m_{02}u^2 + m_{03}u^3  .
\end{align*}
We chose these functional forms because they are the polynomials of lowest degree that the simplest version of our design---with only two cells---cannot recover. With three or more cells, our approach can perfectly recover the true MTR functions. The parameters of these functions are calibrated to match the moments we observe in the data.

Second, we choose the cell-specific eligibility probabilities and propensity scores based on the quantities we observe in the data. We use the simplest version of a multi-cell design, with only two cells.

Third, we generate the additional $\psi$s that would have been observed had this design been implemented through equations (\ref{eq:info1}) and (\ref{eq:info0}). We then combine them with our postulated propensity scores to implement the methods we described in Sections \ref{sec:appr_met} and \ref{sec:bayes}.

Appendix B provides further details of this process. Appendix G explores a more complex DGP that is not a polynomial, examining values of $C \in \{2,3,5\}$. 

\subsection{Results from our proposed approach} \label{sec:app_bmw_stud4}

\begin{figure}
    \centering
    \begin{subfloat}[DGP 1 \label{fig:app_dgp1}]
        {\includegraphics[width=0.495\textwidth]{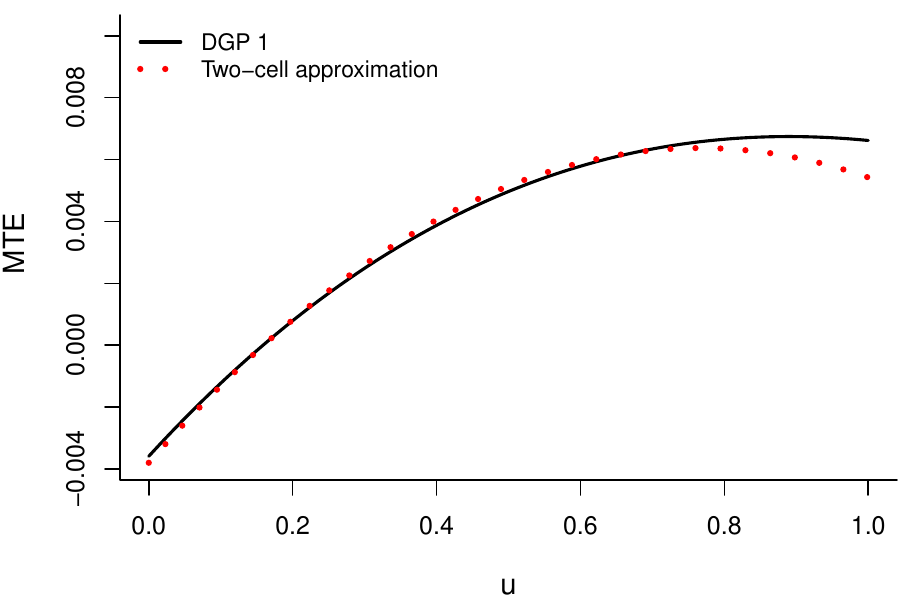}}
    \end{subfloat} 
    \begin{subfloat}[ DGP 2  \label{fig:app_dgp2}]
        {\includegraphics[width=0.495\textwidth]{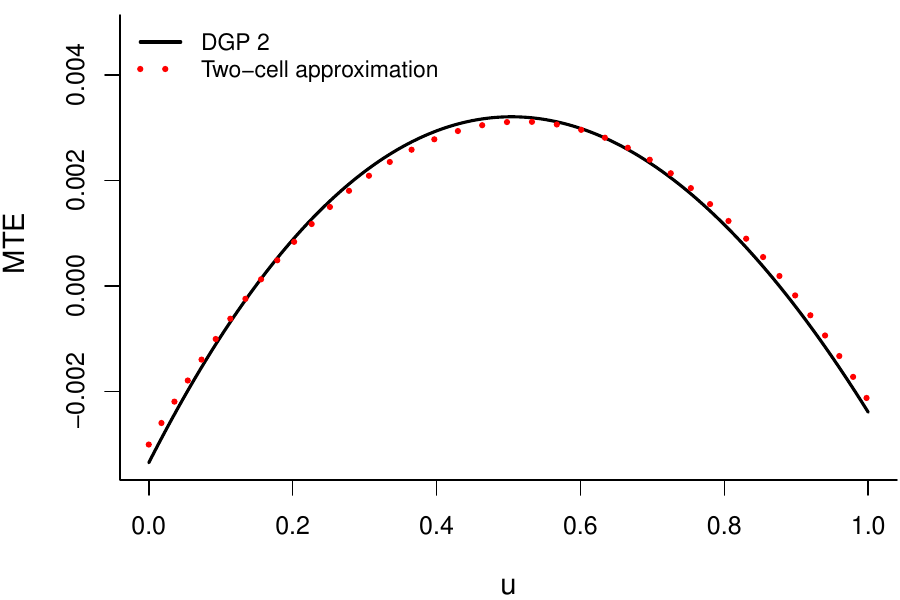}}
    \end{subfloat}
    \floatfoot{\textit{Note:} The figures show the true MTE functions (solid line) and their two-cell approximations (dotted line). The true functions under DGPs 1 and 2 are $\text{MTE}(u)=-0.0036+0.0254u-0.0179u^2 + 0.0027u^3$ and $\text{MTE}(u)=-0.0034+0.0268u-0.0288u^2 + 0.003u^3$, respectively.} 
\caption{Approximations to simulated DGPs from Study 4}
    \label{fig:app_dgps}%
\end{figure}

We use the $\psi$s from above to obtain the approximated MTE functions following the procedure we described in Section \ref{sec:appr_met}. The MTE functions we consider and the resulting approximations are shown in Figure \ref{fig:app_dgps}. 

Although the underlying MTE functions are cubic, the fact that their shapes are close to quadratic implies that a two-cell design yields good approximations. We quantify the quality of these approximations through three metrics. Denote the approximation to the true MTE function by $\text{MTE}_{\text{app}}(\cdot)$. The metrics we consider are:
\begin{align}\label{eq:norms}
    \begin{split}
        &\text{sup-norm}=\max_{u\in[0,1]}\left| \text{MTE}(u)- \text{MTE}_{\text{app}}(u) \right | \\
        &L_2\text{-norm}=\sqrt{ \left ( \int_0^1 \left [ \text{MTE}(u) - \text{MTE}_{\text{app}}(u) \right ]^2 du \right )} \\
        &\text{Relative ATE error}=\frac{\text{ATE}_{\text{app}}-\text{ATE}}{\text{ATE}}
    \end{split}
\end{align}
where $\text{ATE}_{\text{app}}$ is the ATE computed from $\text{MTE}_{\text{app}}(\cdot)$. We consider this last metric as a different way of summarizing the discrepancy between the true and approximated MTE functions because often the ATE is the treatment effect parameter of original interest to the researcher.

The results are given in Table \ref{tab:dist_measures}. Overall, our method generates a small difference between the approximations and their true values. For example, it produces a relative error in the estimated ATEs of about -3\% and 2\% for each of the DGPs, respectively. 

\begin{table}
\begin{threeparttable}
\caption{Distance between true and approximated MTE functions from two-cell design}
\begin{tabular}{c|cc}
    \hline \hline 
 Metric & DGP 1 & DGP 2 \\ 
\hline 
 sup-norm & 0.00019 & 0.00015 \\  
 $L_2$-norm & 0.00036 & 0.00014 \\ 
 Relative ATE error & -0.03009 & 0.02432 \\ 
\hline \hline
    \end{tabular}
    \begin{tablenotes}
    \footnotesize
    \textit{Note:} This table shows the discrepancy between the true and approximated MTE functions from Figure \ref{fig:app_dgps} using the criteria given in equation (\ref{eq:norms}).
    \end{tablenotes}
  \label{tab:dist_measures}
  \end{threeparttable}
\end{table} 

These results are based on the true population moments and thus would be obtained if the researcher had unlimited data. This raises the question of how well our approach performs with finite samples. We assess this by generating 10,000 samples and computing the approximation for each sample, dividing observations equally between the two cells. We consider four sample sizes to study how the performance of the method changes as the number of observations increase.

\begin{figure}
    \centering
    \begin{subfloat}[ $n=1,200$ \label{fig:dgp1_mte_app_ss1}]
        {\includegraphics[width=0.495\textwidth]{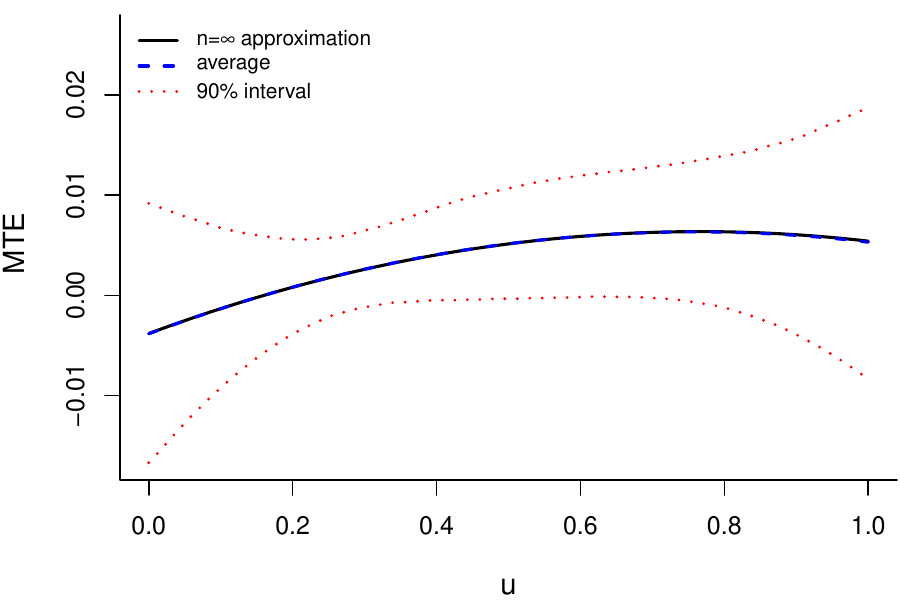}}
    \end{subfloat} 
    \begin{subfloat}[ $n=12,000$ \label{fig:dgp1_mte_app_ss2}]
        {\includegraphics[width=0.495\textwidth]{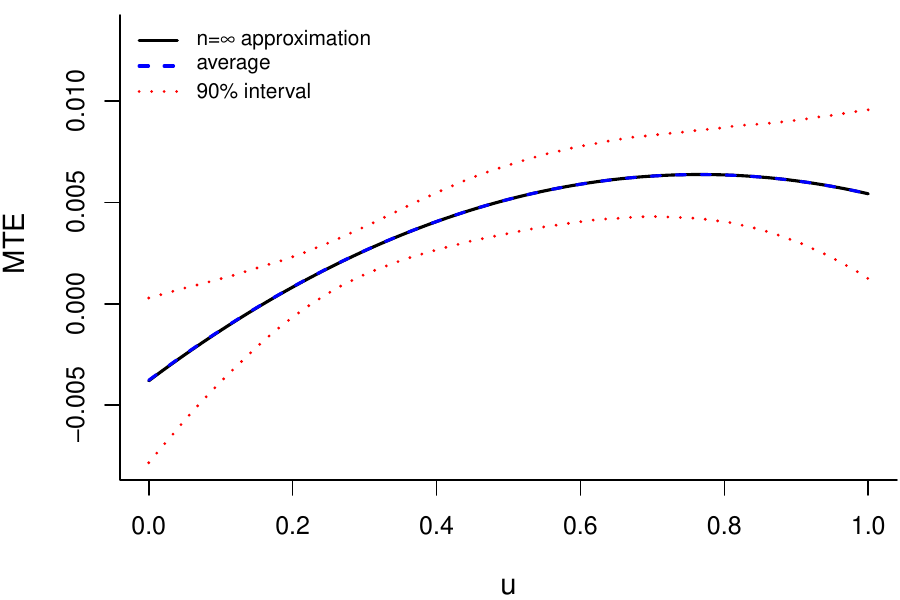}}
    \end{subfloat} \\
    \begin{subfloat}[ $n=120,000$ \label{fig:dgp1_mte_app_ss3}]
        {\includegraphics[width=0.495\textwidth]{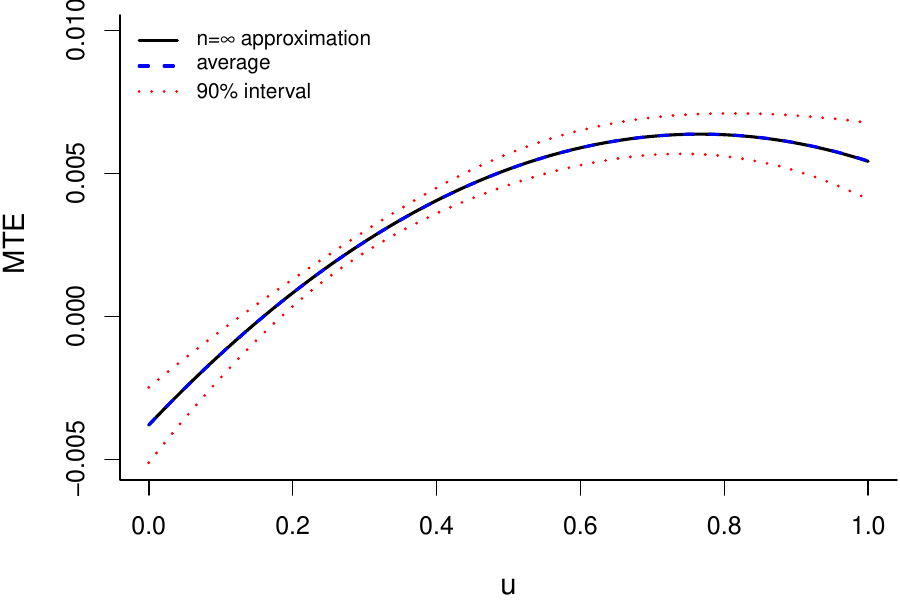}}
    \end{subfloat}
    \begin{subfloat}[ $n=1,200,000$ \label{fig:dgp1_mte_app_ss4}]
        {\includegraphics[width=0.495\textwidth]{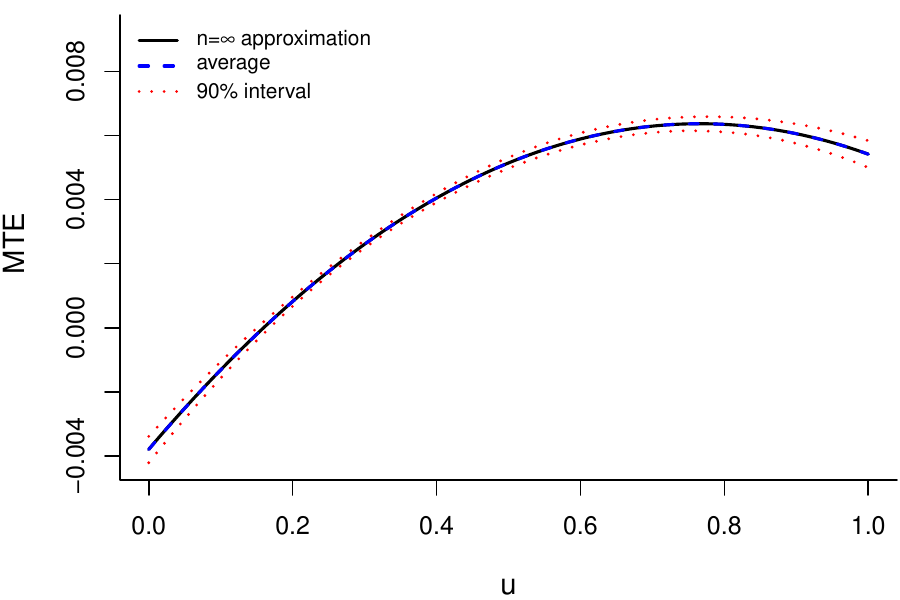}}
    \end{subfloat}
\caption{Two-cell approximation to MTE function under DGP 1 by sample size}
\floatfoot{\textit{Note:} This figure shows the approximations to the MTE function under DGP 1 based on two cells with propensity scores equal to 0.37 and 0.863 with varying sample sizes. The solid black line shows the population approximation obtainable from unlimited data. The dashed blue line, which completely overlaps the solid black line, shows the average approximation obtained from 10,000 random samples, and the dotted red lines show the 5th and 95th percentiles from these same samples.}
    \label{fig:dgp1_mte_app_ss}%
\end{figure}

For DGPs 1 and 2, respectively, Figures \ref{fig:dgp1_mte_app_ss} and \ref{fig:dgp2_mte_app_ss} plot the population level approximation, the average approximation across the 10,000 samples, and the 5th and 95th percentiles of these approximations, with sample sizes ranging from 1,200 to 1,200,000.

We verify that the bias is small across all sample sizes, even when the number of observations is as limited as 1,200. Expectedly, the variance of the approximation decreases as the sample size increases, and the 90\% interval almost collapses to the true approximation when the overall number of observations is 1.2 million. We find this result encouraging as the number of observations of these experiments in practice can be much larger. For instance, Study 4, which we use to calibrate these simulations, involved north of 25 million users.

Those familiar with experimental studies about online advertising may still worry because of the well-known low power issue pervasive in this setting. We believe this is a smaller concern for us because of differences between the tasks of profit maximization and hypothesis testing.

\begin{figure}
    \centering
    \begin{subfloat}[ $n=1,200$ \label{fig:dgp2_mte_app_ss1}]
        {\includegraphics[width=0.45\textwidth]{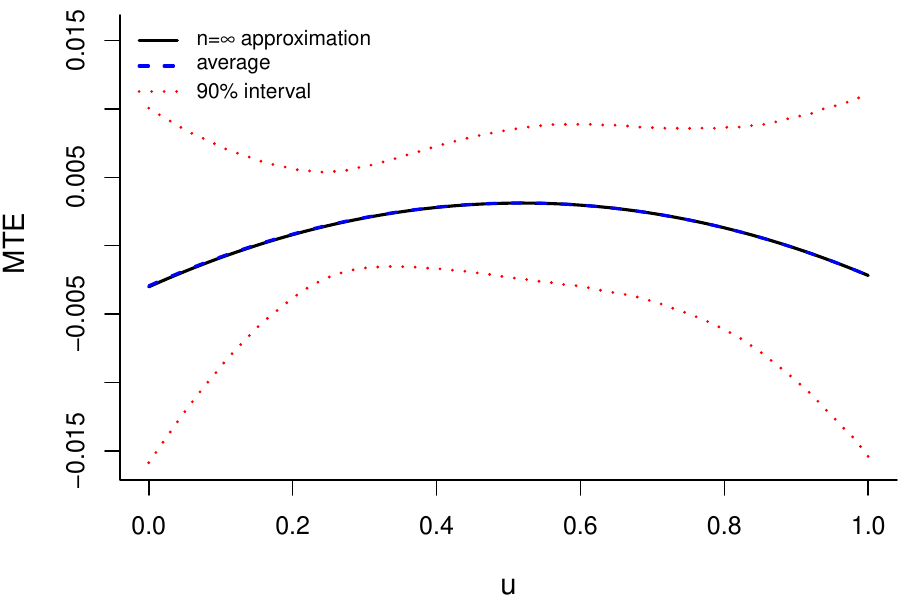}}
    \end{subfloat} 
    \begin{subfloat}[ $n=12,000$ \label{fig:dgp2_mte_app_ss2}]
        {\includegraphics[width=0.45\textwidth]{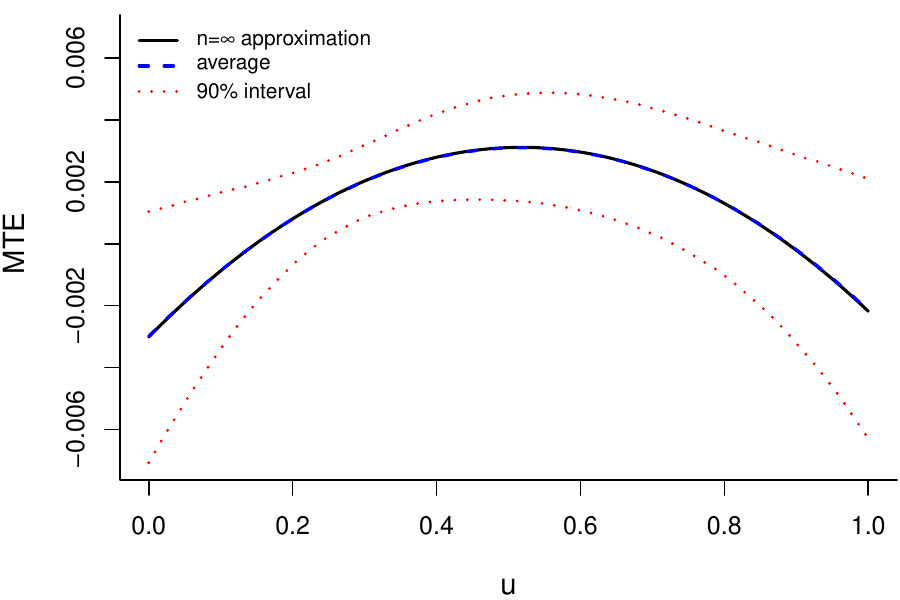}}
    \end{subfloat} \\
    \begin{subfloat}[ $n=120,000$ \label{fig:dgp2_mte_app_ss3}]
        {\includegraphics[width=0.45\textwidth]{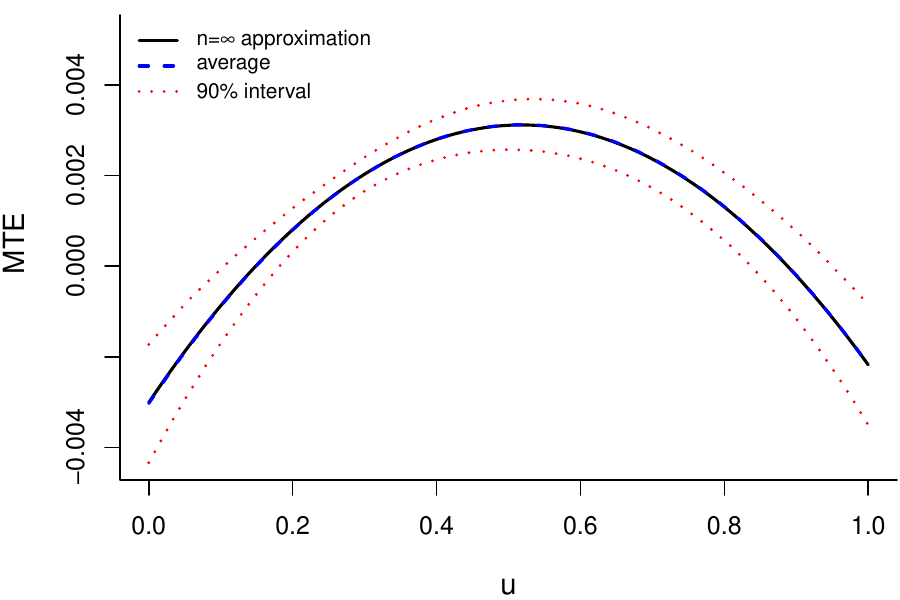}}
    \end{subfloat}
    \begin{subfloat}[ $n=1,200,000$ \label{fig:dgp2_mte_app_ss}]
        {\includegraphics[width=0.45\textwidth]{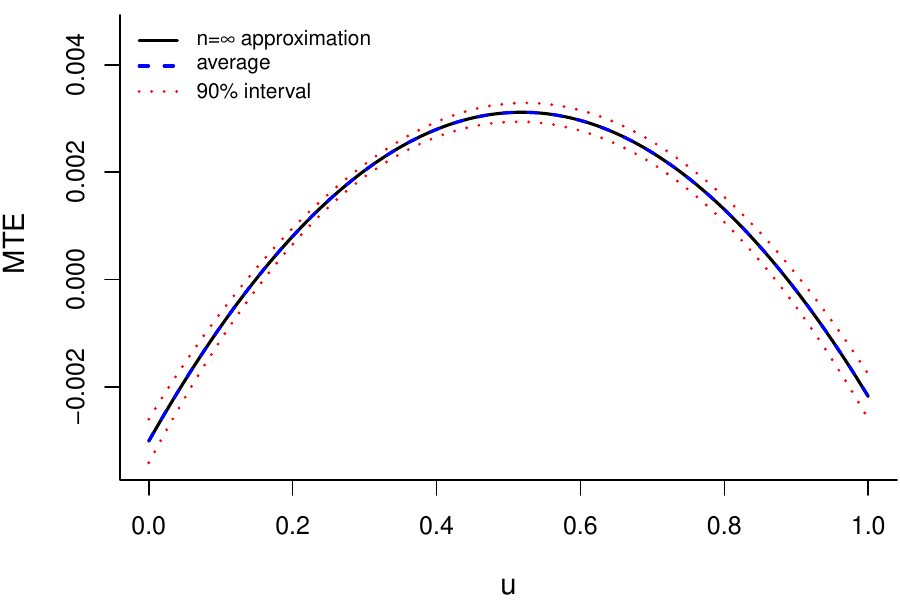}}
    \end{subfloat}
\caption{Two-cell approximation to MTE function under DGP 2 by sample size}
\floatfoot{\textit{Note:} This figure shows the approximations to the MTE function under DGP 2 based on two cells with propensity scores equal to 0.37 and 0.863 with varying sample sizes. The solid black line shows the population approximation obtainable from unlimited data. The dashed blue line, which completely overlaps the solid black line, shows the average approximation obtained from 10,000 random samples, and the dotted red lines show the 5th and 95th percentiles from these same samples.}
    \label{fig:dgp2_mte_app_ss}%
\end{figure}

Our objective in recovering the MTR functions is to solve the optimization problem introduced in Section \ref{sec:dp1}. Estimating these functions requires calculating functions of sample means (the sample analogs of the $\psi$s and $p$s), each of which is precisely estimated. More importantly, as outlined in Section \ref{sec:bayes}, our approach then \textit{averages} over these functions to account for the uncertainty around them when maximizing profits.

On the other hand, consider a hypothesis test procedure to verify whether the ATT, for example, equals zero. Even though the Wald estimator is also a function of sample means, each of which is precisely estimated, performing this test requires \textit{dividing} this estimator, not averaging, by a measure of its uncertainty. In the context of online advertising, ATTs are often small and the measure of uncertainty around the Wald estimator are of similar magnitude, which creates power issues.

The precise nature of statistical power issues when estimating the MTR functions will depend on the true underlying DGP and the data in hand. Although the discussion above is informal, the results in this section suggest that power issues may be of relatively less concern using our approach.

\subsection{Implications for decision-making}\label{sec:dm_imp}

\begin{figure}
    \centering
	\includegraphics[width=0.6\textwidth]{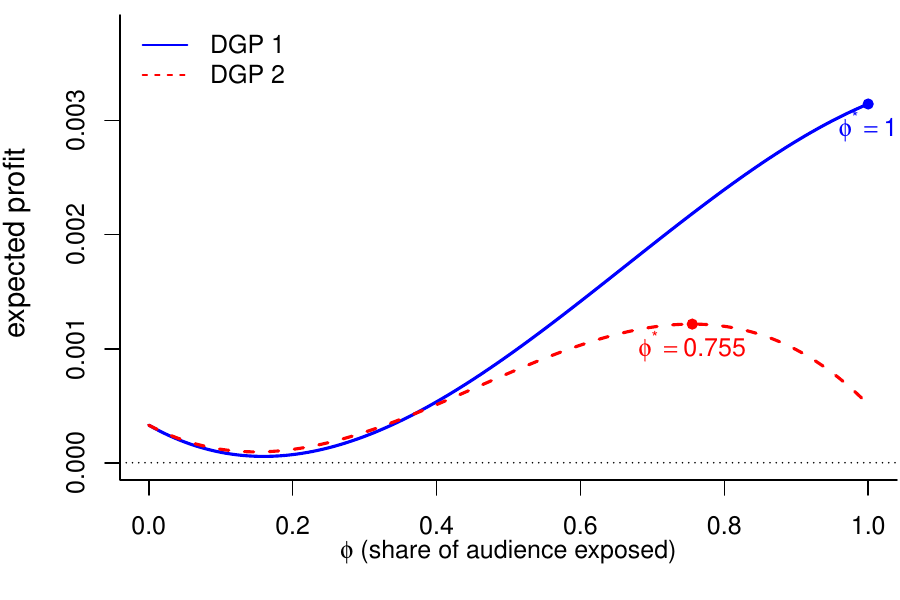}
    \caption{Expected profit functions from simulated DGPs of Study 4}
    \floatfoot{\textit{Note:} This figure plots the resulting expected profit function and optimal exposure rates under the DGPs shown in Figure \ref{fig:app_dgps} and the cost function $\kappa(\phi)=0.001\phi^4$.}
    \label{fig:profs_study4}
\end{figure}

We consider the firm's decision problem given in equation (\ref{eq:dec_problem3}). For the sake of illustration, we set $\delta=1$ and $\kappa(\phi) = 0.001 \phi^4$. A firm can set $\delta$ based on their internal assessment of the value of a conversion event. We specify $\kappa(\phi)$ as convex to capture the notion that reaching the marginal consumer becomes more expensive as overall campaign reach increases. As we note in Section \ref{sec:dp1}, most advertising platforms provide advertisers with campaign planning tools to help them predict how reach is expected to vary as a function of their budget.
Based on the simulated DGPs we outlined above, the resulting expected profit functions are given in Figure \ref{fig:profs_study4}.

The expected profit functions reflect the differences across the different DGPs shown in Figure \ref{fig:app_dgps}. They demonstrate how different MTE functions can affect optimal decisions. In this case, the optimal exposure rates, $\phi^*$, associated with DGPs 1 and 2 are to treat 100\% and 75.5\% of the population, respectively, as Figure \ref{fig:profs_study4} shows.

We now compare the true optimal solutions to what the decision-maker would do if information obtained from our experimental design was available, following the Bayesian estimation procedure we presented in Section \ref{sec:bayes}. In this exercise, we consider a sample size of 25,553,093, which corresponds to that of Study 4, use uniform priors, and take 1,000 draws to estimate the posterior means from equation (\ref{eq:dec_prob_bayes}).


Table \ref{tab:loss_multi} presents the results. The multi-cell approach yields virtually no losses across both DGPs, and is able to identify the true optimal solution correctly under DGP 1. This may be unsurprising given that it was able to approximate the underlying MTE functions well.

\begin{table}
\begin{threeparttable}
\caption{Expected profit losses (\%) from multi-cell design}
\begin{tabular}{c|c|cc}
    \hline \hline 
 DGP & True $\phi^*$ & Estimated $\phi^*$ & Loss \\ 
\hline 
 1 & 1 & 1 & 0.00 \\  
 2 & 0.755 & 0.762 & 0.03 \\ 
\hline \hline
    \end{tabular}
    \begin{tablenotes}
      \footnotesize
      \textit{Note:} This table shows the expected profit losses from using the multi-cell approach under each DGP. The optimal exposure rate, $\phi^*$, is obtained using the procedure described in Section \ref{sec:bayes}. We use a sample size of 25,553,093, which corresponds to that of Study 4, uniform priors, and 1,000 draws to estimate the posterior means from equation (\ref{eq:dec_prob_bayes}).
    \end{tablenotes}
  \label{tab:loss_multi}
  \end{threeparttable}
\end{table} 

\subsection{Comparison to direct expected budget optimization} \label{sec:dir_opt}

An alternative and arguably simpler approach an advertiser can take is direct expected budget optimization: experiment with different budgets, use the observed data to estimate the expected revenue function, and then use it to maximize expected profits. This approach, which we will refer to as the ``direct'' method, circumvents the need to estimate the MTR functions and relies only on observed revenues and exposure rates, $\phi$, but not the different $\psi$s. Focusing on DGP 2, we will assess how this approach can perform vis-\`{a}-vis our proposed method as the number of cells increases, the budget levels under consideration change, and the cost function changes. We provide these assessments based on population moments and under various finite sample sizes.

Suppose that the advertiser considers $c=0,1,\dots,C$ budget levels. Each budget level, $B_c$, given the cost function that held during the experiment, induces a different exposure rate because $B_c=\kappa(\phi_c)$. If the experiment involves a budget of zero, which induces zero exposures, and only one positive budget, then this experiment is equivalent to the single-cell design with one-sided noncompliance. In all our simulations, we include a budget of zero to preserve this correspondence. 

For each budget level, the expected revenue is:
\begin{align}
    \text{Revenue}_c=\phi_c \times \mathbb{E} \left [Y \middle \vert D=1,\phi_c \right] + (1-\phi_c) \times\mathbb{E} \left [Y \middle \vert D=0,\phi_c \right].
\end{align}
Using different $\left \{\phi_c, \text{Revenue}_c \right \}$ pairs, the advertiser can approximate the true expected revenue function and then combine it with the cost function to maximize expected profits.

\subsubsection{Population level analysis}

\begin{figure}
    \centering
    \begin{subfloat}[ Two cells \label{fig:dir_profs2}]
        {\includegraphics[width=0.495\textwidth]{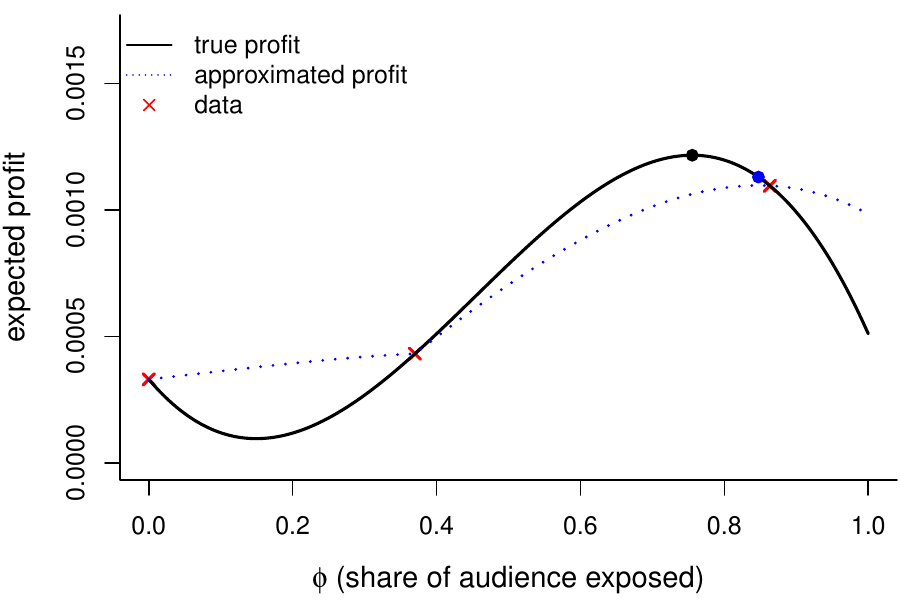}}
    \end{subfloat}
    \begin{subfloat}[ Three cells \label{fig:dir_profs3}]
        {\includegraphics[width=0.495\textwidth]{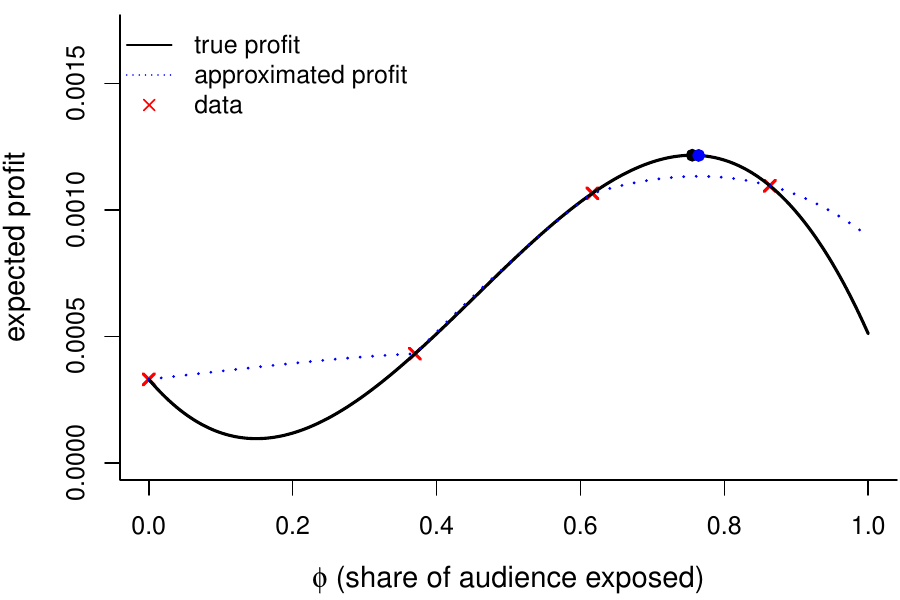}}
    \end{subfloat}\\
    \begin{subfloat}[ Three cells: high costs \label{fig:profs_costs2}]
        {\includegraphics[width=0.495\textwidth]{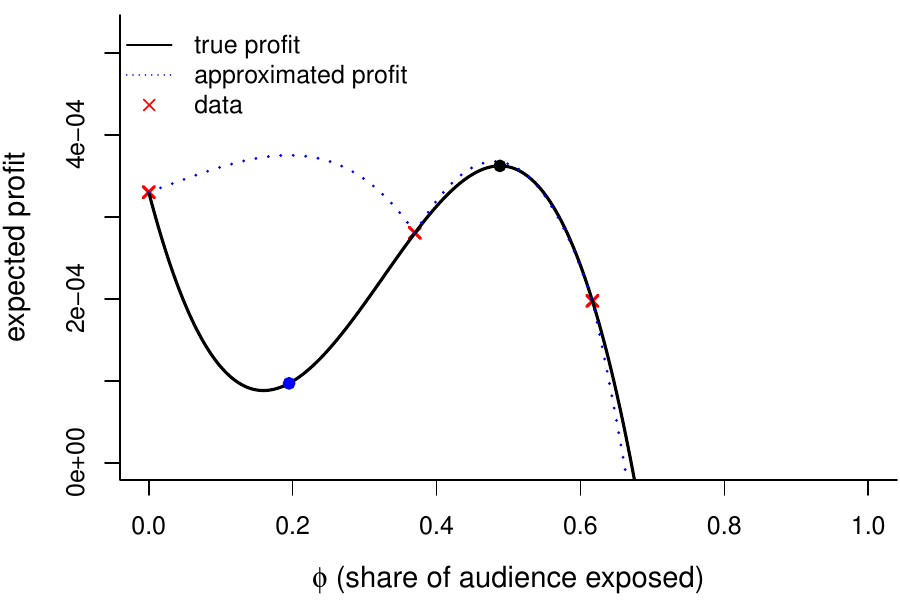}}
    \end{subfloat}\\
        \begin{subfloat}[ Three cells: high $\phi$s \label{fig:profs_knots2}]
        {\includegraphics[width=0.495\textwidth]{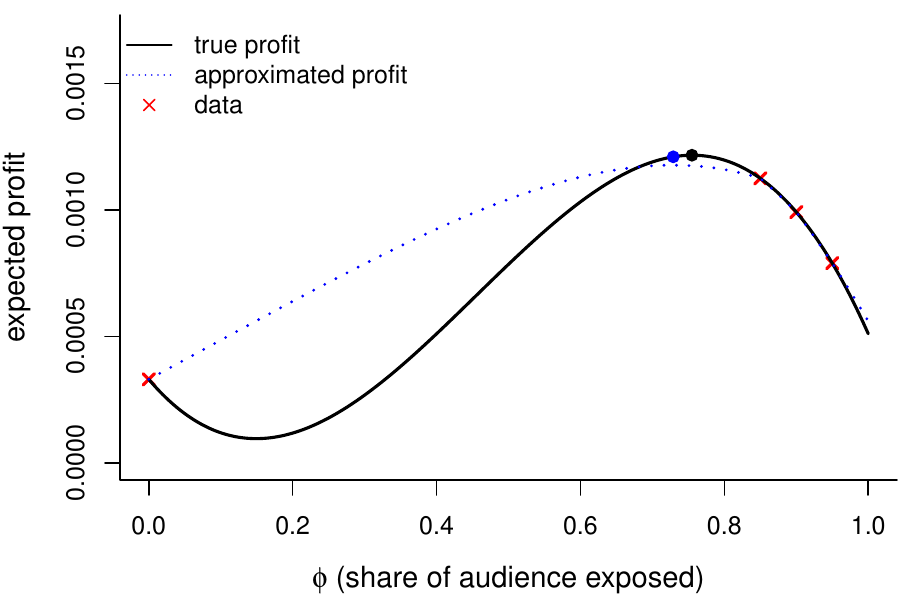}}
    \end{subfloat} 
    \begin{subfloat}[ Three cells: low $\phi$s \label{fig:profs_knots3}]
        {\includegraphics[width=0.495\textwidth]{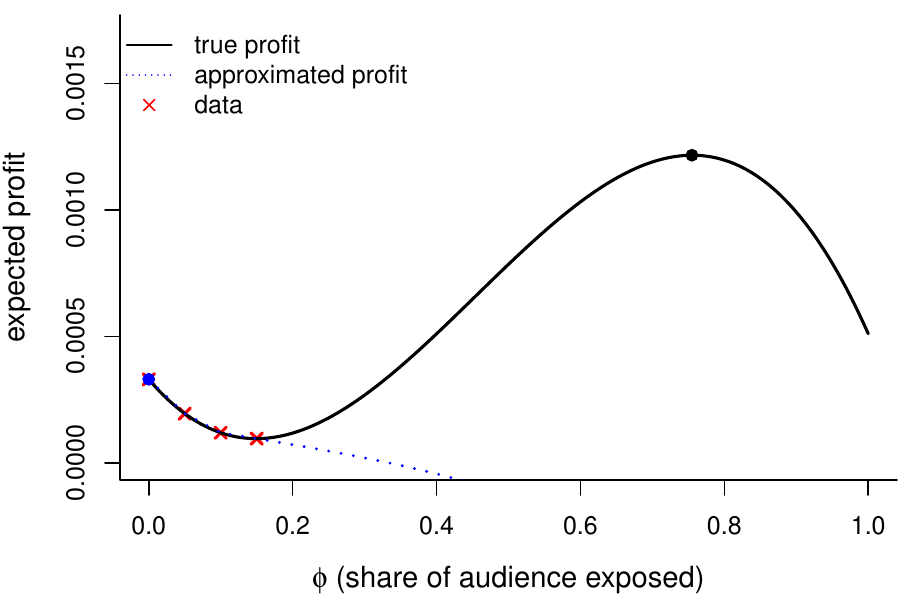}}
    \end{subfloat}
\caption{Direct optimal choices under varying number of cells, exposure rates, and cost functions}
\floatfoot{\textit{Note:} The solid black lines show the true expected profit function using the MTE function from DGP 2 and a given cost function; the baseline cost function is $\kappa(\phi)=0.001\phi^4$. The dotted blue line shows the approximated expected profit function. The black and blue circles show the true and approximated optimal exposure rates, respectively. The red crosses indicate the observed exposure-profit levels from the experiments. Figure (a) shows the estimated expected profit function from the direct method based on two cells with exposure rates equal to 0.37 and 0.863. Figure (b) adds a third cell with exposure rate equal to 0.617, which constitute the baseline exposure rates. Under the baseline $\kappa(\cdot)$, Figures (c) and (d) change the exposure rates to 0.85, 0.89, and 0.95 and 0.05, 0.1, and 0.15, respectively. Under the baseline exposure rates, Figure (e) changes the cost function to $\kappa(\phi)=0.005\phi^{3.5}$.}
    \label{fig:revs_profs}%
\end{figure}

We first consider the same experiment with two cells from Section \ref{sec:app_bmw_stud4}, whose exposure rates are 0.37 and 0.863. This experiment yields three values of revenues, one associated to each of the two positive values plus one associated with zero. We use these three exposure-revenue pairs to approximate the expected revenue function and combine it with the cost function to approximate the expected profit function and then optimize this approximation. We show these results in Figure \ref{fig:dir_profs2}. The estimated optimal exposure rate is $\phi=0.847$ and captures 92.97\% of the optimal expected profit.

We then add a third cell with an exposure rate of 0.617 and repeat the same exercise, whose result is given in Figure \ref{fig:dir_profs3}. The new estimated optimal exposure rate is 0.764 and captures 99.94\% of the optimal expected profit. This reflects the intuition that having more cells allows for a more precise approximation, even when one adopts the direct approach instead of our proposed method. Importantly, remember that, under this DGP, three cells is enough for our approach to perfectly recover the true MTE function with unlimited data and thus the true optimal exposure rate regardless of the cost function.

The performance of the direct optimization approach relies on its approximation to the true expected revenue function and how it interacts with the cost function. Next, we investigate how its performance changes under a new cost function under the same three-cell design as above.

We change the cost function from $0.001\phi^4$ (Figure \ref{fig:dir_profs3}) to $0.0055\phi^{3.5}$ (Figure \ref{fig:profs_costs2}), corresponding to an increase in advertising costs. The results demonstrate that the performance of direct budget optimization can be very sensitive to the cost function: while direct optimization captures 99.94\% of the true expected profit optimization under the original cost function, it only captures 26.75\% under the new cost function.

Finally, we investigate the role the observed exposure rates play in successfully approximating the expected revenue function directly. We consider experiments with three cells with high exposures (0.85, 0.90, and 0.95) and with low exposures (0.05, 0.10, and 0.15). Results are shown in Figures \ref{fig:profs_knots2} and \ref{fig:profs_knots3}, respectively. 

The results are intuitive. The true optimal exposure rate is high, 0.755, as shown in Figure \ref{fig:profs_study4}. Consequently, when experimenting with high exposure rates the direct approach approximates the function well on that region and is thus able to capture most of the true optimal expected profit. On the other hand, the opposite occurs when all exposure rates in the experiment are low and the performance of the direct approach suffers. 

\subsubsection{Finite sample analysis}

We examine how the different approaches perform under different finite sample sizes. A particular concern is the extent to which estimation errors propagate to the budget optimization decision. We address this by performing analogous simulation exercises to those presented in Section \ref{sec:app_bmw_stud4}. We take 10,000 samples for each sample size: 1,200, 12,000, 120,000, and 1,200,000. We divide the observations equally across cells, use uniform priors, and take 1,000 draws to estimate the posterior means from equation (\ref{eq:dec_prob_bayes}).

Table \ref{tab:losses_cells} shows results comparing expected profit losses from our method and the direct approach with two versus three cells in the experiment. The results echo those from Figures \ref{fig:dir_profs2} and \ref{fig:dir_profs3}: across all sample sizes, our method outperforms the direct approach, and the expected profit losses from both methods decrease as the sample size increases. It is interesting to note that, particularly for our method, the convergence to the unlimited data outcome is slower under three cells than under two. This is likely because each sample mean is estimated using a smaller number of observations as the number of cells increase holding the overall sample size fixed.

Table \ref{tab:losses_costs} performs an analogous exercise comparing outcomes from the two different cost functions considered in Figures \ref{fig:dir_profs3} and \ref{fig:profs_costs2}. Once again, the results echo those from before. Our multi-cell approach performs well under both cost functions and its performance improves as the sample size increases. In addition, it always outperforms the direct method, which performs poorly under the cost function that implies high advertising costs.

\begin{table}
\begin{threeparttable}
\caption{Expected profit losses (\%) under different number of cells}
\begin{tabular}{c|c|ccc|ccc}
    \hline \hline 
    \multirow{2}{*}{Sample size} & \multirow{2}{*}{Method} & \multicolumn{3}{c|}{Two cells} & \multicolumn{3}{c}{Three cells}  \\ \cline{3-8}
     &  & 5\% & Median & 95\% & 5\% & Median & 95\%  \\
    \hline 
     \multirow{2}{*}{$n=1,200$} & Multi-cell & 0.13 & 16.97 & 72.88 & 0.47 & 21.09 & 92.08  \\
       & Direct & 0.45 & 58.00 & 72.88 & 8.78 & 26.66 & 72.88 \\
       \hline
     \multirow{2}{*}{$n=12,000$} & Multi-cell & 0.01 & 1.85 & 21.77 & 0.07 & 6.64 & 91.91 \\
       & Direct & 0.09 & 9.99 & 58.00 & 0.46 & 12.36 & 64.55 \\
       \hline
     \multirow{2}{*}{$n=120,000$} & Multi-cell & 0.002 & 0.20 & 1.76 & 0.01 & 0.90 & 22.93 \\
       & Direct & 0.52 & 7.09 & 20.02 & 0.04 & 5.38 & 26.63 \\
       \hline
     \multirow{2}{*}{$n=1,200,000$} & Multi-cell & 0.0003 & 0.04 & 0.26 & 0.0008 & 0.09 & 0.85 \\
       & Direct & 4.17 & 7.11 & 10.68 & 0.005 & 0.54 & 4.55 \\
 \hline \hline
    \end{tabular}
    \begin{tablenotes}
      \footnotesize
       \textit{Note:} This table shows the median and 5th and 95th percentiles of expected profit losses (\%) from 10,000 samples based on the multi-cell approach and on direct optimization. The cost function was $\kappa(\phi)=0.001\phi^4$. The two cell implementation had exposure rates equal to 0.37 and 0.863, and the three-cell implementation added a third cell with exposure rate equal to 0.617. We divide observations equally across cells, use uniform priors, and take 1,000 draws to estimate the posterior means from equation (\ref{eq:dec_prob_bayes})
    \end{tablenotes}
  \label{tab:losses_cells}
  \end{threeparttable}
\end{table} 

\begin{table}
\begin{threeparttable}
\caption{Expected profit losses (\%) under different cost functions}
\begin{tabular}{c|c|ccc|ccc}
    \hline \hline 
\multirow{2}{*}{Sample size} & \multirow{2}{*}{Method} & \multicolumn{3}{c|}{Baseline costs} & \multicolumn{3}{c}{High costs}  \\ \cline{3-8}
     &  & 5\% & Median & 95\% & 5\% & Median & 95\%  \\
    \hline 
     \multirow{2}{*}{$n=1,200$} & Multi-cell & 0.47 & 21.09 & 92.08 & 0.79 & 45.78 & 82.20 \\
       & Direct & 8.78 & 26.66 & 72.88 & 4.70 & 22.61 & 70.70 \\
       \hline
     \multirow{2}{*}{$n=12,000$} & Multi-cell & 0.07 & 6.64 & 91.91 & 0.26 & 18.10 & 75.55 \\
       & Direct & 0.46 & 12.36 & 64.55 & 1.61 & 23.80 & 73.22 \\
       \hline
     \multirow{2}{*}{$n=120,000$} & Multi-cell & 0.01 & 0.90 & 22.93 & 0.03 & 3.07 & 72.71 \\
       & Direct & 0.04 & 5.38 & 26.63 & 1.42 & 55.66 & 75.37 \\
       \hline
     \multirow{2}{*}{$n=1,200,000$} & Multi-cell & 0.001 & 0.09 & 0.85 & 0.003 & 0.31 & 3.27 \\
       & Direct & 0.005 & 0.54 & 4.55  & 61.63 & 72.78 & 75.62 \\
 \hline \hline
    \end{tabular}
    \begin{tablenotes}
      \footnotesize
       \textit{Note:} This table shows the median and 5th and 95th percentiles of expected profit losses (\%) from 10,000 samples based on the multi-cell approach and on direct optimization. The implementations were based on three cells with exposure rates equal to 0.37, 0.617, and 0.863. The baseline cost function is $\kappa(\phi)=0.001\phi^4$ and the high cost function is $\kappa(\phi)=0.0055\phi^{3.5}$. We divide observations equally across cells, use uniform priors, and take 1,000 draws to estimate the posterior means from equation (\ref{eq:dec_prob_bayes})
    \end{tablenotes}
  \label{tab:losses_costs}
  \end{threeparttable}
\end{table} 

\begin{table}
\begin{threeparttable}
\caption{Expected profit losses (\%) under different exposure rates}
\begin{tabular}{c|c|ccc|ccc}
    \hline \hline 
  \multirow{2}{*}{Sample size} & \multirow{2}{*}{Method} & \multicolumn{3}{c|}{High exposures} & \multicolumn{3}{c}{Low exposures}  \\ \cline{3-8}
     &  & 5\% & Median & 95\% & 5\% & Median & 95\%  \\
    \hline 
     \multirow{2}{*}{$n=1,200$} & Multi-cell & 2.58 & 58.01 & 90.92 & 0.30 & 85.46 & 91.88 \\
       & Direct & 0.14 & 12.66 & 72.88 & 33.35 & 84.00 & 90.24 \\
       \hline
     \multirow{2}{*}{$n=12,000$} & Multi-cell & 1.80 & 53.98 & 90.53 & 0.51 & 72.87 & 91.92 \\
       & Direct & 0.02 & 2.21 & 22.23 & 3.67 & 72.88 & 90.24 \\
       \hline
     \multirow{2}{*}{$n=120,000$} & Multi-cell & 1.15 & 18.68 & 90.17 & 1.21 & 58.00 & 91.87 \\
       & Direct & 0.01 & 0.54 & 3.49 & 0.83 & 72.88 & 90.24 \\
       \hline
     \multirow{2}{*}{$n=1,200,000$} & Multi-cell & 0.41 & 11.51 & 88.99 & 0.78 & 58.00 & 91.11 \\
       & Direct & 0.15 & 0.52 & 1.15 & 1.76 & 72.88 & 90.23 \\
 \hline \hline
    \end{tabular}
    \begin{tablenotes}
      \footnotesize
       \textit{Note:} This table shows the median and 5th and 95th percentiles of expected profit losses (\%) from 10,000 samples based on the multi-cell approach and on direct optimization. The cost function was $\kappa(\phi)=0.001\phi^4$. The implementations used three cells. The high exposure rates were 0.85, 0.9, and 0.95, and the low exposure rates were 0.05, 0.1, and 0.15. We divide observations equally across cells, use uniform priors, and take 1,000 draws to estimate the posterior means from equation (\ref{eq:dec_prob_bayes})
    \end{tablenotes}
  \label{tab:losses_scores}
  \end{threeparttable}
\end{table}

Finally, Table \ref{tab:losses_scores} compares the outcomes from experiments that tested only high versus only low exposure rates. The results for the direct method are in alignment with the previous results shown in Figures \ref{fig:profs_knots2} and \ref{fig:profs_knots3}: it performs well with higher exposures but poorly with low, and its performance improves as the sample size grows. However, we obtain new insights regarding the performance of the multi-cell approach.

Although the multi-cell approach, in theory, is able to perfectly recover the underlying MTE with three cells given the DGP we use, its performance is poor. This is because it relies on a polynomial approximation, which, absent variation in exposure rates, is unstable. The performance is less poor when the exposure rates under consideration are high because the true optimum in this case is also high. However, the performance of the multi-cell approach is still much worse than that of the direct approach.

\vspace*{-0.28in}
\paragraph{Summary}\hfill 

\vspace*{-0.1in}
The exercises with finite samples indicate that our proposed multi-cell approach should perform better than the direct approach provided that there is variation in exposure rates. Table \ref{tab:losses_cells} indicates that this will be the case for each sample size and number of cells combination, while demonstrating the tradeoff between the two. In turn, Table \ref{tab:losses_costs} shows that the multi-cell approach is more robust to changes in advertising costs, which often occur.

Table \ref{tab:losses_scores} suggests a specific circumstance under which the direct approach is superior to ours. If the practitioner has reason to believe that the true optimal exposure rate is low, then running an experiment testing only low exposure rates and using the resulting data to approximate the expected profit function directly will likely yield a better outcome than our proposed approach. This is a preferable course of action because testing only low exposure rates is cheaper than inducing more variation across them. 

\section{Implementing the multi-cell experiment} \label{sec:implem}

This section provides some guidance for how researchers (or an ad platform) might implement our experimental design in practice. First, we discuss the main set of implementation requirements for the model. Second, we address the fact that many ad campaigns entail multiple advertising exposures, whereas our model assumes treatment is binary. Third, we explain how certain strong functional form assumptions yield precise prescriptions for the number of cells and propensity score values. We highlight how similar guidance under weaker assumptions is more difficult.

\subsection{Testing requirements} \label{sec:implem:testing}

Implementing our multi-cell experimental design in practice requires knowledge of certain inputs for the model and access to the appropriate experimentation service on an advertising platform. We discuss each of these in more detail below. Importantly, our proposed design does not require additional budget to be implemented relative to the usual experiment design that randomizes eligibility to receive treatment. 

One key input to our model is the monetary value of an outcome, $\delta$. For ``direct response'' campaigns, in which the advertiser has a particular outcome in mind, for example, to increase sales of a product, $\delta$ would be the average profit margin on products sold through the campaign. This quantity should be known, or estimable, to the advertiser. Most ad platforms allow advertisers to programatically connect outcomes with their monetary conversion values so that all reporting reflects this information.\footnote{For example, on Meta, see \url{https://www.facebook.com/business/help/296463804090290?id=561906377587030}, accessed on 12/19/2023.} Although outcomes with direct monetary values are the most natural fit for our model, any outcome that the advertiser deems of value could work provided that the advertiser can assign a monetary value to the outcome. 

The second key input is the cost of treating a fraction $\phi$ of the target audience, $\kappa(\phi)$. Platforms share information with advertisers that can be used to estimate $\kappa(\cdot)$ when they set up their campaigns. Common campaign planning tools present predicted campaign reach (in terms of users) as a function of budget, conditional on audience targeting parameters. This allows the advertiser to understand the fraction of the audience that they can expect to reach for a particular budget choice, $B$, or effectively, to understand $\phi=\kappa^{-1}(B)$.  

The advertising platform must be capable of implementing multi-cell experiments with sufficient flexibility in their configuration. If the platform creates separate test/control splits within each cell (as Meta does), it must allow for \textit{either} different splits across cells while keeping the budget per cell fixed \textit{or} different budgets across cells. This flexibility is necessary to create the appropriate variation in $p(Z_c=1)$. Similarly, if the platform creates one control group and $C$ test groups, which is equivalent to our design, then our method requires flexibility in the relative size of each group or the budget allocation across groups.

To create variation in $p(Z_c=1)$ across cells, the goal is to create variation in the (expected) budget per user. Although ad platforms do not provide explicit control of the budget per user, advertisers can use several levers to indirectly affect it. For example, all platforms allow advertisers to set overall campaign budgets and provide information on the expected campaign audience size, given targeting parameters and budget levels. Furthermore, all major ad platforms enable ad frequency limits. Together, these tools allow advertisers to roughly control the (expected) budget per user across cells.

For simplicity, normalize the size of the target audience to one and let a fraction $\Pr(\mathcal{C}=c)$ be randomly allocated to cell $c$. Denote the total budget allocated to this cell by $B_c$. Then, the budget per user in cell $c$ is $\tilde{B}_c=\frac{B_c}{\Pr(Z_c=1|\mathcal{C}=c)\times \Pr(\mathcal{C}=c)}$, such that $p(Z_c=1)=\kappa^{-1} \left ( \tilde{B}_c \right)$. Hence, the experimenter can vary the budget per user by (i) allocating different fractions $B_c$ of the original budget, $B$, (ii) choosing different values for $\Pr(Z_c=1|\mathcal{C}=c)$, or (iii) assigning different fractions of users to the different cells, $\Pr(\mathcal{C}=c)$. Thus, the experimenter is able to generate variation in the budget per user across cells, which, in turn, generates variation in $p(Z_c=1)$ through the function $\kappa(\cdot)$. We discuss this intuition in more detail in Appendix F and the challenge of optimally choosing $C$ and $p(Z_c)$ in Section \ref{sec:implem:cells}.

After the experiment is complete, the platform must report $\psi_{dzc}$ and $p(Z_c=1)$. Note that these quantities are aggregated, such that they do not require access to any individual-level data.

\subsection{Addressing multiple advertising exposures} \label{sec:implem:multi}

The approach we took casts advertising as a binary treatment variable and assumes away the existence of multiple ad exposure effects. If these effects exist, then this approach is inadequate. One possible alternative is to incorporate the number of previous exposures as a covariate into the model, which we do in Appendix C and where we discuss the circumstances under which this approach is valid. Here, we provide a more informal discussion of the consequences of ignoring multiple exposures. 

When these exposures are relevant but ignored, the exclusion restriction given in Assumption \ref{assum:basic} is violated. To see this, notice that, if ignored, these exposures become part of the error term associated with the potential outcomes. At the same time, they are correlated with the instrument: if different cells are associated with different budgets, which can thus be seen as the instrument, then higher budgets should be associated with higher number of impressions. However, as we illustrate in Appendix C, ignoring the total number of impressions can be inconsequential when their effect is negligible. Whether this is the case depends on the specific setting.

One example where ignoring repeated exposures might be reasonable are settings with low frequency caps, because they induce a low number of impressions per user overall. Even though all major ad platforms make frequency caps available (e.g., The Trade Desk, Google’s DV 360, and Amazon Advertising\footnote{The Trade Desk: \url{https://partner.thetradedesk.com/v3/portal/api/doc/FrequencyConfigurationBasicCaps}; Google: \url{https://support.google.com/displayvideo/answer/2696786?hl=en}; Amazon: \url{https://advertising.amazon.com/library/guides/frequency-capping}.}), there is little agreement in the industry on whether this number should be high or low. For example, an analysis by Meta found that ``a frequency cap of at least 1 to 2 per week was able to capture a substantial portion of the total potential brand impact.''\footnote{\url{https://www.facebook.com/business/news/insights/effective-frequency-reaching-full-campaign-potential}.} This is roughly consistent with a paper by \cite{ywz2013}. The Trade Desk offers substantially different guidance, while cautioning that there is no one-size-fits-all approach to setting frequency caps.\footnote{\url{https://www.thetradedesk.com/us/resource-desk/ideal-frequency-optimization.}} Ultimately, we think advertisers should test out different frequency caps to understand what is best for them \citep{forbes2019}.

Whether repeated exposures are significant also depends on the specific outcome variable. Research on this topic remains relatively scarce. \cite{sahni2015} finds significant effects on calls to restaurants. In turn, \cite{jlr2016} finds significant effects on sales when imposing certain functional forms, but more flexible specifications cast doubt on this finding. \cite{lewis2014} examines 30 ad campaigns on Yahoo and finds mixed evidence of repeated ad exposure effects on click-through rates. \cite{snk2019} find some evidence of frequency on visits at one online retailer. 

Overall, it is difficult to explicate the specific conditions of an ad campaign that are likely to generate repeated exposure effects. To our knowledge, the literature has not been able to detect effects of repeated exposures on purchases, which is the outcome variable of the Facebook experiment we use in our simulation. This suggests that the approach that casts advertising as a binary treatment may be appropriate when this is the outcome of interest. Otherwise, if a researcher feels multiple exposures are important, then so long as they (or the ad platform) can collect the necessary data, they could instead apply our extended model that conditions on exposure count to maintain the validity of the exclusion restriction.

\subsection{Choosing the number of cells and propensity score values} \label{sec:implem:cells}

The choice of $C$ and $p(Z_c=1)$ are key design elements of our approach, taking the budget for the experiment as given. Based on the results from the simulations in Section \ref{sec:dir_opt}, we offer some general suggestions on how researchers can best proceed.

As we discuss in Appendix D, a common approach is to assume that the MTR functions are linear, or to approximate them with a linear function. The requirement to obtain such an approximation is to observe two values of $p(\cdot)$ that are different from each other and strictly between zero and one. This can be accomplished with a two-cell design with an unequal allocation of the overall budget to each cell. In theory, any two values of $p(\cdot)$ suffice to obtain this linear approximation. However, our exercises with finite samples from Section \ref{sec:dir_opt} suggest that more variation in the $p(\cdot)$ is still necessary to obtain credible approximations to the MTR functions.

A different approach that can also be implemented with just two different values of $p(\cdot)$ is the one used, for example, by \cite{htv2001,htv2003}: to assume an underlying DGP featuring a normal distribution. Under this assumption, the method to compute the MTR functions is not that of BMW; however, our proposed multi-cell design can still be used with this different estimation method. We consider this specific case in Appendix D, which, like the linear case, features monotone MTR functions.

Without such strong functional form assumptions, however, it becomes difficult to obtain clear guidelines on how to choose $C$ or $p(Z_c=1)$. In Appendix H, we consider MTE functions that are monotonic or that satisfy the assumption of monotone treatment response \citep{manski1997}. We find that these assumptions do not necessarily produce sufficiently ``well-behaved'' MTE functions so as to enable precise guidance. We leave to additional work on how to best leverage such assumptions.

Another practical concern is estimation precision. With a finite number of units (consumers), the choice of the number of cells in the experiment creates a type of bias-variance trade-off. More cells generate more values of the propensity score, theoretically enabling a more flexible approximation of the MTE function, and thus decreasing bias. But as the number of units per cell decreases, the estimates of the approximating function will become noisier, and thus increasing variance. Hence, the number of cells can be seen as somewhat akin to the bandwidth in nonparametric estimation. Without strong assumptions on the underlying MTE function, it is not possible to establish how to best choose the number of cells given the available sample size. Nevertheless, as we illustrate in Section \ref{sec:dir_opt}, sample sizes commonly used in digital advertising settings tend not to be limiting; for instance, the experiment we considered in our simulations featured more than 25 million observations. Therefore, we expect this bias-variance trade-off concern to be secondary.

\section{Conclusion} \label{sec:conc}

Randomized experiments are considered an attractive tool to estimate the impacts of treatments. When treatment assignment cannot be randomized, a common approach is to randomize eligibility to receive treatment instead, leading to one-sided noncompliance. Nevertheless, decision-makers who conduct experiments are often interested in obtaining information to assist them in making specific decisions, and not just measuring the effects of treatment per se. Unfortunately, the typical experimental design with one-sided noncompliance does not provide enough information to assist with many decisions.

This paper proposes an approach to obtain such information. This approach combines a novel multi-cell experimental design and modern estimation techniques, where the former leads to the collection of data that contain more information about treatment effects and the latter exploits this information. Our method to estimate the MTE functions draws inspiration from \cite{bmw2017}. However, our approach differs significantly by being tailored specifically to leverage a multi-cell experimental design characterized by one-sided noncompliance, which, by construction, provides a suitable instrumental variable. We point out that, in a single-cell experiment with one-sided noncompliance, direct application of \cite{bmw2017} is infeasible without additional strong assumptions.

Using data from an online advertising experiment at Facebook, we addressed the performance of our proposed multi-cell experimental design vis-\'{a}-vis that of the typical experimental design and of direct budget optimization. To do so, we implemented the aforementioned estimators on simulated data calibrated based on this experiment. We found that the decisions obtained from our design yield lower losses in expected profit than those from these alternatives.

\subsection*{Limitations and future research}

Three natural questions arise in the context of many approximations. First, to what extent can the approximated MTE function be used be generalized beyond the specific audience and ad platform on which it was obtained? In Appendix C, we extend our model to condition on a scalar $X$ that we interpret as being the number of previous exposures for a user. However, it might be possible to further generalize this specification to make $X$ vector valued. Allowing for a rich enough set of conditioning variables in $X$ might be one way to help generalize the estimated MTE function to other advertising contexts. 

Second, is there a way to intelligently choose the number of cells and propensity score values? Intuitively, the higher the number of cells and the more variation there is in the propensity score values, the better. Nevertheless, the quality of the resulting approximations depends crucially on the underlying DGP. Without strong restrictions, such as the normality assumption discussed in Section \ref{sec:implem:cells}, it is difficult to obtain specific guidance for these choices. The choice of propensity score values is akin to the choice of knot values for numerical integration, with the added component that the obtainable values depend on the budget; for example, a higher budget per user is required to obtain a higher exposure rate, that is, a high value for the propensity score. Incorporating this additional component to the problem adds yet another layer of complexity.

Third, are there additional reasonable restrictions that might improve the quality of the approximation? Imposing theory-based restrictions on the underlying DGP might be a way to make progress on obtaining theoretical bounds on the quality of the approximation even when ignoring estimation and monetary concerns. Ideally, such restrictions would impose enough structure to imply a ``well-behaved'' MTE function, whose properties could then be leveraged for approximation. To this end, in Appendix H we consider two commonly made and interpretable assumptions, monotone treatment response and monotonicity of the MTE function, but find that they are insufficient to generate a DGP whose properties can be exploited for approximation. One possible alternative for future research is to replace the polynomial approximation with more flexible functional forms that can better leverage and incorporate these restrictions.

It is possible that having access to one or more real multi-cell tests could help guide us to solutions to some of these questions.  However, lacking a real multi-cell test, we do not have any information about what a true DGP would look like in terms of the resulting MTE. This makes it impossible for us to assess the true quality of our approximation, since we do not know if our assumed DGP bears any resemblance to a true DGP. We chose polynomials because this method is consistent with \cite{bmw2017}, possesses analytic integrals, and has favorable approximating properties. Any approximation method that is well-defined on the unit interval could potentially work (e.g., Bernstein polynomials). However, without any knowledge of a real-world DGP, it is hard to assess the relative accuracy of one approximation technique over the other. After running a sufficient number of multi-cell tests, an advertising platform could attempt to characterize common features and functional forms of the resulting MTEs to provide some guidance on preferred approximation methods.

\thispagestyle{empty}

\newpage

\maketitle

\appendix


\addcontentsline{toc}{section}{Appendix } 
\part{Appendix} 
\parttoc 

\newpage

\definecolor{okorange}{RGB}{230,159,0}
\definecolor{okblue}{RGB}{86,180,233}
\definecolor{okgreen}{RGB}{0,158,115}
\definecolor{okyellow}{RGB}{240,228,66}
\definecolor{okdarkblue}{RGB}{0,114,178}
\definecolor{okred}{RGB}{213,94,0}
\definecolor{okmagenta}{RGB}{204,121,167}
\providecommand\blue[1]{{\color{okblue}#1}}
\providecommand\dkblue[1]{{\color{okdarkblue}#1}}
\providecommand\orange[1]{{\color{okorange}#1}}
\providecommand\green[1]{{\color{okgreen}#1}}
\providecommand\red[1]{{\color{okred}#1}}
\providecommand\magenta[1]{{\color{okmagenta}#1}}
\providecommand\gray[1]{{\color{gray}#1}}
\providecommand\black[1]{{\color{black}#1}}
\usetikzlibrary{shapes,decorations,arrows,calc,arrows.meta,fit,positioning}
\tikzset{
    -Latex,auto,node distance =1 cm and 1 cm,semithick,
    box/.style ={rectangle, draw, inner sep=0.08cm,fill,node contents={}},
    open/.style = {circle, draw, inner sep=0.05cm,node contents={}},
    bidirected/.style={Latex-Latex,dashed},
    unidashed/.style={-Latex,dashed},
    gray80/.style={color=black!20!white},
    gray50/.style={color=black!50!white},
    red/.style={color=okred},
    dkblue/.style={color=okdarkblue},
    blue/.style={color=okblue},
    magenta/.style={color=okmagenta},
    green/.style={color=okgreen},
    orange/.style={color=okorange},
    el/.style = {inner sep=2pt, align=left, sloped},
    square/.style = {rectangle, fill, draw, inner sep=0.07cm,node contents={}},
    point/.style = {circle, draw, inner sep=0.05cm,fill,node contents={}},
    every label/.style = {label distance=0mm}
}

\renewcommand{\thesubsection}{\Alph{subsection}}
\counterwithin{table}{section}   
\counterwithin{figure}{section} 
\numberwithin{equation}{section}
\numberwithin{subsection}{section}
\counterwithin{assum}{section}   

\section{Rewriting the firm's decision problem}\label{app:dec_prob}

We demonstrate how the firm's decision problem given in equation (2) can be rewritten as in equation (7). We begin by deriving the terms in equation (6). First, we have that:
\begin{align}\label{eq:exp1_detailed}
      \mathbb{E} \left [Y_1 \middle \vert  U \leq \phi \right ]&= \int_0^{\phi} \int_{y_1\in\mathcal{Y}_1}  y_1 \frac{f(y_1,u)}{\Pr(U \leq \phi)}dudy_1 \nonumber \\
	&=\int_0^{\phi} \left ( \int_{y_1\in\mathcal{Y}_1}  y_1 f(y_1|u)dy_1 \right )\frac{f(u)}{\phi}du \nonumber \\ 
		&=\int_0^{\phi} \mathbb{E} \left [Y_1 \middle \vert U=u \right ] \times \frac{1}{\phi}du \nonumber \\ 
	&\equiv\int_0^{\phi} m_1(u) \frac{1}{\phi} du, 
\end{align}
where in the third equality we used that $U \sim U(0,1)$ and defined $m_1(u) \equiv \mathbb{E} \left [Y_1 \middle \vert U=u \right ]$. Second,
\begin{align}\label{eq:exp0_detailed}
      \mathbb{E} \left [Y_0 \middle \vert  U > \phi \right ]&= \int_{\phi}^1 \int_{y_0\in\mathcal{Y}_0}  y_0 \frac{f(y_0,u)}{\Pr(U > \phi)}dudy_0 \nonumber \\
	&=\int_{\phi}^1 \left ( \int_{y_0\in\mathcal{Y}_0}  y_0 f(y_0|u)dy_0 \right )\frac{f(u)}{1-\phi}du \nonumber \\ 
	&=\int_{\phi}^1 \mathbb{E} \left [Y_0 \middle \vert U=u \right ]  \times  \frac{1}{1-\phi}du \nonumber \\ 
	&\equiv\int_{\phi}^1 m_0(u) \frac{1}{1-\phi} du,
\end{align}
where in the third equality we also used that $U \sim U(0,1)$ and defined $m_0(u) \equiv \mathbb{E} \left [Y_0 \middle \vert U=u \right ]$.

Plugging equations (\ref{eq:exp1_detailed}) and (\ref{eq:exp0_detailed}) back into equation (2) yields:
\begin{align*}
	&\max_{\phi\in[0,1]} \left ( \delta  \times \left \{ \int_0^{\phi} m_1(u)du + \int_{\phi}^1 m_0(u)du \right \} - \kappa(\phi) \right ) \\  
	&\max_{\phi\in[0,1]} \left ( \delta \times \int_{0}^1 m_0(u)du +   \delta  \times \left \{ \int_0^{\phi}  \left [ m_1(u) -  m_0(u) \right ]du   \right \} - \kappa(\phi) \right ) \\
	&\max_{\phi\in[0,1]} \left ( \delta \times \mathbb{E} \left [ Y_0 \right ] +   \delta  \times  \int_0^{\phi} \text{MTE}(u)   du  - \kappa(\phi) \right ) \\
	&\max_{\phi\in[0,1]} \left ( \delta  \times  \int_0^{\phi} \text{MTE}(u)   du  - \kappa(\phi) \right ),	
\end{align*}
which establishes equation (7).

\section{Details of calibrated simulations} \label{app:cal_sim}

Below we provide more details about how we simulate data based on Study 4 from \cite{gordon_zettelmeyer_2019} for our analysis in Section 4.

\begin{enumerate} 
\item \textbf{MTR functions specification. } We specify true MTR functions of the form:\footnote{Here we follow the approach from \cite{mst2018}, who directly specified the MTR functions in their simulations. An alternative and commonly used approach, employed, for instance, in \cite{hv2007}, is to specify the joint distribution of $Y_1$, $Y_0$ and $U$, which is often normal, and work with the implied MTR functions. We follow this alternative approach in Appendix \ref{app:observables}, where we consider the role of multiple ad exposures.}
\begin{align*}
m_1(u) & = m_{10} + m_{11}u + m_{12}u^2 \\
m_0(u) & = m_{00} + m_{01}u + m_{02}u^2 + m_{03}u^3  
\end{align*}
We chose these functional forms because they are the polynomials of lowest degree that the simplest version of our design---with only two cells---cannot  recover. With three or more cells, our approach can perfectly recover the true MTR functions. These forms imply that the MTE is a cubic polynomial.

\item \textbf{Implied moments. } We need to choose parameter values such that the implied moments match the observed $\psi_{11}$, $\psi_{01}$ and $\psi_{00}$. Under these functional forms, from equations (10), (11), and (12) it follows that:
\begin{align*}
    \psi_{11} &= m_{10} + \frac{m_{11}}{2}p(1) + \frac{m_{12}}{3}p(1)^2 \\
    \psi_{01} &= m_{00} + \frac{m_{01}}{2} \left [1 + p(1) \right ] + \frac{m_{02}}{3} \left [ 1 + p(1) + p(1)^2 \right ] + \frac{m_{03}}{4} \left [ 1 + p(1) + p(1)^2 + p(1)^3 \right ] \\
    \psi_{00} &= m_{00} + \frac{m_{01}}{2} + \frac{m_{02}}{3} + \frac{m_{03}}{4}
\end{align*}

\item \textbf{Choosing parameter values. } Based on the observed moments, we need to choose seven parameters, $\left \{ \left \{m_{1c} \right\}_{c=0}^2, \left \{m_{0c} \right\}_{c=0}^3 \right\} $, to satisfy the three constraints above, plus the additional constraint that the MTR functions must be between 0 and 1, because the outcome of Study 4 is binary (purchase). We consider two sets of parameters to illustrate our proposed approach, labeled DGP1 and DGP2. For each DGP, we choose four parameters and then solve for the remaining three by plugging the observed $\psi_{11}$, $\psi_{01}$, $\psi_{00}$, and $p(1)$ into the system of equations above. We choose different set of values for the DGPs to generate different patterns of the MTE function. 
Since both DGPs correspond to functions that are very close to quadratic, we expect that a two-cell design will suffice to obtain good approximations.

\item \textbf{Simulating the multi-cell experiment. } We use these MTE functions to simulate data from our experimental design with two cells. This design limits us to a linear approximation to $m_1(\cdot)$ and a quadratic approximation to $m_0(\cdot)$. First, we set the propensity scores in our simulated multi-cell experiment. Because Study 4 assigned users to be eligible to receive treatment with probability 0.7, we consider this to be one cell and add a second cell in which this probability equals 0.3. For the first cell we keep the propensity score at the original value, $p(Z_1=1)=0.37$, and set the propensity score for the second cell so that $\Pr(Z_1=1 | C=1) \times p(Z_1=1) = \Pr(Z_2=1 | C=2) \times p(Z_2=1)$, implying that $p(Z_2=1) \approx 0.863 $. We use these propensity scores and the underlying MTE functions to compute the $\psi$s from equations (10) and (11) that would have been observed had our design been implemented. 

\end{enumerate} 

\section{Dealing with multiple ad exposures}\label{app:observables}

\subsection{General incorporation of observable characteristics}\label{sec:app_genobs}

We now demonstrate how to incorporate observable characteristics, captured in a vector $X$, into the model above. Three changes have to be made. First, equation (3) is replaced with:
\begin{align}\label{eq:sel_eq_X}
D&=\mathbbm{1} \left \{p(Z,X)\geq U \right \}.
\end{align}
Importantly, notice that the error term, $U$, remains additively separable.

Second, Assumption 1 is replaced with:
\begin{assum}\label{assum:basic_X} \hfill\break
	\subasu $U \independent Z \vert X$, where $\independent$ denotes statistical independence. \label{assum:uncorr_X} \\
	\subasu $\mathbb{E} \left [Y_d \middle \vert Z,X, U \right ] = \mathbb{E} \left [Y_d \middle \vert X, U \right ]$ and $\mathbb{E} \left [Y_d^2 \right ]< \infty$ for $d\in\{0,1\}$. \label{assum:excl_X} \\
	\subasu $U$ is continuously distributed conditional on X. \label{assum:cont_X} 
\end{assum}

Finally, we replace Assumption 2 with:
\begin{assum}\label{assum:exp_X} \hfill\break
	\subasu $\Pr (Z_c = z | X,\mathcal{C}=c) \in (0,1) $ for all $X$, $c$ and all $z$. \label{assum:exp1_X} \\
	\subasu $p(Z_c=1|X)\equiv\Pr \left (D=1 \middle \vert X,Z_c=1 \right ) \in (0,1)$ for all $c$ and $X$. \label{assum:exp2_X} \\
 	\subasu $p(Z_c=1|X) \neq p(Z_{c^\prime}=1|X)$ for all $c \neq c^\prime$ and $X$. \label{assum:exp3_X} 
\end{assum}

In summary, Assumptions \ref{assum:basic_X} and \ref{assum:exp_X} simply add conditioning on $X$ to Assumptions 1 and 2. The MTR and MTE functions become $m_d(u,x)\equiv\mathbb{E} \left [Y_d \middle \vert U=u, X=x \right ]$, where $d\in\{0,1\}$, and $\text{MTE}(u,x)\equiv \mathbb{E} \left [Y_1 - Y_0 \middle \vert X=x, U=u \right ]= m_1(x,u) - m_0(x,u)$ , respectively.

\subsection{Number of previous ad exposures as a covariate}\label{app:model_with_s}

In certain cases, one can expect that multiple exposures to an ad can have a significant impact on outcomes. We now show how multiple ad impression opportunities can be incorporated as a covariate into our setup.

To this end, assume that each user can be exposed to the ad $S$ times. We index all variables by $s=1,\dots,S$, to indicate that they are associated with impression opportunity $s$. We define $X_s$ as the number of times the user was exposed to the ad \textit{prior} to impression opportunity $s$. 

In this specific case in which previous exposures is a covariate, we use the budget designated to each user as the instrument that generates variation in the probability of exposures and in the number of previous ad exposures. We denote the budget originally designated to a user by $B$. As the advertiser shows their ad over impression opportunities, this budget diminishes due to the auction payments. We denote the remaining available budget at impression opportunity $s$ by $B_s$. 

We follow the model with essential heterogeneity from \cite{huv2006}:\footnote{Conceptually, this model could be further generalized by letting $Y_d=\mu_d(X,\epsilon_d)$, where $d\in\{0,1\}$. This may be desirable to explicitly model, for example, binary potential outcomes. However, we maintain the additively separable structure for simplicity.}
\begin{align}
    \begin{split}
        Y_{1s} &= \mu_1(X_s)+\epsilon_{1s} \\
        Y_{0s} &= \mu_0(X_s)+\epsilon_{0s} \\
        D_s&=\mathbbm{1} \left \{p(B_s,X_s)\geq U_s \right \}. 
    \end{split}
\end{align}

We complete this model with the following assumption. As we discuss below, this assumption is key for the analysis we describe next.
\begin{assum}\label{assum:iid} \hfill\break
    Conditional on $X_s$, $\epsilon_{1s}$, $\epsilon_{0s}$, and $U_{s}$ are i.i.d.~across $s$.
\end{assum}

The i.i.d.~condition across $s$ deserves attention, especially regarding $U_s$. We interpret the variable $U$ as the ease with which a user can be exposed to the ad. Thus, it would be reasonable to expect that this term would be correlated across $s$. In other words, users may have a persistent unobservable that determines how easy it is to expose them to an ad. Consequently, Assumption \ref{assum:iid} maintains that conditioning on the $X_s$ eliminates this persistent term. The DAG presented in Figure \ref{fig:dag} illustrates a violation of Assumption \ref{assum:iid}.\footnote{We thank an anonymous reviewer for providing us with this figure.}

\begin{figure}
\caption{DAG showing violation of Assumption \ref{assum:iid}}
\label{fig:dag}
\begin{tikzpicture}[scale=.85]
\node (D)  at (4, 0) [point, dkblue, label=below right:{$\dkblue{D}$}] ;
\node (Y)  at (6, 0) [point, red, label=below right:{$\red{Y}$}] ;
\node (Bs) at (2, 0) [point, label=below:{$B_{s}$}] ;
\node (Us) at (4, 3) [open, label=above right:{$U_{s}$}] ;
\node (Xs) at (2, 2) [square, green, label=above:{$\green{X_{s}}$}] ;
\node (B) at (0, 0) [point, label=below left:{$B$}] ;
\node (U0) at (0, 3) [open, label=above left:{$U_{0}$}] ;

\draw (Bs) edge [] (D);
\draw (D)  edge [] (Y);
\draw (Us) edge [unidashed] (D);
\draw (Us) edge [unidashed] (Y);
\draw (Xs) edge [] (Bs);
\draw (Xs) edge [] (D);
\draw (Xs) edge [] (Us);
\draw (B) edge [] (Bs);
\draw (B) edge [] (Xs);
\draw (U0) edge [unidashed] (Xs);
\draw (B) edge [bend left=-60] (D);
\draw (U0) edge [bidirected, bend left=30] (Us);

\end{tikzpicture}

\end{figure}

In Figure \ref{fig:dag}, Assumption \ref{assum:iid} is violated: even conditional on $X_s$, the unobservables $U_s$ remain correlated across $s$. As a result, the available budget, $B_s$, becomes an invalid instrument for $D_s$. This could happen if among people exposed to many ads (high $X_s$), users who are hard to reach (high $U_s$) had a high initial budget (large $B$) relative to users who are easier to reach (low $U_s$). This would invalidate the available budget, $B_s$, as an instrument for $D_s$.

Assumption \ref{assum:iid} restores this validity by effectively assuming away the dependence of $X_s$ on $U_s$. This arguably is a strong assumption, which could become more palatable if additional variables, such as user and impression opportunity characteristics, were incorporated into $X$. However, this is a caveat that should be taken into consideration.  

\subsection{Budget optimization problem with multiple exposures}\label{app:mult_budg_opt}

We now present the advertiser's optimization problem in the presence of multiple ad exposures, which can be seen as a generalized version of the problem presented in Section 2.3. Unlike the original problem, it will be more convenient to cast it as a function of $B$. Hence, the problem is given by:
\begin{align}\label{eq:comp_op}
	\max_{{B}\in\mathbb{R}_+} \delta\times\mathbf{E} \left [  Y \middle \vert B \right ] - B
\end{align}
where all variables are as previously defined. Notice that:
\begin{align}\label{eq:obs_cexp}
    \mathbf{E} \left [  Y \middle \vert B \right ] &= \mathbf{E} \left \{ \mathbf{E} \left [  Y \middle \vert X, B \right ] \middle \vert B \right \} \nonumber \\
    &= \sum_{x=0}^{S-1} \Pr \left (X=x \middle \vert B \right ) \times \mathbf{E} \left [  Y \middle \vert X=x, B \right ] \nonumber \\
     &= \sum_{x=0}^{S-1} \Pr \left (X=x \middle \vert B \right ) \times \left \{ \mathbf{E} \left [  D Y_1 \middle \vert X=x, B \right ] + \mathbf{E} \left [  (1-D) Y_0 \middle \vert X=x, B \right ] \right \} \nonumber \\
     &= \sum_{x=0}^{S-1} \Pr \left (X=x \middle \vert B \right ) \times \left \{ \int_0^{p(B,x)}m_1(x,u)du + \int_{p(B,x)}^1m_0(x,u)du \right \}.
\end{align}
Note that we are slightly abusing notation by denoting the propensity score as a function of the original budget, $B$, instead of the available budget after $x$ impressions. Conditional on $x>0$, we expect that the available budget to be lower than $B$ but this difference depends on the particular ad payment structure. We will illustrate this in our simulations below.

Plugging (\ref{eq:obs_cexp}) into (\ref{eq:comp_op}) yields:
\begin{align}\label{eq:comp_op_}
	\max_{{B}\in\mathbb{R}_+}  \left \{ \delta\times\sum_{x=0}^{S-1} \Pr \left (X=x \middle \vert B \right ) \times \left [ \int_0^{p(B,x)}m_1(x,u)du + \int_{p(B,x)}^1m_0(x,u)du \right ]  - B \right \}.
\end{align}

Whether there is a unique solution to this problem depends on the functions $\Pr \left (X=x \middle \vert B \right )$ and $p(B,x)$, which are either known (i.e., the platform reports them to the advertiser) or estimable since $X$, $B$, and $D$ are observable. Hence, the remaining pieces that have to be estimated in order for the advertiser to solve this budget optimization problem are the MTR functions. We now discuss how to accomplish this estimation task.

\subsection{Sequential approximations to MTE function}\label{app:seq_appr}

We propose a sequential approach that approximates the MTE function at each value of $X_s$ separately. This is motivated by the fact that: (i) online advertising experiments involve a large number of users; (ii) the data are often stored sequentially as they arrive for the duration of the experiment; and (iii) the only variable we are including in $X_s$ is the previous number of impressions. Should other variables be included in the model, alternative approaches can also be considered.

Under Assumption \ref{assum:iid}, the MTR functions take the form:
\begin{align}\label{eq:mtr_mult}
    m_d(x,u)=\mu_d(x)+\mathbb{E} \left [\epsilon_{ds} \middle \vert X_s=x, U_s=u \right ],
\end{align}
where $d\in\{0,1\}$. Hence, using a multi-cell experimental design with sequential storage of data allows us to recover the following moments:
\begin{align}\label{eq:psi1_seq}
    \psi_{1bx}&=\mathbb{E} \left [Y \middle \vert D_s=1, X_s=x, B_s=b \right ] \nonumber \\
    &=\mathbb{E} \left [Y_1 \middle \vert U_s \leq p(b,x), X_s=x, B_s=b \right ] \nonumber \\
    &= \mu_1(x)+\frac{1}{p(b,x)}\int_0^{p(b,x)}\mathbb{E} \left [\epsilon_{1s} \middle \vert U_s=u \right ]du
\end{align}
and 
\begin{align}\label{eq:psi0_seq}
    \psi_{0bx}&=\mathbb{E} \left [Y \middle \vert D_s=0, X_s=x, B_s=b \right ] \nonumber \\
    &=\mathbb{E} \left [Y_1 \middle \vert U_s > p(b,x), X_s=x, B_s=b \right ] \nonumber \\
    &= \mu_0(x)+\frac{1}{1-p(b,x)}\int_{p(b,x)}^1\mathbb{E} \left [\epsilon_{0s} \middle \vert U_s=u \right ]du.
\end{align}
We can approximate the MTE function at each value of $x$ separately using $\psi_{1bx}$, $\psi_{0bx}$, and $p(b,x)$ following the procedure described in Section 3.3.

This procedure has an important feature. As impressions are obtained, the available budget decreases; that is, the number of different values of $B_s$ decreases as $X_s$ increases. Consequently, the degree of the approximating polynomial is lower for higher values of previous exposures to the ad, rendering this approximation, at least theoretically, poorer.

\subsection{Calibrated simulations with multiple exposures}\label{app:cal_sim_with_s}

We now describe the specification we chose for our calibrated simulations with multiple ad impression opportunities. This specification is motivated solely for simplicity, but the parameter values are chosen based on the data from the Facebook experiment we consider.

\subsubsection{Distributional and parametric assumptions}

To implement our calibrated simulations, we make the following normality assumption:
\begin{align}\label{eq:normal}
    \begin{bmatrix}
        \epsilon_{1s} \\ \epsilon_{0s} \\ V_s
    \end{bmatrix}\Big\vert X_s \sim N \left ( \begin{bmatrix}
        0 \\ 0 \\ 0
    \end{bmatrix}, \begin{bmatrix}
        \sigma_1^2 & \rho_{10}\sigma_1 \sigma_0 & \rho_{1V}\sigma_1 \\
        \rho_{10}\sigma_1 \sigma_0 & \sigma_0^2 & \rho_{0V}\sigma_0 \\
        \rho_{1V}\sigma_1 & \rho_{0V}\sigma_0 & 1
    \end{bmatrix} \right )
\end{align}
where $V_s=\Phi^{-1}(U_s)$ and where $\Phi(\cdot)$ denotes the cdf of the standard normal distribution. We further assume that:
\begin{align}\label{eq:mu_quad}
    \mu_d(x)=\mu_{d0}+\mu_{d1}x+\mu_{d2}x^2
\end{align}
where $d\in\{0,1\}$. Consequently, we obtain:
\begin{align}\label{eq:mtr_mult_norm}
    m_d(x,u)=\mu_{d0}+\mu_{d1}x+\mu_{d2}x^2 + \rho_{dV}\sigma_d \Phi^{-1}(u).
\end{align}

The normality assumption in (\ref{eq:normal}) is oftentimes maintained and is of great aid in identifying and estimating the MTE function. As discussed, for example, in \cite{mt2018}, all that is necessary to recover $\rho_{dV}\sigma_d$, where $d\in\{0,1\}$, is for the propensity score to have two distinct values between 0 and 1 at just one value of $x$. Consequently, with enough variation in $x$ to estimate $\mu_d(\cdot)$ it is possible to recover the entire MTR functions under this weaker condition.

However, this convenience also imposes non-trivial restrictions on the model. For instance, it imposes that the MTR functions are monotonic in $u$, ruling out the DGPs we considered in Section 4, and that they tend to plus and minus infinity at the extremes of the unit interval, as depicted in Figure \ref{fig:mte_mult}. On the other hand, the approach of approximating the MTR functions with polynomials from BMW is agnostic and does not impose such restrictions on the DGP. 

\subsubsection{Features of simulated experiment}

First, we set $S=3$. In practice, the number of potential ad impressions can be much larger and is a function of the duration of the experiment. We choose a low value solely for simplicity. In addition, such reduction is sometimes implemented in practice via bracketing. For instance, $s=1$ can correspond to the first five impressions, $s=2$ to the following five, and so on.

Second, we assume that:
\begin{align}
    p(B_s,X_s)=\frac{B_s}{B_s+t},
\end{align}
where $t>0$. This is a simple way of ensuring that the probability of exposing a user to the ad is zero when there is no budget left, that is, when $B_s=0$. We choose the parameter $t$ to generate variation in the propensity score across the different values of $B_s$. Also, notice that this probability does not depend on $X_s$, a choice we made solely for simplicity. $X_s$ could be easily incorporated into this propensity function, for instance, via frequency caps.

Third, we assume that the experiment randomizes users into five groups determined by the initial budget allocated to them. In particular, we set $B\in\left \{0,1,2,3,4 \right \}$.

Finally, when the advertiser wins an impression opportunity, they make a payment and the available budget decreases. We simply assume that:
\begin{align}
  B_{s+1} = \begin{cases} B_s -1 & \text{if } D_s=1 \\ B_s & \text{if } D_s = 0  \end{cases}
\end{align}
We choose the budget to decrease by one to facilitate the simulation by aligning the evolution of budget with the initial budget allocation. In practice, decreases in the available budget are determined by auction payments, which almost always vary across impression opportunities.

\subsubsection{Choice of parameter values}

Under this setup, we have to choose values for ten parameters: $\mu_{10}$, $\mu_{11}$, $\mu_{12}$, $\mu_{00}$, $\mu_{01}$, $\mu_{02}$, $\sigma_1$, $\sigma_0$, $\rho_{1V}$, and $\rho_{0V}$.\footnote{Notice that the parameter $\rho_{10}$ is irrelevant for the purposes of computing the MTE function.} To do so, we use data from the same Study 4 we considered in Section 4. It is important to recall that Study 4 only randomized eligibility to ad exposure, denoted by the dummy variable $Z$, and not budget levels.

We set $\sigma_d^2= \text{Var} \left (Y \middle \vert Z=d \right )$, where $d\in\{0,1\}$, and obtain $\sigma_1=0.0185584$ and $\sigma_0=0.0158245$. Nevertheless, it is important to note that, in practice, these equalities are unlikely to hold. 

The data to which we have access to calibrate $\mu_d\equiv \left [\mu_{d0}, \mu_{d1}, \mu_{d2} \right ]'$, where $d\in\{0,1\}$, consist of the following moments: $\mathbb{E} \left[ Y \middle \vert D=0,Z=0 \right ]$, $\mathbb{E} \left[ Y \middle \vert D=0,Z=1 \right ]$, $\mathbb{E} \left[ Y \middle \vert X=1 \right ]$, $\mathbb{E} \left[ Y \middle \vert X=1 \right ]$, $\mathbb{E} \left[ Y \middle \vert X=2 \right ]$, and $\mathbb{E} \left[ Y \middle \vert X=3 \right ]$.\footnote{We aggregated the quantities for $X\geq3$ into a single quantity.} These expressions and their interpretation warrant a few comments.

We only have access to aggregate results at the end of the experiment, but not to the sequential evolution of results. Thus, $Y$ and $X$ are the observed outcomes and total number of ad exposures, respectively, at the end of the experiment. As a consequence, we omit the conditioning on $X$ when $D=0$ because, by construction, it is equal to zero. Furthermore, also by construction, $X>0$ if and only if $D=Z=1$, so we omit the conditioning on $D=1$ and $Z=1$ when $X>0$. Given this structure, we only have two moments informative of $m_0(\cdot)$, $\mathbb{E} \left[ Y \middle \vert D=0,Z=0 \right ]$ and $\mathbb{E} \left[ Y \middle \vert D=0,Z=1 \right ]$, and three moments informative of $m_1(\cdot)$, $\mathbb{E} \left[ Y \middle \vert X=1 \right ]$, $\mathbb{E} \left[ Y \middle \vert X=2 \right ]$, and $\mathbb{E} \left[ Y \middle \vert X=3 \right ]$.

Given (\ref{eq:mtr_mult_norm}), evaluating (\ref{eq:psi0_seq}) at $x=0$ yields two equations that allow us to estimate $\mu_{00}$ and $\rho_{0V}$. This is because we already calibrated $\sigma_0$ and because $p(0,0)$ and $p(1,0)$ are observed in the data from the numbers of observations associated with each $(Z,D,X)$ combination. Solving for these parameters yields $\mu_{00}=0.0002505$ and $\rho_{0V}=-0.006081128$. We then arbitrarily set $\mu_{01}=0.001$ and $\mu_{02}=-0.00075$ so that $m_0(\cdot)$ becomes concave in $x$.

Similarly, given (\ref{eq:mtr_mult_norm}), by equation (\ref{eq:psi1_seq}) we have three equations. Since we already calibrated $\sigma_1$, we have four unknown parameters associated with $m_1(\cdot)$. We set $\rho_{1V}=0.3$ to induce endogeneity in treatment because we obtained a very low value for $\rho_{0V}$, which enables us to use these three equations to solve for $\mu_1$. We obtain $\mu_{10}=0.002443738$, $\mu_{11}=0.007943073$, and $\mu_{12}=-0.001791018$. Notice that these values imply that $m_1(\cdot)$ is also concave in $x$.

\begin{figure}
    \centering
        {\includegraphics[width=0.6\textwidth]{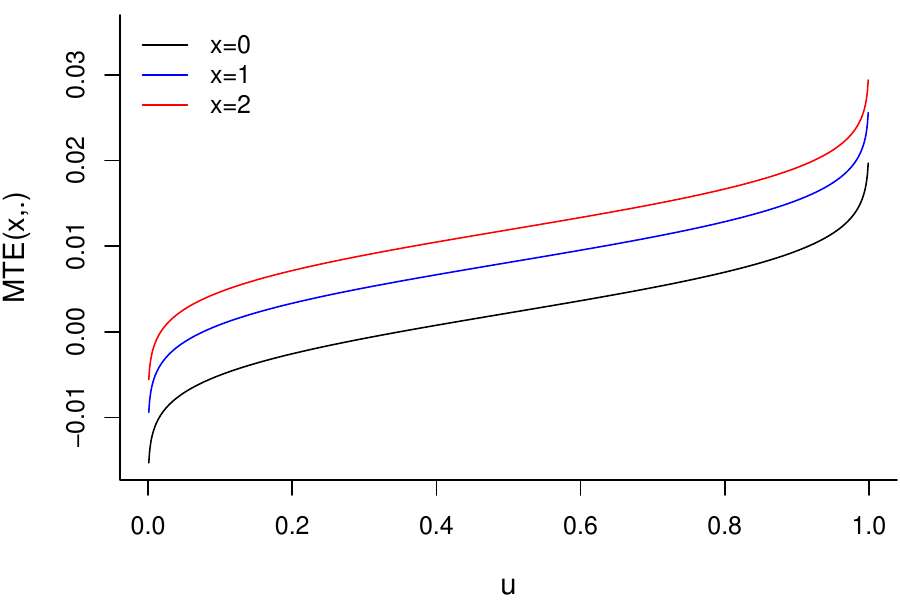}}
\caption{MTE function with multiple exposures}
\floatfoot{\textit{Note:} This figure shows the MTE function implied by the MTR functions from equation (\ref{eq:mtr_mult_norm}). Because we consider $S=3$ impression opportunities and because $x$ is the number of exposures prior to an impression opportunity, the MTE functions takes only three values in $x$.}
    \label{fig:mte_mult}%
\end{figure}

The resulting MTE function at $x\in\{0,1,2\}$ is displayed in Figure \ref{fig:mte_mult}. Because of additive separability, the MTE function does not cross at different values of $x$. While it is increasing in $x$, we can see that it is also concave.

\subsubsection{Results}

\begin{figure}
    \centering
    \begin{subfloat}[ $x=0$ \label{fig:mult_seq0}]
        {\includegraphics[width=0.495\textwidth]{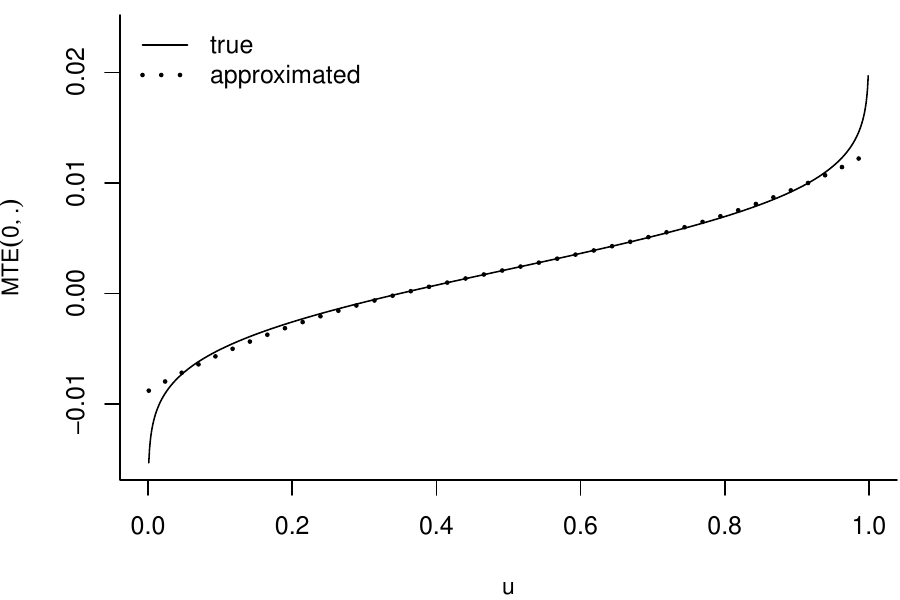}}
    \end{subfloat} 
    \begin{subfloat}[ $x=1$ \label{fig:mult_seq1}]
        {\includegraphics[width=0.495\textwidth]{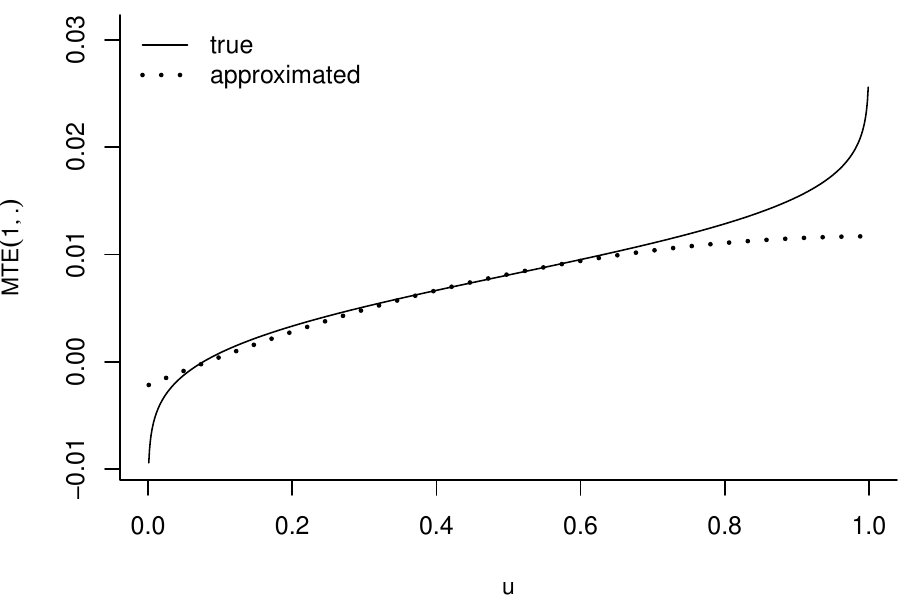}}
    \end{subfloat} \\
    \begin{subfloat}[ $x=2$ \label{fig:mult_seq2}]
        {\includegraphics[width=0.495\textwidth]{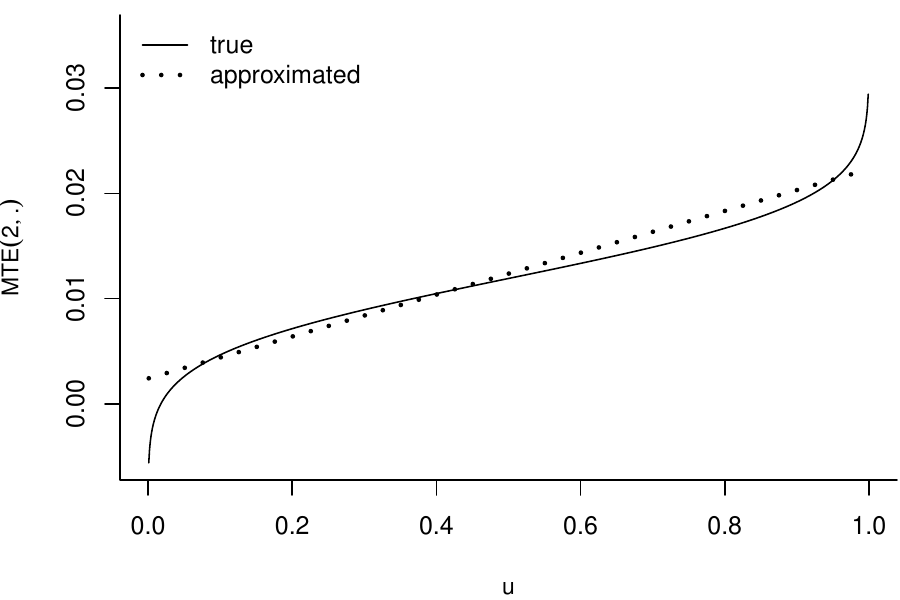}}
    \end{subfloat}
    \begin{subfloat}[ Naive \label{fig:mult_naive}]
        {\includegraphics[width=0.495\textwidth]{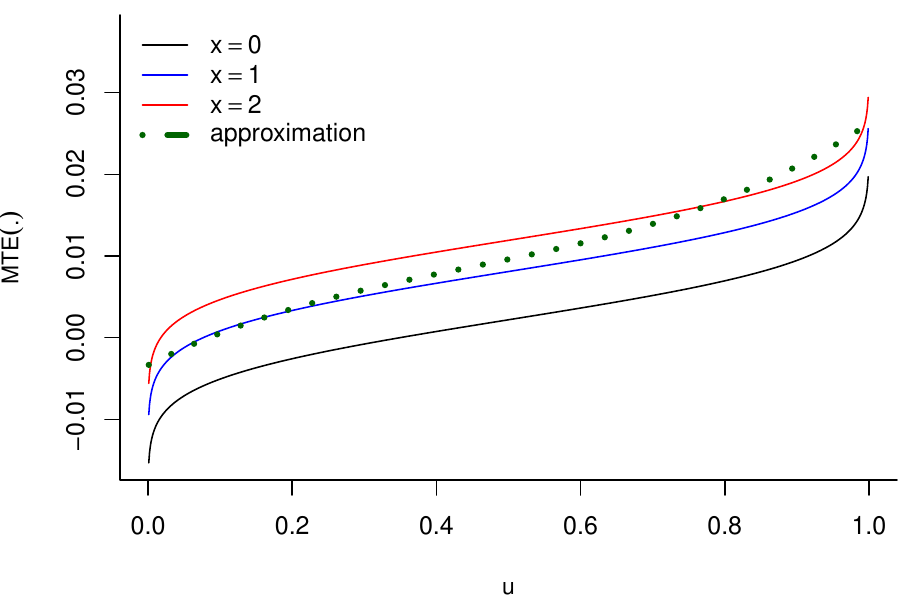}}
    \end{subfloat}
\caption{Approximations to MTE function with multiple exposures}
\floatfoot{\textit{Note:} Figures \ref{fig:mult_seq0}$-$\ref{fig:mult_seq2} plot the true MTE function for each value $x$ with a solid black line and its sequential approximation with a dotted line. Figure \ref{fig:mult_naive} replicates Figure \ref{fig:mte_mult} and adds the naive approximation with a dotted line.}
    \label{fig:app_mult_seq}%
\end{figure}

We now present the results of our calibrated simulations. We separate these results into two groups. First, we show the results from the sequential approach we described above. Second, we present the results from the procedure we described in Section 4, which ignores multiple ad exposures. We refer to this as the ``naive approach.'' All results are shown in Figure \ref{fig:app_mult_seq}.

\paragraph{Sequential approach}\hfill \break
Figures \ref{fig:mult_seq0}$-$\ref{fig:mult_seq2} show the approximations to the MTE function at different values of $x$ following the sequential procedure described above, depicting the MTE function with a solid line and the approximating polynomial with a dotted line. The method is able to approximate the MTE function well across all values of $x$ even though, as discussed above, the degree of the approximating polynomial decreases as $x$ increases.

\paragraph{Naive approach}\hfill \break
Figure \ref{fig:mult_naive} shows the MTE functions at different values of $x$ in solid lines and the naive approximation with a dotted line. This naive approximation, which ignores multiple exposures, resembles an approximation to an ``averaged'' over $x$ MTE function. Because the true function does not vary significantly between $x=1$ and $x=2$, this approximation is close to the true MTE function at these values, and especially at $x=1$. However, the approximation to the MTE function at $x=0$ is arguably poor.

The overall quality of the naive approximation is a reflection of the extent to which multiple exposures affect outcomes. Given our assumed DGP and calibration approach, repeated exposures do not seem to have a sizable impact. Since the outcome variable is an indicator for a purchase, this result is in accordance with results of the extant literature, which we discussed in Section 5.2. However, it is important to note that it is possible to choose a different DGP, also consistent with the observed data, where repeated exposures would have a strong impact on the outcome variable.

\section{Direct application of \cite{bmw2017} with one cell} \label{app:dir_bmw}

The method in \cite{bmw2017} (``BMW'') \textit{can} be applied to data obtained from the typical experimental design. However, because this design yields one-sided noncompliance, $p(0)=0$ while $p(1)\in(0,1)$. Thus, equations (14) and (15) imply that the only moments identified from the data it provides are $\psi_{11}$, $\psi_{01}$, and $\psi_{00}$, where we omit the subscript $c$ to ease notation since there is only one cell. Based on the logic from equations (16) and (17), these moments allow us to approximate $m_1(\cdot)$ with a constant function and $m_0(\cdot)$ with a linear function. Consequently the MTE function itself can be approximated with a linear function.

The ability to approximate the MTE function with a linear function might seem attractive, especially because it is not uncommon to maintain the assumption that the MTE function is indeed linear.\footnote{Examples of studies that maintained this linearity assumption are \cite{olsen1980}, \cite{moffitt2008}, \cite{fs2014}, \cite{bmw2017}, and \cite{kowalski2021}.} Nevertheless, the missingness of $\psi_{10}$ implies that this approximation inherently features a restriction that nontrivially impacts not only the quality of the approximation, but also the structure of the endogeneity of treatment. 

To see this, suppose for simplicity that $m_d(u)=\lambda_{d0} + \lambda_{d1} u$ for $d\in\{0,1\}$. It follows that:
\begin{align*}
    \mathbb{E} \left [Y \middle \vert D=1, Z=z \right ] &=   \lambda_{10} + \frac{\lambda_{11}}{2}p(z) \\
    \mathbb{E} \left [Y \middle \vert D=0, Z=z \right ]  &=  \left ( \lambda_{00} + \frac{\lambda_{01}}{2} \right ) + \frac{\lambda_{01}}{2}p(z)  .
\end{align*}

The data obtained from the typical experimental design thus enable us to recover $\lambda_{00}$ and $\lambda_{01}$. Nevertheless, they do not allow us to recover $\lambda_{10}$ and $\lambda_{11}$ separately because we do not observe $\psi_{10}$. This is an underidentification problem: there are four parameters to be estimated ($\lambda_{10}$, $\lambda_{11}$, $\lambda_{00}$, and $\lambda_{01}$) but only three moments available to estimate such parameters ($\psi_{11}$, $\psi_{01}$, and $\psi_{00}$). To make progress, therefore, the researcher must impose an additional restriction on the parameters.

Unfortunately, we are unaware of any general theory or methodology that can provide clear guidance on such restriction, although progress has been recently made in this direction (e.g., \citealp{kowalski2023}). Whichever restriction is chosen, a key point is that such additional constraint must \textit{always} be imposed to implement BMW's estimator using data collected from an experiment with one-sided noncompliance, which, to our knowledge, has not been noted in the literature. Given the pervasiveness of such experimental designs, we hope our approach provides a valuable solution to recover the MTE function more credibly. 

For the purposes of comparing our approach to a direct application of BMW, we consider four possible restrictions. First, in accordance with the approximation method from Section 3.3, we consider approximating $m_1(\cdot)$ with a constant, that is, imposing that $\lambda_{11}=0$, which enables the estimation of $\lambda_{10}$. From a purely mechanical perspective, the higher $|\lambda_{11}|$ is, the lower the quality of the approximation. However, the constraint $\lambda_{11}=0$ also has a deeper structural implication. It implies that all endogeneity stems from $Y_0$. This rules out certain forms of self-selection, such as the classic case of $D=\mathbb{1} \left \{Y_1 \geq Y_0 \right \}$, where only units that benefit from treatment are treated. In our setting of online advertising, this precludes ``perfect'' ad exposure where units are exposed to the ad only if this benefits the advertiser. 

The second restriction we consider it to set $\lambda_{11}=\lambda_{01}$, thereby ruling out endogeneity altogether but also allowing us to recover $\lambda_{10}$. However, this is often considered implausible, including in the context of online advertising.

Third, we impose $\lambda_{10}=0$, which allows us to recover $\lambda_{11}$. Given linearity, this restriction implies that $m_1(\cdot)$ is either always negative or always positive, which can be justified in cases where $Y_1$ is bounded either from above or below at 0. The same can be achieved for the MTE function by imposing that $\lambda_{10}=\lambda_{00}$, while also enabling the estimation of $\lambda_{11}$.

Table \ref{tab:prof_dgps} also shows the optimal decisions and expected profit losses from using BMW's method directly under each of these constraints. We obtain the decisions using the procedure described in Section 3.4. We consider a sample size of 25,553,093, which corresponds to that of Study 4, use uniform priors, and take 1,000 draws to estimate the posterior means from equation (19).

The results do not show a systematic pattern. Notice that setting $\lambda_{10}=0$ eliminates expected profit losses under DGP 1, which the multi-cell approach also does. 

However, the decisions implied by the different versions of BMW do not vary across DGPs. To the extent that DGPs differ, and thus their implied optimal decisions, so will the performance of these different versions of BMW depending on what the true DGP is. As Figure 5 in the paper shows, the optimal solutions under DGPs 1 and 2 are quite different. Imposing $\lambda_{10}=0$ would yield high losses under DGP 2, as expected. Under such DGP, the best version of the direct application of BMW among the ones we consider is the one that sets $\lambda_{10}=\lambda_{00}$; however, this version, unsurprisingly, performs poorly under DGP 1.

\begin{table}
\resizebox{\textwidth}{!}{\begin{threeparttable}
\caption{Expected profit losses (\%) from different direct applications of BMW and from the multi-cell approach}
\begin{tabular}{c|c|cc|cc|cc|cc|cc}
    \hline \hline 
\multirow{2}{*}{DGP} & \multirow{2}{*}{True $\phi^*$}  &  \multicolumn{2}{c|}{Multi-cell}  &  \multicolumn{2}{c|}{$\lambda_{11}=0$}  & \multicolumn{2}{c|}{$\lambda_{11}=\lambda_{01}$} &  \multicolumn{2}{c|}{$\lambda_{10}=0$}  &  \multicolumn{2}{c}{$\lambda_{10}=\lambda_{00}$}  \\ \cline{3-12}
 &  & $\phi^*$ & Loss & $\phi^*$ & Loss & $\phi^*$ & Loss & $\phi^*$ & Loss & $\phi^*$ & Loss \\
 \hline 
 & & & & & & & & & & &  \\
1 & 1 & 1 & 0 & 0.484 & 72.30 & 0.433 & 79.04 & 1 & 0 & 0.661 & 45.38 \\
 & & & & & & & & & & &  \\
2 & 0.755 & 0.762 & 0.03 & 0.484 & 39.10 & 0.433 & 50.70 & 1 & 58.00 & 0.662 & 5.92 \\
 & & & & & & & & & & &  \\
 \hline \hline
    \end{tabular}
    \begin{tablenotes}
      \footnotesize
      \textit{Note:} The optimal exposure rate, $\phi^*$, is obtained using the procedure described in Section 3.4. We use a sample size of 25,553,093, which corresponds to that of Study 4, uniform priors, and 1,000 MCMC draws.
    \end{tablenotes}
  \label{tab:prof_dgps}
  \end{threeparttable}}
\end{table}

\section{Posterior distribution of approximation parameters}\label{app:post_lambda}

We now describe how to obtain draws from $f\left (\lambda_1,\lambda_0 \middle \vert \text{data} \right )$. To this end, we need to set priors over $\psi$ and $p$, denoted by $q(\psi,p)$, and the likelihood function of the data, $\ell \left (Y,D\middle \vert \mathcal{C}, Z;\psi,p \right )$. We condition on $Z$ and $\mathcal{C}$ because they are randomly chosen. 

We need to consider two cases. First, notice that because $D=0$ when $Z_c=0$ for all $c$, we can pool all observations such that $Z_c=0$ from all cells and use them to obtain the posterior distribution of $\psi_{00}$, as given in equation (12), conditional on the data. More precisely, given a prior distribution $q(\psi_{00})$ and the likelihood $\ell(Y|Z=0;\psi_{00})$, we can derive the posterior $f(\psi_{00}|Y,Z=0)$. The form of this distribution clearly depends on what type of variable $Y$ is. In our application, $Y$ is binary. Hence, for convenience we set:
\begin{align}\label{eq:model_y00}
\begin{split}
 & Y|Z=0;\psi_{00} \sim \text{Bernoulli}\left ( \psi_{00} \right )      \\
 &\psi_{00} \sim \text{Beta}(\alpha_0,\beta_0)
\end{split}
\end{align}
which implies that
\begin{align}\label{eq:post_psi00}
\begin{split}
 &\psi_{00}|Y,Z=0 \sim \text{Beta}\left (\alpha_0 + \sum_{i:Z_{ic}=0} Y_i ,n_0 - \sum_{i:Z_{ic}=0} Y_i + \beta_0 \right ),
\end{split}
\end{align}
where $n_0$ is the total number of ineligible users.

The second case conditions on $\mathcal{C}=c$ and $Z_c=1$. Denote the number of observations in this set by $n_c$. For convenience, we proceed in two steps, relying on the factorization:
\begin{align}\label{eq:likel_factor}
\begin{split}
        \ell \left (Y,D\middle \vert \mathcal{C}=c, Z_c=1;\psi_{11c},\psi_{01c},p(z_c) \right )&=\ell \left (Y\middle \vert D, \mathcal{C}=c, Z_c=1;\psi_{d1c} \right ) \\
        & \qquad \times \ell \left (D\middle \vert \mathcal{C}=c, Z_c=1;p(z_c) \right ).
\end{split}
\end{align}
First, recall that $D$ is binary, so we proceed as above:
\begin{align}\label{eq:model_d1c}
\begin{split}
 & D|Z_c=1,\mathcal{C}=c; p(z_c) \sim \text{Bernoulli}\left (p(z_c) \right )      \\
 & p(z_c) \sim \text{Beta}(\alpha_{Dc},\beta_{Dc}) \\
  &p(z_c)|D,Z_c=1,\mathcal{C}=c \sim \text{Beta}\left (\alpha_{Dc} + \sum_{i:Z_{ic}=1,\mathcal{C}_i=c} D_i ,n_c - \sum_{i:Z_{ic}=1,\mathcal{C}_i=c} D_i + \beta_{Dc} \right ).
\end{split}
\end{align}

The second step consists of obtaining the posterior distribution of $\psi_{d1c}$, as given in equations (10) and (11), conditional on the data. Once again, the form of this distribution clearly depends on what type of variable $Y$ is. Given our application, we proceed as in (\ref{eq:model_y00}) and (\ref{eq:post_psi00}). Denote the number of observations such that $D=d$, $Z_c=1$ and $\mathcal{C}=c$ by $n_{d1c}$. Then:
\begin{align}\label{eq:model_yd1c}
\begin{split}
 & Y|D=d,Z_c=1,\mathcal{C}=c;\psi_{d1c} \sim \text{Bernoulli}\left ( \psi_{d1c} \right )      \\
 &\psi_{d1c} \sim \text{Beta}(\alpha_{d1c},\beta_{d1c}) \\
  &\psi_{d1c}|Y,D=d,Z_c=1,\mathcal{C}=c \sim \text{Beta}\left (\alpha_{d1c} + \sum_{i:D_i=d,Z_{ic}=1,\mathcal{C}_i=c} Y_i ,n_{d1c} - \sum_{ i:D_i=d,Z_{ic}=1,\mathcal{C}_i=c} Y_i + \beta_{d1c} \right ). 
\end{split}
\end{align}

The approach we take is sequential: first, we obtains draws of $p$ and then, conditional on them, we draw $\psi$. Consequently, we avoid the feedback issue studied in \cite{zwywcd2013}.

\section{Distributing experiment budget across multiple cells}\label{app:dist_budget}

We formally describe how the experimenter can distribute the experiment budget across multiple cells to generate the variation necessary to estimate the parameters of interest. 

First, consider a two-cell experiment, and fix $\Pr(\mathcal{C}=1)$ and $\Pr(Z_1=1|\mathcal{C}=1)$. Let $\tau_1$ be the fraction of the original budget, $B$, allocated to Cell 1, so that $B_1=\tau_1 \times B$. The effective budget for Cell 1 is then $\frac{B_1}{\Pr(Z_1=1|\mathcal{C}=1)\times\Pr(\mathcal{C}=1)}=\frac{\tau_1}{\Pr(Z_1=1|\mathcal{C}=1)\times\Pr(\mathcal{C}=1)}\times B$, so that $p(Z_1=1)=\kappa^{-1} \left ( \frac{\tau_1}{\Pr(Z_1=1|\mathcal{C}=1)\times\Pr(\mathcal{C}=1)}\times B \right )$. Equivalently, we then obtain $B_2=(1-\tau_1) \times B$, so that the effective budget for Cell 2 becomes $\frac{B_2}{\Pr(Z_2=1|\mathcal{C}=2)\times(1-\Pr(\mathcal{C}=1))}=\frac{1-\tau_1}{\Pr(Z_2=1|\mathcal{C}=2)\times(1-\Pr(\mathcal{C}=1))}\times B$ and $p(Z_2=1)=\kappa^{-1} \left ( \frac{1-\tau_1}{\Pr(Z_2=1|\mathcal{C}=2)\times(1-\Pr(\mathcal{C}=1))}\times B \right )$. 

Hence, when running a two-cell design, given $B$ and $\kappa(\cdot)$, the experimenter has four decision variables: $\Pr(\mathcal{C}=1)$, $\tau_1$, $\Pr(Z_1=1|\mathcal{C}=1)$, and $\Pr(Z_2=1|\mathcal{C}=2)$. For the desired variation in the propensity score to be generated, it is necessary that $\frac{\tau_1}{\Pr(Z_1=1|\mathcal{C}=1)\times\Pr(\mathcal{C}=1)} \neq \frac{1-\tau_1}{\Pr(Z_2=1|\mathcal{C}=2)\times(1-\Pr(\mathcal{C}=1))}$. Thus, the experimenter can always guarantee that this constraint is satisfied.

This framework can be generalized to $C$ cells in a straightforward manner. In this case, the experimenter has $3C-2$ variables: $\left \{\tau_c, \Pr(\mathcal{C}=c) \right \}_{c=1}^{C-1}$ and $\left \{\Pr(Z_c=1|\mathcal{C}=c) \right \}_{c=1}^{C}$, under the constraints that, for all $c=1,\cdots,C$, $\tau_c \in [0,1]$, $\Pr(Z_c=1|\mathcal{C}=c) \in [0,1]$ and $\Pr(\mathcal{C}=c) \in [0,1]$, plus $\sum_{c=1}^C \tau_c = 1$ and $\sum_{c=1}^C \Pr(\mathcal{C}=c) = 1$. To ensure that the propensity scores differ from one another, it is then required that $\frac{\tau_c}{\Pr(Z_c=1|\mathcal{C}=c)\times\Pr(\mathcal{C}=c)} \neq \frac{\tau_{c^{\prime}}}{\Pr(Z_{c^{\prime}}=1|\mathcal{C}=c^{\prime})\times\Pr(\mathcal{C}=c^{\prime})}$ for all $c \neq c^{\prime}$.

\section{A more complex DGP}\label{app:complex}

We consider a more complex MTE function to dig deeper into the ability of our proposed multi-cell experimental design to provide a good approximation of this function and to inform decision-making. In particular, we assess how the performance of our approach changes as the number of cells increases.

\subsection{New MTE and expected profit functions}

\begin{figure}
    \centering
	\includegraphics[width=0.6\textwidth]{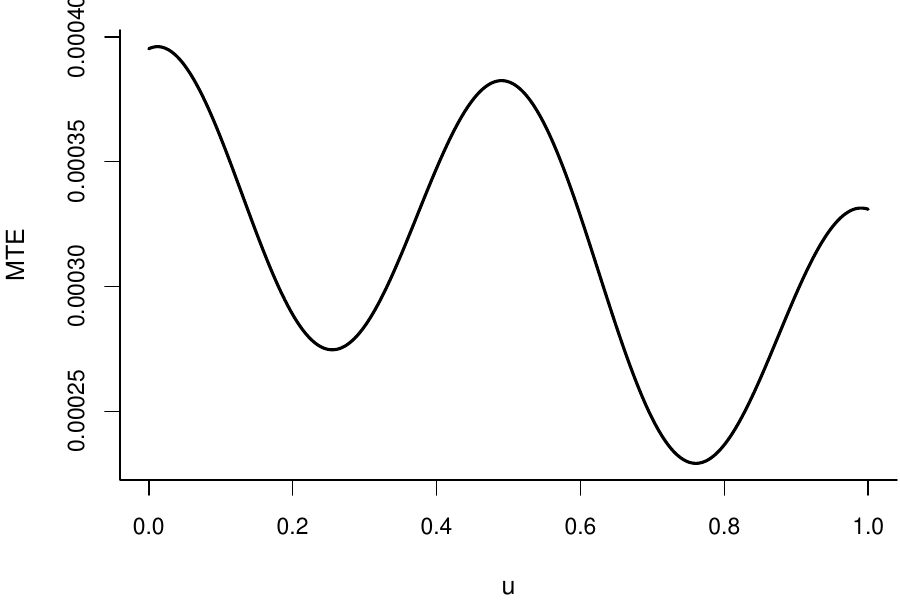}
    \caption{$\text{MTE}(u)=\daleth \frac{1}{1+u} -\gimel\frac{1}{(1+u)^2} - \beth \sin^2(2 \pi u)$}
    \floatfoot{\textit{Note:} }
    \label{fig:complex_mte}
\end{figure}

We choose $m_1(u)=\daleth \frac{1}{1+u}$ and $m_0(u)=\gimel \frac{1}{(1+u)^2} + \beth \sin^2(2 \pi u)$. As before, the parameters $\daleth$, $\beth$, and $\gimel$ are computed to match the observed $\psi_{11}$, $\psi_{01}$, and $\psi_{00}$. Figure \ref{fig:complex_mte} depicts the resulting MTE function. We consider this function to be ``complex'' because is not monotonic, concave, or convex over the entire domain.

\begin{figure}
    \centering
	\includegraphics[width=0.6\textwidth]{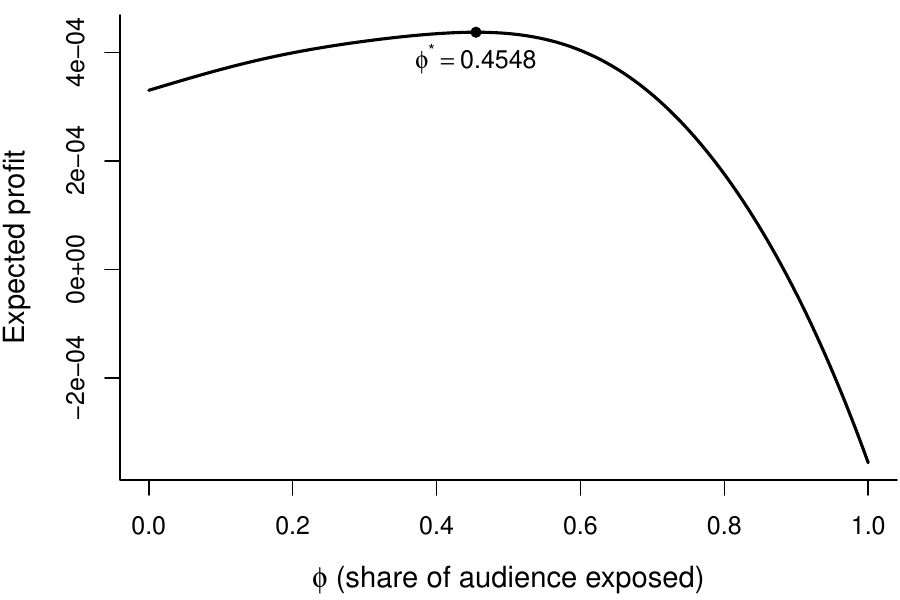}
    \caption{Expected profit function under complex MTE}
    \floatfoot{\textit{Note:} This figure shows the expected profit function implied by the MTE function shown in Figure \ref{fig:complex_mte} combined with the cost function $\kappa(\phi)=0.001\phi^4$.}
    \label{fig:complex_mte_prof}
\end{figure}

Figure \ref{fig:complex_mte_prof} shows the implied expected profit function under our original cost function: $\kappa(\phi)=0.001\phi^4$. Under this DGP, the optimal decision is to treat 45.5\% of the population.

\subsection{Approximations based on different numbers of cells}

\begin{figure}
    \centering
	\includegraphics[width=0.6\textwidth]{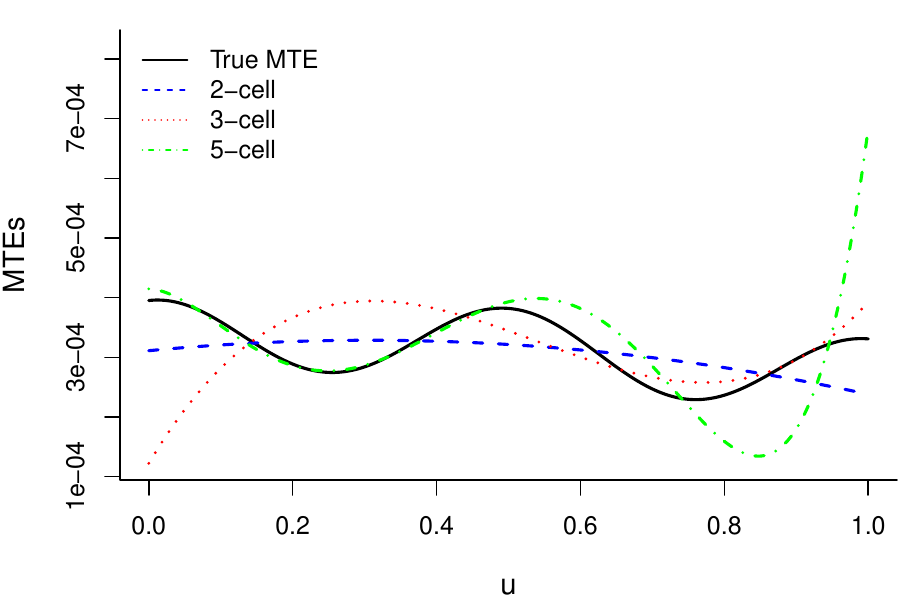}
    \caption{Approximations to complex MTE function}
    \floatfoot{\textit{Note:} This figure shows approximations to the MTE function shown in Figure \ref{fig:complex_mte} for different numbers of cells. The solid black line is the true MTE function, the dashed blue line is the two-cell approximation, the dotted red line is the three-cell approximation, and the dashed-dotted green line is the five-cell approximation.}
    \label{fig:complex_mte_approx}
\end{figure}

We now study how the quality of approximation changes as more cells are included into the experiment. In particular, in addition to the two-cell design, we consider three-cell and five-cell designs, which allow us to obtain cubic and quintic approximations to the MTE function, respectively. Cells 1 and 2 display the same eligibility probabilities and propensity scores as in Section 4. Cell 3 has $\Pr(Z_3=1|\mathcal{C}=3)=0.5$ and $p(Z_3=1)$, chosen so that $\Pr(Z_1=1|\mathcal{C}=1)\times p(Z_1=1)=\Pr(Z_3=1|\mathcal{C}=3)\times p(Z_3=1)$. We set $\Pr(Z_4=1|\mathcal{C}=4)=0.25$ and $\Pr(Z_5=1||\mathcal{C}=5)=0.9$, and set $p(Z_5=1)$ so that $\Pr(Z_1=1|\mathcal{C}=1)\times p(Z_1=1)=\Pr(Z_5=1|\mathcal{C}=5)\times p(Z_5=1)$, obtaining $p(Z_5=1)\approx 0.288$. Finally, we pick $p(Z_4=1)=0.17$ to increase the range of values covered by the propensity scores. Figure \ref{fig:complex_mte_approx} shows the four approximated MTE functions along with the true one.

Seemingly, as the number of cells increases, so does the quality of the approximation. Nevertheless, we note that at the extreme points of the domain these approximations might worsen because of the extrapolation imposed by the polynomial functional form. To quantify and assess its impact, we compute the distance between the true and approximated MTE functions using the sup-norm, $L_2$-norm, and the Relative ATE error. These quantities are shown in Table \ref{tab:ate_complex_prof}, along with the optimal decisions implied by these approximations and the losses in expected profit they imply. 

\begin{table}
\begin{threeparttable}
\caption{Closeness to MTE function and expected profit losses (\%) as a function of the number of cells}
\begin{tabular}{c|ccc|cc}
    \hline \hline 
    Method & sup-norm & $L_2$-norm & Relative ATE error & $\phi^*$ & Loss \\
    \hline 
     & & & & \\
    Two-cell & 0.00009 & 0.00005 & -0.02445  & 0.434 & 0.10  \\
     & & & & \\
     Three-cell & 0.00027 & 0.00008 & 0.00585  & 0.545 & 2.60 \\
      & & & & \\
     Five-cell & 0.00035 & 0.00006 & 0.00144 & 0.696 & 25.51 \\
      & & & & \\
      \hline 
       & & & & \\
      True & ---  & --- & --- &  0.455 & --- \\
      & & & & \\
 \hline \hline
    \end{tabular}
    \begin{tablenotes}
      \footnotesize
      \textit{Note:} This table shows the three distance metrics between the true and the approximated MTE functions given in equation (20). The optimal exposure rate, $\phi^*$, is obtained using the procedure described in Section 3.4. We use a sample size of 25,553,093, which corresponds to that of Study 4, uniform priors, and 1,000 draws to estimate the posterior means from equation (19).
    \end{tablenotes}
  \label{tab:ate_complex_prof}
  \end{threeparttable}
\end{table} 

The different criteria demonstrate that, in this example, no single approximation strictly dominates the others. Interestingly, according to the criteria that measure discrepancy between the entire MTE function and its approximation, the simplest approximation, with only two cells, performs best. In turn, the five-cell approximation is the one whose implied ATE is closest to the true ATE and that implies the smallest loss in expected profit. Caution is required if wishing to generalize these results because we do not know whether the assumed MTE functional form bears any resemblance to one that might be encountered in the real-world.

These results further highlight that, even though they are connected, the tasks of approximating the MTE function and approximating the maximum of the expected profit function are not perfectly aligned. The extent to which these approximations are misaligned depends on the underlying MTE and cost functions.

\section{MTE function under different assumptions}\label{app:mte_assums}

We consider two commonly made assumptions on the DGP to assess whether they impose enough structure on the resulting MTE function to imply specific guidance on how to choose the number of cells or values for the propensity score during the experiment. Under each assumption, we present an example such that no specific guidance is obtained.

\subsection{Monotonic MTE function}

One possible assumption the researcher might be willing to make is that the MTE function is monotonic. This is necessarily satisfied when the MTE function is linear or when the joint distribution of potential outcomes and the selection unobservable is jointly, which are commonly made assumptions. We now provide an example where monotonicity of the MTE function does not necessarily aid in choosing $C$ or $p(Z_c=1)$.

To this end, assume that $\text{MTE}(u)=\frac{2.7}{2+2^{10 u}}$. This function is not only monotonic, but it also is strictly convex and nonnegative. Hence, it is a fairly ``well-behaved'' function.

In particular, the monotonicity might suggest that a linear approximation might be satisfactory. However, \ref{fig:ex_approx} shows that not to be the case because a linear approximation might cover a wide range of negative values, which is a marked difference from the true MTE function. Furthermore, a linear approximation is highly susceptible to values of the propensity score. As Figure \ref{fig:ex_approx} shows, depending on these values the resulting linear curve can have very distinct slopes.

\begin{figure}
\begin{tikzpicture}[scale=6]
  \draw[->,ultra thick] (0, 0) -- (1, 0) node[right] {\large $u$};
  \draw[->,ultra thick] (0, 0) -- (0,1) node[left] {\large MTE};
  \draw[-,ultra thick] (0, 0) -- (0,-.5);
  \draw[-,dotted] (0.1, 0) -- (0.1,2.7/4);
  \draw[-,dotted] (0.4, 0) -- (.4,2.7/18);
  \draw[-,dotted] (.6, 0) -- (.6,2.7/66);
  \draw[-,dotted] (.9, 0) -- (0.9,2.7/514);
  
  \draw[-,  ultra thick, black,domain=0:1]   plot (\x,{ 2.7/(2 + 2^(10*\x))});
  
  \draw[-, dashed, ultra thick, blue,domain=0:1]   plot (\x,{ 2.7/4 -0.2*(2.7/(2^6+2) - 2.7/(2^1+2)) + 2*(2.7/(2^6+2) - 2.7/(2^1+2))*\x });
  
  \draw[-, dotted, ultra thick, red,domain=0:1]   plot (\x,{ 2.7/(2^4+2)- 0.8*(2.7/(2^9+2) - 2.7/(2^4+2)) + 2*(2.7/(2^9+2) - 2.7/(2^4+2))*\x });
  
  \draw[fill,blue] (0.1,2.7/4) circle [radius=0.015];
  \draw[fill,blue] (0.6,2.7/66) circle [radius=0.015];
  
  \draw[fill,red] (0.4,2.7/18) circle [radius=0.015];
  \draw[fill,red] (0.9,2.7/514) circle [radius=0.015];
   
  \end{tikzpicture}
\caption{Different linear approximations for an MTE curve}
\label{fig:ex_approx}
\end{figure}

Given this MTE function, it might seem like having one additional cell might suffice to obtain a satisfactory approximation. Nevertheless, this need not be the case. Figure \ref{fig:ex_approx2} depicts two quadratic approximations to the MTE function. As we can see, neither approximation is particularly good, and one fails to capture the monotonicity of the MTE function. This is a result of the range of values taken by the propensity score, which, in both cases, is limited. 

\begin{figure}
\begin{tikzpicture}[scale=6]
  \draw[->,ultra thick] (0, 0) -- (1, 0) node[right] {\large $u$};
  \draw[->,ultra thick] (0, 0) -- (0,1) node[left] {\large MTE};
  
  \draw[-,  ultra thick, black,domain=0:1]   plot (\x,{ 2.7/(2 + 2^(10*\x))});
  
  \draw[-, dotted, ultra thick, purple,domain=0:1]   plot (\x,{ 0.99 -3.3*\x + 3*(\x^2) });

    \draw[-, dashed, ultra thick, orange,domain=0:1]   plot (\x,{ 0.2354717 -0.4849349*\x + 0.2545958*(\x^2) });
  
  \draw[fill,purple] (0.3,2.7/10) circle [radius=0.015];
    \draw[fill,purple] (0.4,2.7/18) circle [radius=0.015];
      \draw[fill,purple] (0.5,2.7/34) circle [radius=0.015];
  \draw[fill,orange] (0.9,2.7/514) circle [radius=0.015];
  \draw[fill,orange] (0.8,2.7/258) circle [radius=0.015];
  \draw[fill,orange] (0.7,2.7/130) circle [radius=0.015];
   
  \end{tikzpicture}
\caption{Different quadratic approximations for an MTE curve}
\label{fig:ex_approx2}
\end{figure}

On the other hand, as perhaps expected, when the propensity score covers a wider range of values, the quality of the approximation can be higher, as we show in Figure \ref{fig:ex_approx3}. This results echoes the ones from our analysis in Section 4.4. Note, however, that this approximation, unlike the true MTE function, displays negative values.

\begin{figure}
\begin{tikzpicture}[scale=6]
  \draw[->,ultra thick] (0, 0) -- (1, 0) node[right] {\large $u$};
  \draw[->,ultra thick] (0, 0) -- (0,1) node[left] {\large MTE};
  
  \draw[-,  ultra thick, black,domain=0:1]   plot (\x,{ 2.7/(2 + 2^(10*\x))});
  
  \draw[-, dashed, ultra thick, blue,domain=0:1]   plot (\x,{ 0.9053704 -2.4666507*\x + 1.6294668*(\x^2) });
  
  \draw[fill,blue] (.1,2.7/4) circle [radius=0.015];
      \draw[fill,blue] (0.5,2.7/34) circle [radius=0.015];
  \draw[fill,blue] (0.9,2.7/514) circle [radius=0.015];
   
  \end{tikzpicture}
\caption{Different quadratic approximations for an MTE curve}
\label{fig:ex_approx3}
\end{figure}

A priori, it is unclear, however, the curvature of the function, and therefore what the exact range of propensity score values should be. 

\subsection{Monotone treatment response}

A different assumption is that of monotone treatment response \citep{manski1997}. It implies that the treatment effects themselves always have the same sign, which implies that so does the MTE function. However, the previous example suggests that this will not suffice to make this MTE function sufficiently ``well-behaved'' for us to obtain precise guidance for the specific design of the experiment. Indeed, the example below confirms this.

\begin{figure}
    \centering
	\includegraphics[width=0.6\textwidth]{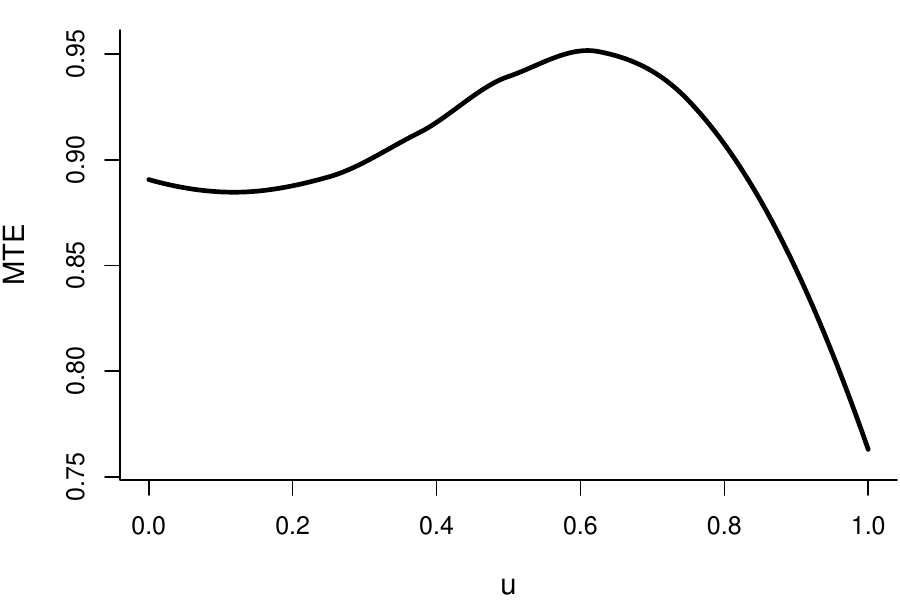}
    \caption{MTE function under monotone treatment response}
    \label{fig:mte_manski}
\end{figure}

Assume that $Y_0 | U=u \sim N \left (0.9u -3.8u^2 + 3.3 u^3,1 \right )$ and $Y_1|Y_0\sim TN \left (0,1,Y_0, +\infty \right)$, so that $Y_1$ follows a standard normal distribution truncated from below at $Y_0$. The resulting MTE function is shown in Figure \ref{fig:mte_manski}. Unsurprisingly, monotone treatment response ensures that the function always has the same sign, but is not even enough to impose, for instance, monotonicty. Consequently, on its own this assumption is not helpful in informing specifically how the experiment should be designed and implemented.

\newpage

\bibliography{mte}

\begin{thebibliography}{}

\bibitem[Baardman et~al., 2019]{bfpp2019}
Baardman, L., Fata, E., Pani, A., and Perakis, G. (2019).
\newblock Learning optimal online advertising portfolios with periodic budgets.
\newblock {\em SSRN Working Paper 3346642}.

\bibitem[Barajas and Bhamidipati, 2021]{bb2021}
Barajas, J. and Bhamidipati, N. (2021).
\newblock Incrementality testing in programmatic advertising: Enhanced
  precision with double-blind designs.
\newblock In Leskovec, J., Grobelnik, M., Najork, M., Tang, J., and Zia, L.,
  editors, {\em Proceedings of the Web Conference 2021 (WWW '21)}, pages
  2818--2827, New York, USA. ACM.

\bibitem[Basu and Batra, 1988]{bb1988}
Basu, A.~K. and Batra, R. (1988).
\newblock {ADSPLIT}: A multi-brand advertising budget allocation model.
\newblock {\em Journal of Advertising}, 17(2):44--51.

\bibitem[Bj\"{o}rklund and Moffitt, 1987]{bm1987}
Bj\"{o}rklund, A. and Moffitt, R. (1987).
\newblock The estimation of wage gains and welfare gains in self-selection
  models.
\newblock {\em Review of Economics and Statistics}, 69(1):42--49.

\bibitem[Brinch et~al., 2017]{bmw2017}
Brinch, C.~N., Mogstad, M., and Wiswall, M. (2017).
\newblock Beyond {LATE} with a discrete instrument.
\newblock {\em Journal of Political Economy}, 125(4):985--1039.

\bibitem[Brodersen et~al., 2015]{bgkrs2015}
Brodersen, K.~H., Gallusser, F., Koehler, J., Remy, N., and Scott, S.~L.
  (2015).
\newblock Inferring causal impact using {B}ayesian structural time-series
  models.
\newblock {\em Annals of Applied Statistics}, 9(1):247--274.

\bibitem[Carneiro et~al., 2010]{chv2010}
Carneiro, P., Heckman, J.~J., and Vytlacil, E. (2010).
\newblock Evaluating marginal policy changes and the average effect of
  treatment for individuals at the margin.
\newblock {\em Econometrica}, 78(1):377--394.

\bibitem[Chetty et~al., 2016]{chk2016}
Chetty, R., Hendren, N., and Katz, L.~F. (2016).
\newblock The effects of exposure to better neighborhoods on children: New
  evidence from the {M}oving to {O}pportunity experiment.
\newblock {\em American Economic Review}, 106(4):855--902.

\bibitem[Cr\'{e}pon et~al., 2015]{cddp2015}
Cr\'{e}pon, B., Devoto, F., Duflo, E., and Parient\'{e}, W. (2015).
\newblock Estimating the impact of microcredit on those who take it up:
  Evidence from a randomized experiment in {M}orocco.
\newblock {\em American Economic Journal: Applied Economics}, 7(1):123--150.

\bibitem[Cr\'{e}pon et~al., 2013]{cdgrt2013}
Cr\'{e}pon, B., Duflo, E., Gurgand, M., Rathelot, R., and Zamora, P. (2013).
\newblock Do labor market policies have displacement effects? {E}vidence from a
  clustered randomized experiment.
\newblock {\em Quarterly Journal of Economics}, 128(2):531--580.

\bibitem[Daljord et~al., 2023]{dmrsy2022}
Daljord, {\O}., Mela, Carl F.~and~Roos, M.~J., Sprigg, J., and Yao, S. (2023).
\newblock The design and targeting of compliance promotions.
\newblock {\em Marketing Science}, 42(5):866--891.

\bibitem[Deng et~al., 2023]{dys2021}
Deng, A., Yuan, L.-H., Kanai, N., and Salama-Manteau, A. (2023).
\newblock Zero to hero: Exploiting null effects to achieve variance reduction
  in experiments with one-sided triggering.
\newblock In Chua, T.-S. and Lauw, H., editors, {\em Proceedings of the 16th
  ACM International Conference on Web Search and Data Mining(WDSM '23)}, pages
  823--831, New York, USA. ACM.

\bibitem[Forbes, 2019]{forbes2019}
Forbes (2019).
\newblock Marketing best practices for frequency capping.

\bibitem[French and Song, 2014]{fs2014}
French, E. and Song, J. (2014).
\newblock The effect of disability insurance receipt on labor supply.
\newblock {\em American Economic Journal: Economic Policy}, 6(2):291--337.

\bibitem[Geng et~al., 2021]{gswznl2021}
Geng, T., Sun, F., Wu, D., Zhou, W., Nair, H.~S., and Lin, Z. (2021).
\newblock Automated bidding and budget optimization for performance advertising
  campaigns.
\newblock {\em SSRN Working Paper 3913039}.

\bibitem[Gordon et~al., 2023]{gordon_moakler_zettelmeyer_2022}
Gordon, B.~R., Moakler, R., and Zettelmeyer, F. (2023).
\newblock Close enough? {A} large-scale exploration of non-experimental
  approaches to advertising measurement.
\newblock {\em Marketing Science}, 42(4):768--793.

\bibitem[Gordon et~al., 2019]{gordon_zettelmeyer_2019}
Gordon, B.~R., Zettelmeyer, F., Bhargava, N., and Chapsky, D. (2019).
\newblock A comparison of approaches to advertising measurement: Evidence from
  big field experiments at {F}acebook.
\newblock {\em Marketing Science}, 38(2):193--225.

\bibitem[Green et~al., 2003]{ggn2003}
Green, D.~P., Gerber, A.~S., and Nickerson, D.~W. (2003).
\newblock Getting out the vote in the local elections: Results from six
  door-to-door canvassing experiments.
\newblock {\em Journal of Politics}, 65(4):653--663.

\bibitem[Gui et~al., 2021]{gnn2021}
Gui, G., Nair, H.~S., and Niu, F. (2021).
\newblock Auction throttling and causal inference of online advertising
  effects.
\newblock {\em arXiv preprint arXiv:2112.15155}.

\bibitem[Heckman et~al., 2001]{htv2001}
Heckman, J.~J., Tobias, J.~L., and Vytlacil, E. (2001).
\newblock Four parameters of interest in the evaluation of social programs.
\newblock {\em Southern Economic Journal}, 68(2):210--223.

\bibitem[Heckman et~al., 2003]{htv2003}
Heckman, J.~J., Tobias, J.~L., and Vytlacil, E. (2003).
\newblock Simple estimators for treatment parameters in a latent-variable
  framework.
\newblock {\em Review of Economics and Statistics}, 85(3):748--755.

\bibitem[Heckman et~al., 2006]{huv2006}
Heckman, J.~J., Urzua, S., and Vytlacil, E. (2006).
\newblock Understanding instrumental variables in models with essential
  heterogeneity.
\newblock {\em Review of Economics and Statistics}, 88(3):389--432.

\bibitem[Heckman and Vytlacil, 2005]{hv2005}
Heckman, J.~J. and Vytlacil, E. (2005).
\newblock Structural equations, treatment effects, and econometric policy
  evaluation.
\newblock {\em Econometrica}, 73(3):669--738.

\bibitem[Heckman and Vytlacil, 2007]{hv2007}
Heckman, J.~J. and Vytlacil, E. (2007).
\newblock Econometric evaluation of social programs, part {II}: Using the
  marginal treatment effect to organize alternative econometric estimators to
  evaluate social programs, and to forecast their effects in new environments.
\newblock In Heckman, J.~J. and Leamer, E.~E., editors, {\em Handbook of
  Econometrics, Vol. 6B}, pages 4875--5143. Elsevier.

\bibitem[Hermle and Martini, 2022]{hm2022}
Hermle, J. and Martini, G. (2022).
\newblock Valid and unobtrusive measurement of returns to advertising through
  asymmetric budget split.
\newblock {\em arXiv preprint arXiv:2207.00206}.

\bibitem[Holthausen~Jr. and Assmus, 1982]{ha1982}
Holthausen~Jr., D.~M. and Assmus, G. (1982).
\newblock Advertising budget allocation under uncertainty.
\newblock {\em Management Science}, 28(5):487--499.

\bibitem[Imbens and Angrist, 1994]{ia1994}
Imbens, G.~W. and Angrist, J.~D. (1994).
\newblock Identification and estimation of local average treatment effects.
\newblock {\em Econometrica}, 62(2):467--475.

\bibitem[Johnson, 2023]{johnson2022}
Johnson, G.~A. (2023).
\newblock Inferno: A guide to field experiments in online display advertising.
\newblock {\em Journal of Economics and Management Strategy}, 32(3):469--490.

\bibitem[Johnson et~al., 2017a]{jln2017a}
Johnson, G.~A., Lewis, R.~A., and Nubbemeyer, E.~I. (2017a).
\newblock The online ad effectiveness funnel \& carryover: Lessons from 432
  field experiments.
\newblock {\em SSRN Working Paper 2701578}.

\bibitem[Johnson et~al., 2016]{jlr2016}
Johnson, G.~A., Lewis, R.~A., and Reiley, D.~H. (2016).
\newblock Location, location, location: Repetition and proximity increase
  advertising effectiveness.
\newblock {\em SSRN Working Paper 2268215}.

\bibitem[Johnson et~al., 2017b]{jlr2017}
Johnson, G.~A., Lewis, R.~A., and Reiley, D.~H. (2017b).
\newblock When less is more: Data and power in advertising experiments.
\newblock {\em Marketing Science}, 36(1):43--53.

\bibitem[Kitagawa and Tetenov, 2018]{kt2018}
Kitagawa, T. and Tetenov, A. (2018).
\newblock Who should be treated? {E}mpirical welfare maximization methods for
  treatment choice.
\newblock {\em Econometrica}, 86(2):591--616.

\bibitem[Kowalski, 2023a]{kowalski2023}
Kowalski, A.~E. (2023a).
\newblock Behavior within a clinical trial and implications for mammography
  guidelines.
\newblock {\em Review of Economic Studies}, 90(1):432--462.

\bibitem[Kowalski, 2023b]{kowalski2021}
Kowalski, A.~E. (2023b).
\newblock Reconciling seemingly contradictory results from the {O}regon
  {H}ealth {I}nsurance {E}xperiment and the {M}assachussets {H}ealth {R}eform.
\newblock {\em Review of Economics and Statistics}, 105(3):646--664.

\bibitem[Lewis, 2014]{lewis2014}
Lewis, R.~A. (2014).
\newblock Worn-out or just getting started: The impact of frequency in online
  advertising.
\newblock {\em Mimeo}.

\bibitem[Lewis and Reiley, 2014]{lr2014}
Lewis, R.~A. and Reiley, D.~H. (2014).
\newblock Online ads and offline sales: Measuring the effect of retail
  advertising via a controlled experiment on {Y}ahoo!
\newblock {\em Quantitative Marketing and Economics}, 12(3):43--53.

\bibitem[Lewis and Wong, 2018]{lewis_wong_2018}
Lewis, R.~A. and Wong, J. (2018).
\newblock Incrementality bidding \& attribution.

\bibitem[Manski, 1997]{manski1997}
Manski, C.~F. (1997).
\newblock Monotone treatment response.
\newblock {\em Econometrica}, 65(6):1311--1334.

\bibitem[Mealli et~al., 2004]{mifb2004}
Mealli, F., Imbens, G.~W., Ferro, S., and Biggeri, A. (2004).
\newblock Analyzing a randomized trial on breast self-examination with
  noncompliance and missing outcomes.
\newblock {\em Biostatistics}, 59(2):207--222.

\bibitem[Miguel and Kremer, 2004]{mk2004}
Miguel, E. and Kremer, M. (2004).
\newblock Worms: Identifying impacts on education and health in the presence of
  treatment externalities.
\newblock {\em Econometrica}, 72(1):159--217.

\bibitem[Moffitt, 2008]{moffitt2008}
Moffitt, R. (2008).
\newblock Estimating marginal treatment effects in heterogeneous populations.
\newblock {\em Annales d'\'{E}conomie et Statistique}, 91/92:239--261.

\bibitem[Mogstad et~al., 2018]{mst2018}
Mogstad, M., Santos, A., and Torgovitsky, A. (2018).
\newblock Using instrumental variables for inference about policy relevant
  treatment parameters.
\newblock {\em Econometrica}, 86(5):1589--1619.

\bibitem[Mogstad and Torgovitsky, 2018]{mt2018}
Mogstad, M. and Torgovitsky, A. (2018).
\newblock Identification and extrapolation of causal effects with instrumental
  variables.
\newblock {\em Annual Review of Economics}, 10:577--613.

\bibitem[Olsen, 1980]{olsen1980}
Olsen, R.~J. (1980).
\newblock A least squares correction for selectivity bias.
\newblock {\em Econometrica}, 48(7):1815--1820.

\bibitem[Pani et~al., 2017]{prs2017}
Pani, A.~A., Raghavan, S., and Sahin, M. (2017).
\newblock Large-scake advertising portfolio optimization in online marketing.

\bibitem[Sahni, 2015]{sahni2015}
Sahni, N.~S. (2015).
\newblock Effect of temporal spacing between advertising exposures: Evidence
  from online field experiments.
\newblock {\em Quantitative Marketing and Economics}, 13(3):203--247.

\bibitem[Sahni et~al., 2019]{snk2019}
Sahni, N.~S., Narayanan, S., and Kalyanam, K. (2019).
\newblock An experimental investigation of the effects of retargeted
  advertising: The role of frequency and timing.
\newblock {\em Journal of Marketing Research}, 56(3):401--418.

\bibitem[Sasaki and Ura, 2020]{su2020}
Sasaki, Y. and Ura, T. (2020).
\newblock Welfare analysis via marginal treatment effects.
\newblock {\em arXiv preprint arXiv:2012.07624}.

\bibitem[Schochet et~al., 2008]{sbm2008}
Schochet, P.~Z., Burghardt, J., and Sheena, M. (2008).
\newblock Does {J}ob {C}orps work? {I}mpact findings from the {N}ational {J}ob
  {C}orps {S}tudy.
\newblock {\em American Economic Review}, 98(5):855--902.

\bibitem[Sethi, 1977]{sethi1977}
Sethi, S.~P. (1977).
\newblock Optimal advertising for the {N}erlove-{A}rrow model under a budget
  constraint.
\newblock {\em Journal of the Operational Research Society}, 3(28):683--693.

\bibitem[Simon, 1982]{simon1982}
Simon, H. (1982).
\newblock {ADPULS}: An advertising model with wearout and pulsation.
\newblock {\em Journal of Marketing Research}, 19(3):352--363.

\bibitem[Sommer and Zeger, 1991]{sz1991}
Sommer, A. and Zeger, S.~L. (1991).
\newblock On estimating efficacy from clinical trials.
\newblock {\em Statistics in Medicine}, 10(1):45--52.

\bibitem[Vytlacil, 2002]{vytlacil2002}
Vytlacil, E. (2002).
\newblock Independence, monotonicity, and latent index models: An equivalence
  result.
\newblock {\em Econometrica}, 70(1):331--341.

\bibitem[Waisman et~al., 2024a]{wnc2022}
Waisman, C., , Nair, H.~S., and Carrion, C. (2024a).
\newblock Online causal inference for advertising in real-time bidding
  auctions.
\newblock {\em arXiv preprint arXiv:1908.08600}.

\bibitem[Waisman et~al., 2024b]{wsnl2022}
Waisman, C., Sahni, N.~S., Nair, H.~S., and Lin, X. (2024b).
\newblock Parallel experimentation on advertising platforms.
\newblock {\em arXiv preprint arXiv:1903.11198}.

\bibitem[Yuan et~al., 2013]{ywz2013}
Yuan, S., Wang, J., and Zhao, X. (2013).
\newblock Real-time bidding for online advertising: Measurement and analysis.
\newblock In Saka, E., Shen, D., Gao, B., Yan, J., and Li, Y., editors, {\em
  Proceedings of the 7th ACM ADKDD International Workshop on Data Mining for
  Online Advertising (ADKDD '13)}, pages 3:1--8, New York, USA. ACM.

\bibitem[Zhao et~al., 2019]{zhyzxy2019}
Zhao, K., Hua, J., Yan, L., Zhang, Q., Xu, H., and Yang, C. (2019).
\newblock A unified framework for marketing budget allocation.
\newblock In Teredesai, A. and Kumar, V., editors, {\em Proceedings of the 25th
  ACM SIGKDD International Conference on Knowledge Discovery \& Data Mining(KDD
  '19)}, pages 1820--1830, New York, USA. ACM.

\bibitem[Zigler et~al., 2013]{zwywcd2013}
Zigler, C.~M., Watts, K., Yeh, R.~W., Wang, Y., Coull, B.~A., and Dominici, F.
  (2013).
\newblock Model feedback in {B}ayesian propensity score estimation.
\newblock {\em Biometrics}, 69(1):263--273.

\end{thebibliography}

\end{document}